\documentclass[a4paper,11pt]{article}
\usepackage{jheppub} 
\usepackage{lineno}
\usepackage{amsmath}
\usepackage{comment}
\usepackage[x11names]{xcolor}
\usepackage{graphicx}
\usepackage[export]{adjustbox}
\graphicspath{ {./images/} }
\usepackage{hyperref}
\hypersetup{colorlinks=true, linkcolor=blue, filecolor=magenta, urlcolor=cyan,}
\usepackage{bbold}
\usepackage{tikz}
\usetikzlibrary{decorations.markings}
\usetikzlibrary{decorations.pathmorphing}
\usetikzlibrary{decorations.markings}
\usetikzlibrary{decorations.pathmorphing}
\usetikzlibrary{patterns.meta}
\usepackage[utf8]{inputenc}
    \usepackage{tikz}
    \usepackage{sectsty}
    \usepackage{titlesec}
    \usetikzlibrary{patterns}

\preprint{TIFR/TH/24-25}
\title{\boldmath Grey Galaxies in $AdS_5$}

\author[a]{Kabir Bajaj,\note{kabir.bajaj@iitb.ac.in}}
\author[b,2]{Vipul Kumar,\note{vipul.kumar@tifr.res.in}}
\author[b,3]{Shiraz Minwalla,\note{minwalla.theory@tifr.res.in}}
\author[b,4]{Jyotirmoy Mukherjee,\note{jyotirmoy.mukherjee\_119@tifr.res.in}}

\author[b,5]{Asikur Rahaman.\note{asikur.rahaman@tifr.res.in}}
\affiliation[a]{Department of Physics,
Indian Institute of Technology Bombay, Powai, Mumbai 400076, India}
\affiliation[b]{Department of Theoretical Physics,
Tata Institute of Fundamental Research, Homi Bhabha Rd, Mumbai 400005, India}

\abstract{It has recently been conjectured
\cite{Kim:2023sig} that the end point of the rotational superradiant instability of black holes in $AdS_4$  is a Grey Galaxy: an $\omega=1$ black hole sitting at the centre of $AdS_4$, surrounded by a large disk of rapidly rotating gravitons and other bulk fields. In this paper we study Grey Galaxies in $AdS_5$. In this case, the rotational group is of rank 2, and so has two distinct angular velocities $\omega_1$ 
and $\omega_2$. We demonstrate that $AdS_5$ hosts two qualitatively distinct Grey Galaxy phases: the first with either  $\omega_1\approx 1$ or $\omega_2\approx 1$,  and the second with both angular velocities $\approx 1$. We use these results to present a conjecture for a part of the phase diagram of ${\cal N}=4$ Yang-Mills (as a function of energy and the two angular momenta) that displays several phase transitions between regular black holes and various Grey Galaxy phases. We present an explicit gravitational construction of the phases in which $\omega_1$ and $\omega_2$ are both parametrically close to unity, and demonstrate that the corresponding boundary stress tensor is the sum of two pieces. The first is the stress tensor of the central black hole. The second  - the contribution of the bulk gas - has the spatial distribution of the stress tensor of an equilibrated boundary conformal fluid,  rotating at the given angular speeds $\omega_i$. We also briefly comment on the structure of Grey Galaxies in  $AdS_D$ for $D > 5$.}

\begin{document}
\maketitle
\flushbottom

\newpage
\let\svthefootnote\thefootnote
\newcommand\blfootnotetext[1]{%
  \let\thefootnote\relax\footnote{#1}%
  \addtocounter{footnote}{-1}%
  \let\thefootnote\svthefootnote%
}

\let\svfootnotetext\footnotetext
\renewcommand\footnotetext[2][?]{%
  \if\relax#1\relax%
    \ifnum\value{footnote}=0\blfootnotetext{#2}\else\svfootnotetext{#2}\fi%
  \else%
    \if?#1\ifnum\value{footnote}=0\blfootnotetext{#2}\else\svfootnotetext{#2}\fi%
    \else\svfootnotetext[#1]{#2\fi%
  \fi
}
}

\section{Introduction}

It has been known for over twenty years that spinning black holes with angular velocities greater than unity are unstable in $AdS$ space \cite{Cardoso:2004hs, Dias:2015rxy, Ishii:2018oms, Chesler:2018txn, Ishii:2020muv, Chesler:2021ehz, Kunduri:2006qa, Murata:2008xr,Cardoso:2013pza}. The endpoint of this instability has recently been conjectured \cite{Kim:2023sig} to be a Grey Galaxy: an $\omega=1$ AdS black hole surrounded by a gas of fast rotating gravitons and other bulk fields, with a total energy of order $1/G_N$ ($G_N$ is Newton's constant). The authors of \cite{Kim:2023sig} focused on the study of $AdS_4$ and demonstrated that the graviton gas arranges itself into a large flat disk with a radius of order $1/G_N^\frac{1}{4}$ and a proper thickness of order unity. They also demonstrated that the gas contribution to the boundary stress tensor takes a striking form; it is a delta function, localized on the equator of the boundary $S^2$. \footnote{In contrast, the boundary stress tensor from the black hole part of the Grey Galaxy is smoothly distributed over the boundary $S^2$. }

Spinning $AdS_D$ black holes with $\omega>1$ are also unstable for $D>4$. The thermodynamical arguments of \cite{Kim:2023sig} suggest that Grey Galaxies also constitute the end point of this instability - and dominate the microcanonical ensemble at large angular momentum. Grey Galaxies in $D>4$, thus, likely compute the entropy (as a function of energy and angular momentum, at large angular momentum) in much studied theories like ${\cal N}=4$ Yang-Mills and the $(0,2)$ theory, and so are of great interest.

\subsection{Grey Galaxies of higher rank} \label{gghr}

When $D\geq 5$ the 
rotational group $SO(D-1)$ has  
\footnote{$SO(D-1)$ is the maximal compact subgroup of the conformal group $SO(D-1,2)$.}
rank greater than unity. Consequently,  black holes in such dimensions 
are characterized by 
multiple angular momenta $J_i$ and multiple angular velocities $\omega_i$. Grey Galaxies are made up of a black hole with $|\omega_i| \approx 1$ \footnote{Here the symbol $\approx$ means `parametrically near to'.} for at least one value of $i$.\footnote{All other $|\omega_j|$ are necessarily less than unity.} In $D \geq 5$ these solutions appear in several distinct families, parameterized by the number of $i$ for which  $|\omega_i|\approx 1$.
The chiral gas of gravitons in a Grey Galaxy carries angular momentum only in those two planes that have $|\omega_i| \approx 1$; consequently, the rank $r$ of the $SO(D-1)$ angular momentum matrix of the gas is generically twice this number. Through this paper, 
we label distinct Grey Galaxy phases with the rank $r$ of the angular momentum of their gas component.

$AdS_D$ Grey Galaxies of rank 2 turn out to be qualitatively very similar to their 4-dimensional cousins. The bulk gas in these solutions is sharply localized around a two-dimensional disk of radius of order $1/G_N^\frac{1}{4}$. And its contribution to the boundary stress tensor is sharply localized around an equator of $S^{D-2}$.  

In contrast, both the bulk gas and its boundary stress tensor are less sharply localized in Grey Galaxies of rank $r >2$. Instead, the bulk gas is concentrated around an $r$ dimensional surface\footnote{The `radius' of this region turns out to be of order $(1/G_N)^{\frac{1}{r + 2}}$.} while its boundary stress tensor is concentrated round an $r-1$ dimensional submanifold of $S^{D-2}$. This difference is most pronounced when $D$ is odd and $r$ takes its maximal value $D-1$; in this case, the gas is smoothly distributed over the full spatial bulk, and its boundary stress tensor is smoothly distributed over the full boundary sphere.

Grey Galaxy solutions of rank $r$ 
enjoy invariance under $U(1)^\frac{r}{2}$ rotations in each of their $\frac{r}{2}$ two planes. Thus while  their  boundary stress tensor is spread over a $r-1$ dimensional submanifold, it is a nontrivial function of only 
$(r-1)-(\frac{r}{2}) =\frac{r}{2}-1$ coordinates on this surface.
When $r=2$, $\frac{r}{2}-1= 0$, so the boundary stress tensor is 
distributed uniformly around a one-dimensional equator (as in \cite{Kim:2023sig}). When $r>2$, however, $\frac{r}{2}-1>0$, and the distribution of the boundary stress tensor along these $\frac{r}{2}-1>0$ coordinates is not determined by symmetry considerations. A prediction for the precise form of the spatial distribution of the boundary stress tensor of Grey Galaxies 
with rank $r>2$  is one of the key goals of the current paper. 

\subsection{Microcanonical Phase diagram in $AdS_5$}

 The simplest $AdS_D$ space with $D>4$ is $AdS_5$. In this case, Grey Galaxies are either of 
\begin{itemize}
	\item rank 2 (when the central black hole has either $\omega_1\approx 1$ with $\omega_2<1$ or $\omega_2\approx 1$ with $\omega_1<1$ \footnote{In this case the gas carries a macroscopic amount of angular momentum only in the plane in which $\omega\approx 1$.}).
	\item rank 4 (when the central black hole has  $(\omega_1 , \omega_2) \approx 1$.\footnote{In this case the gas carries macroscopic angular momentum in both planes})
\end{itemize}

It seems likely \footnote{See \ref{glins} for some discussion of this assumption.} that the thermodynamically dominant bulk phase - at any value of $E$, $J_1$ and $J_2$ - is always either a vacuum black hole (see subsection \ref{adskerr} for a review) or one of the Grey Galaxies described above.  Making this assumption, in section \ref{pd} we construct a fully quantitative\footnote{Our phase diagram is quantitative, in the sense that we have precise equations for each of the phase boundaries, and an equation for the entropy as a function of energy and angular momenta in each phase.} phase diagram of ${\cal N}=4$ Yang-Mills theory as a function of energy and angular momenta in the microcanonical ensemble. Our phase diagram exhibits several phase transitions. At high enough energies (at every value of angular momenta) we find in section \ref{pd} that the dominant phase is always the usual Kerr black hole. If we lower energies at fixed $J_i$ with $J_i >J_j$, the system undergoes a phase transition to a
 rank 2 with $\omega_i \approx 1$ at a critical energy. At a lower critical energy, the system then makes a second phase transition to the rank 4 Grey Galaxy case and stays in this phase all the way down to the unitarity bound. In the special case $J_1=J_2$, the system undergoes only one phase transition, directly from the vacuum black hole to the rank 4 Grey Galaxy phase, and stays in this phase down to the unitarity bound.\footnote{In qualitative terms the phase diagram is similar (but simpler than) the DDBH phase diagram for ${\cal N}=4$ Yang-Mills theory as a function of energy and the three $SO(6)$ charges, see \cite{Choi:2024xnv}.}

 The analysis of section \ref{pd} also applies to $AdS_5 \times S^5$ (the bulk dual to ${\cal N}=4$ Yang-Mills theory at large $N$ and strong coupling) except for one complication \footnote{We thank E. Lee for a very useful discussion on this point.} . Small 5d black holes in $AdS_5 \times S^5$ are unstable to a Gregory Laflamme clumping in the $S^5$ direction \cite{Hubeny:2002xn}. The analysis of this paper is blind to the $S^5$, and so to these new phases.
 Once this effect is taken into account, we expect the `high energy' (compared to a critical value, of order unity in units of $N^2$) part of the  ${\cal N}=4$ Yang Mills phase diagram to follow the analysis of this paper. At lower energies the dominant solutions will be given by 10 dimensional black holes. \footnote{This will certainly be the case when the black holes in the dominant phase of this paper turn out to be  $S^5$ Gregory Laflamme unstable. The Grey Galaxy phase extends all the way down to this instability curve if the phase transition to localized black holes is of second order. If this phase transition is of first order, on the other hand, 10d black holes will dominate the ensemble even before the dominant 5d (Grey Galaxy) black holes go unstable.} We leave a careful demarcation of these two different families of phases (and a detailed study of the clumped phases) to future work.   
 
\subsection{Rank 2 Grey Galaxies in $AdS_5$}

The metric at the boundary of $AdS_5$ is Weyl equivalent to $S^3 \times $ time. It is useful to picture the $S^3$ as embedded in an $\mathbb{R}^4$ with coordinates $x_i$ $(i=1\ldots 4)$. Rank 2 Grey Galaxies carry an angular momentum in a single plane - let's say the $x_1 x_2$ plane of this $\mathbb{R}^4$. As in $AdS_4$ 
(see \cite{Kim:2023sig}) the bulk centrifugal force flattens rank 2 Grey Galaxies into disks that are tightly localized around the plane $x_3=x_4=0$. The boundary stress tensor of this Grey Galaxy is (parametrically sharply) localized about the equator on which this plane intersects the boundary $S^3$. If we let $\theta$ be a coordinate that vanishes on the equator, then the boundary stress tensor of the gas component of a Grey Galaxy takes the form $ A \delta(\theta)  (dt-d\phi)^2$ (note the similarity with Eq 5.33 of \cite{Kim:2023sig}) where $A$ is a constant \footnote{That $A$ is a constant along the equator follows from the $U(1)$ invariance of the solution, as explained at the end of subsection \ref{gghr}} whose value can be read off from gas thermodynamics (see section \ref{pf}). As anticipated at the end of subsection \ref{gghr}, it follows that the boundary stress tensor of rank 2 Grey Galaxies can be determined by simple thermodynamical considerations,   without a detailed bulk construction of the relevant solution.

\subsection{A Bulk Solution for a rank 4 Grey Galaxy} \label{absagg}

In contrast (and as already mentioned at the end of subsection \ref{gghr}), the boundary stress tensor of a rank 4 Grey Galaxy is not immediately determined directly from thermodynamical and symmetry considerations alone. In order to evaluate this boundary stress tensor, we have determined its bulk solution. In this subsection, we describe this solution within a simple model. We take our bulk to be $AdS_5 \times S^5$, and take the bulk matter to consist of a single 10d massless scalar field (let us call it the dilaton). We discuss the generalization to models with more realistic bulk matter in the next subsection. 

Our construction of the bulk solution follows the method presented in \cite{Kim:2023sig}. We first calculate the bulk stress tensor of the bulk gas and then use Einstein equations to evaluate the backreaction of this stress tensor on the metric. We now describe each of these steps in more detail. 

At radial coordinates $r \gg 1$, the black hole spacetime is indistinguishable from thermal $AdS_5 \times S^5$. As the dominant contribution to the bulk stress tensor comes at large values of the radius, we work in this simpler background spacetime.
We compute the bulk stress tensor by first taking the appropriate derivatives of the thermal dilaton propagator and then taking the coincident limit.\footnote{See Section 4.3 of \cite{Kim:2023sig}. As explained in \cite{Kim:2023sig} it is sufficient to work at quadratic when studying Grey Galaxies, as terms in the stress tensor that are of higher homogeneity than quadratic are subleading in inverse powers of $1/N$.}   The thermal 
$AdS_5 \times S^5$ Greens function is constructed from the vacuum propagator via the method of images. Happily, this vacuum propagator turns out to be extremely simple; it is simply given by a Weyl transformation of the flat space 10d massless scalar propagator \cite{Dorn:2003au,Dai:2009zg}. The train of facts mentioned above allows for a simple computation of the dilaton contribution to the bulk gas stress tensor.\footnote{Though the final result for this bulk stress tensor is the sum over the stress tensor, the five-dimensional Kaluza Klein modes of the dilaton, the full answer obtained from the sum turns out to be much simpler than any individual component.} 
Quite remarkably, this bulk stress tensor turns out to take the form expected of an equilibrated perfect fluid in the bulk (see \S \ref{impbulk}). This is surprising because we are working at temperatures that are not necessarily large, so one would not, naively, have expected a hydrodynamical description of the bulk gas to hold.

The gravitational backreaction caused by this bulk gas can also be computed rather simply.\footnote{The computation presented here is, in fact,  significantly simpler than the computation presented in \cite{Kim:2023sig}, see below.} The stress tensor of the bulk gas of the previous paragraph turns out to live dominantly at an $AdS_5$ radial coordinate of order 
$r=N^\frac{1}{3}$. The `scale transformation' coordinate change   
\begin{equation}\label{sctcc}
 r \rightarrow  r' N^{1/3}, ~~~~x^\mu \rightarrow \frac{x^{'\mu}}{N^{1/3}}
\end{equation} 
brings the gas to $r'$ of order unity, but has it varying (in the coordinates $x^{'\mu}$) over parametrically large scales of order  $N^\frac{1}{3}$ \footnote{This follows because the gas varies over 
distances of order unity in the original $x^\mu$ coordinates.}.  To leading order in the large $N$ limit, therefore, the spatial variation can simply be ignored:\footnote{In the rank 2 case, in contrast, the bulk gas varies very rapidly in the original $x^\mu$ coordinates, in such a manner that it varies on scale unity in the $x^{'\mu}$ coordinates. In this case, the back reaction cannot be computed  `point by point', but must be computed in a more complicated manner, see \cite{Kim:2023sig}.}
one computes the gravitational backreaction `point by point' (in $x^\mu$), and finds the final solution by sewing together these point-wise solutions. \footnote{ This is highly reminiscent of the derivation of the  Fluid Gravity correspondence presented in \cite{Bhattacharyya:2007vs,Bhattacharyya:2008ji, Bhattacharyya:2008mz,Bhattacharyya:2008xc}. The analogy is quite close: in fact the equation that appears in the computation of this paper is almost identical to the differential equation in the `tensor sector' of \cite{Bhattacharyya_2008}.}

Once we have the bulk solution, the determination of its boundary stress tensor is a simple exercise. We find 
\begin{equation}\label{bcssi}
T_{\mu\nu}= \frac{ 2 h_{\phi}(\beta)}{ \beta \pi^2}  \gamma^4(\theta) \left( 4 u_\mu u_\nu + g_{\mu \nu} \right)
\end{equation} 
where $g_{\mu\nu}$ is the boundary metric, $u^\mu$ is a `velocity' vector field on this space, and  $\gamma(\theta)$ is the usual special relativistic `gamma factor' for this velocity field 
\begin{equation}\label{bmet}
\begin{split}
ds^2&=g_{\mu\nu} dx^\mu dx^\nu = - dt^2 +  d \theta^2 + \sin^2 \theta  d \phi_1^2 + \cos^2 \theta d \phi_2^2 \\
\end{split}
\end{equation}
\begin{equation} \label{umusp}
\begin{split}
u^\mu \partial_\mu &= \gamma \left( \partial_t - \omega_1 \partial_{\phi_1} - \omega_2 \partial_{\phi_2} \right)\\
\end{split}
\end{equation}
\begin{equation}\label{gammafac}
    \begin{split}
    \gamma(\theta)&= \frac{1}{\sqrt{ 1- \omega_1^2 \sin^2 \theta -\omega_2^2 \cos^2 \theta}}\\
    \end{split}
\end{equation}
 \footnote{Note that $\gamma(\theta)$ is defined in a manner that ensures that $u$ is a velocity vector field, i.e. that $u^2=-1$.}
The function $h_{\phi}(\beta)$ in  \eqref{bcssi} is thermodynamical in origin. This function may be read off from the expression of the partition function of the dilaton gas in $AdS_5\times S_5$. When $\omega_1\approx 1$ and $\omega_2 \approx 1$, this partition function turns out to take the form
\begin{equation}\label{tpbg}
Z_{\rm gas}(\beta, \omega_i) = \frac{ 4h_{\phi}(\beta)}{(1-\omega_1^2) (1-\omega_2^2)} \end{equation} 

so $h_\phi(\beta)$ may be read off from \eqref{tpbg}.

While the stress tensor \eqref{bcssi} has no explicit dependence on $N$,  \footnote{The function $h_{\phi}(\beta)$ (see \eqref{hphi} for an explicit formula) is an order unity function of the temperature.} it nonetheless evaluates to a value of order $N^2$ when $\omega_1$ and $\omega_2$ differ from unity only at order $1/N^\frac{2}{3}$, \footnote{This follows from the fact $\gamma$ evaluates to an expression of order $N^\frac{1}{3}$ at such values of $\omega_i$.} as is the case for Grey Galaxies of rank 4.

\subsection{Fluid Interpretation of the Boundary Stress Tensor}

Equation \eqref{bcssi} has a striking physical interpretation: it is the equilibrium stress tensor of a `conformal fluid' rotating on $S^5$ with fluid velocity $u^\mu$ \cite{Bhattacharyya:2007vs}. 
The reader may find the emergence of a hydrodynamical stress tensor at every $\beta$ surprising, as one usually expects hydrodynamics to be quantitatively accurate only at high temperatures in units of the radius of the sphere. In Appendix \ref{s1s3}, we investigate this point by directly evaluating the stress tensor of a free conformal scalar on $S^3\times S^1$ at nonzero angular velocities. We verify that while the stress tensor does indeed take the completely hydrodynamical form 
\eqref{bcssi} at high temperatures, it also takes the form 
\eqref{bcssi} in the limit $\omega_i \to 1$. \footnote{The main difference between the high temperature and $\omega$ near one limits lies in the analogue of the function $h_{\phi}(\beta)$. While this function is proportional the simple 
$\frac{1}{\beta^4}$ at high temperatures, it has a complicated dependence on $\beta$ in the limit $\omega \rightarrow 1$. }
We offer a physical explanation for why this works (in terms of Lorentz contractions, or, equivalently, redshift factors). 

While we have honestly derived \eqref{bcssi} only in the context of a simple model of bulk matter (consisting only of a single dilaton), the physical interpretation above suggests that the structural properties of our result are, in fact, universal. We are led to conjecture: 
\begin{itemize}
\item The boundary stress tensor of Grey Galaxies of rank 4 in $AdS_5$ - with arbitrary bulk matter content - always takes the form \eqref{bcssi}, but with the function $h_\phi(\beta)$ replaced by the function that is read off from the gas via the analogue of 
\eqref{tpbg}.
\end{itemize}

Early in this introduction, we pointed out that the boundary stress tensor of rank 2 Grey Galaxies is completely fixed by
symmetry and {\it thermodynamical} considerations.
The conjecture of this section postulates a generalization of this fact to Grey Galaxies of higher rank; their boundary stress tensors are completely determined by symmetry and {\it hydrodynamical} considerations.

It is now straightforward to 
apply this conjecture to the bulk dual of $AdS_5\times S^5$. The bulk gas, in this case, consists of all the fields of IIB Supergravity on $AdS_5 \times S^5$
(this is roughly 256 times the bulk stress tensor of the dilaton of the previous subsection).

Much to our surprise, we were unable to find an expression for the thermal partition function of this 10d gas, so we performed this computation ourselves; and used it to read off $h_{\rm{YM}} (\beta)$, obtaining the result presented in \eqref{hym}. 

The conjecture of this section thus gives us a natural guess for the boundary stress tensor of Grey Galaxies in ${\cal N}=4$ Yang-Mills theory: 
  \begin{equation}\label{bcssn}
T_{\mu\nu}= \frac{2 h_{\rm{YM}}(\beta)}{\beta \pi^2}  \gamma^4(\theta) \left( 4 u_\mu u_\nu + g_{\mu \nu} \right)
\end{equation} 
with $h_{\rm{YM}}$ listed in \eqref{hym}.
Note we were able to obtain this result without performing the laborious task of actually evaluating the bulk stress tensor of the thermal IIB supergravity gas in $AdS_5\times S^5$.

\subsection{`$AdS_D$ Gregory Laflamme' and Black Rings?}\label{glins}

To end this introduction, we note that black holes in $AdS_D$ (for $D\geq 5$) display some features with no counterpart in $D=4$. In $D\geq 6$, highly spinning black holes sometimes flatten out into a thin pancake which is then subject to Gregory-Laflamme type instabilities \cite{Emparan_2003,  Dias:2010gk}. Also, in  $D\geq 5$ we have new solutions (with new horizon topologies) like black rings \cite{Caldarelli_2008, Emparan_2008}. As we discuss in Appendix \ref{appins}, neither of these phenomena appears to be relevant to the construction of the phase diagram.
The Gregory-Laflamme instability always occurs at values of $\omega$ that are greater than unity, and so for black holes that were already super radiant unstable. 
The black holes at the center of Grey Galaxies are never Gregory Laflamme unstable.  Moreover (at least in the case that they are small, and so amenable to analytic analysis) black rings always have $\omega>1$, and so, themselves, display a superradiant instability. Although our analysis of these points is not completely definitive, it appears that black rings - or black holes at the edge of a Gregory Laflamme instability - 
never appear as dominant phases.

In order to forestall confusion, we emphasize that the Gregory-Laflamme type instabilities discussed in this subsection are those that lead to clumping in the $AdS_D$ directions. In many situations of physical interest, AdS spaces are accompanied by an internal manifold (e.g. the $S^5$ in $AdS_5 \times S^5$). In such contexts, small black holes typically do undergo Gregory Laflamme instabilities in the internal manifold (e.g. $S^5$), and the resultant
black holes are expected to dominate the phase diagram 
in appropriate parameter regimes (roughly when black holes are small). We never study this phenomenon in this paper. 

\subsection{Structure of this Paper}\label{sp}

The rest of this paper is organized as follows. In section \ref{adskerr} we review relevant aspects of Kerr-$AdS_5$ Black Hole solutions and their thermodynamics. In section \ref{pd} we construct the microcanonical phase diagram of ${\cal N}=4$ Yang-Mills theory (as a function of energy and angular momenta) under the assumption that black holes and Grey Galaxies are the only relevant phases. In section \ref{pf} we study the thermodynamics of the bulk gas. In \ref{section5} we focus on the model of a single 10d massless scalar in $AdS_5 \times S^5$, and 
construct the bulk stress tensor at finite temperature assuming $\omega_1$ and $\omega_2$ to be near unity. We also point out that stress tensor of the bulk gas is that of an equilibriated bulk perfect fluid. In section \ref{section6} we compute the resultant back reaction of the metric, and read off the resultant boundary stress tensor. In section \ref{cfbtst} we present our conjecture for the gas contribution to the boundary stress tensor for a bulk gas with arbitrary matter content (in particular for the dual of ${\cal N}=4$ Yang-Mills theory). We end this paper in section \ref{disc} with a discussion of our results. Several Appendices contain material that supports the analysis of the main text.

\section{Kerr - ${\rm AdS_{5}}$ Black Holes} \label{adskerr}

In this section, we review Kerr-$AdS_5$ Black hole solutions and some of their properties. 

\subsection{The Metric}
The metric of Kerr $AdS_5$ Black hole is given by \cite{Hawking:1998kw,Caldarelli:1999xj,Gibbons:2004ai} 
\begin{equation}\label{metad}
    ds^2=-\Delta_{\tilde\theta}(1+r^2)\rho^2 dt^2+\frac{f}{\rho^4}\left(\frac{\Delta_{\tilde{\theta}}dt}{\Xi_a\Xi_b}-\omega\right)^2+\frac{\rho^2dr^2}{\Delta_r}+\frac{\rho^2d\tilde{\theta}^2}{\Delta_{\tilde{\theta}}}+\frac{r^2+a^2}{\Xi_a}\sin^2\tilde{\theta} d\phi_1^2+\frac{r^2+b^2}{\Xi_b}\cos^2\tilde{\theta} d\phi_2^2
    \end{equation}
where 
\begin{align}
    &\Xi_a=1-a^2\\
    &\Xi_b=1-b^2\\
    &\Delta_{\tilde{\theta}}=1-a^2\cos^2{\tilde{\theta}}-b^2\sin^2{\tilde{\theta}}\\
    &\Delta_r=\frac{(r^2+a^2)(r^2+b^2)(1+r^2)}{r^2}-2m\\
    &\rho^2=r^2+a^2\cos^2{\tilde{\theta}}+b^2\sin^2{\tilde{\theta}}\\
    &f=2m\rho^2\\
    &\omega=a\sin^2\tilde{\theta}\frac{d\phi_1}{\Xi_a}+b\cos^2\tilde{\theta}\frac{d\phi_2}{\Xi_b}
\end{align}

Roughly speaking, the constants  $a$ and $b$ determine the angular momentum of the black hole in the two planes (say $\phi_1$ and $\phi_2$ respectively), and the parameter $m$ determines the mass of the black hole.
The parameters $a$ and $b$ lie in the range : 
\begin{equation}\label{abrange}
    a,b \in [-1,1]
\end{equation}
The parameter $m$ is always positive but is further constrained by the 
requirement that the solution \eqref{metad} should have an event horizon (and so describes a black hole rather than a naked singularity). 

For Kerr-AdS\(_5\) black holes, the positions of the event horizon are determined by solving the following algebraic equation:
\begin{equation}\label{horizoneq}
(r_+^2 + a^2)(r_+^2 + b^2)(r_+^2 + 1) - 2mr_+^2 = 0
\end{equation}
This is a cubic equation in \(r_+^2 = x\).
Let us study the nature of the roots of this cubic polynomial in \(x\):
\begin{equation}\label{horizon}
   x^3 + x^2 (a^2 + b^2 + 1) + x (a^2 b^2 + a^2 + b^2 - 2m) + a^2 b^2 = 0
\end{equation}
Let the roots \eqref{horizon} be denoted by  \(x_1\), \(x_2\), and \(x_3\). Recall that, on physical grounds, we are interested in the positive real roots of \eqref{horizon}. The coefficients of $x^2$ and the constant, in the cubic equation \eqref{horizon} are manifestly positive. The coefficient of $x$, namely \(a^2 + b^2 + a^2 b^2-2m\), could be either positive or negative. Clearly no positive real roots exist if this coefficient is positive. In this case, the `black hole' solution has a 
naked singularity, and is unphysical. In the physically interesting case
\begin{equation}\label{ineqq}
2m > a^2 + b^2 + a^2 b^2
\end{equation}
In Appendix \ref{horapp} we demonstrate the following. When \eqref{ineqq}
is just obeyed (i.e. when $m$ is just larger than the critical value in 
\eqref{ineqq}), \eqref{horizon} has two complex (and complex conjugate) solutions and one negative real solution. In this situation, consequently, the `black hole' solution continues to have a naked singularity (and so 
continues to be unphysical). As $m$ is further increased (at fixed $a$ and $b$) the two complex conjugate solutions approach the real axis. At a critical value of $m=m_{\rm ext}$ given by 
\begin{equation} \label{mextremal}
    \begin{split}
        m_{ext} &=\frac{1}{2} \left(\frac{1}{6} \sqrt{\left(a^2+b^2+1\right)^4+216 a^2 b^2 \left(a^2+b^2+1\right)} \cos \left(\frac{\alpha }{3}\right)+a^2 \left(b^2+1\right)-\frac{1}{12} \left(a^2+b^2+1\right)^2+b^2\right)\\
    \end{split}
\end{equation}
where $\alpha$ is given by
\begin{equation}
    \begin{split}
        \cos(\alpha) &=\left(\frac{1}{\left(a^2+b^2+1\right)^4+216 a^2 b^2 \left(a^2+b^2+1\right)}\right)^{3/2}\\&\times \left(5832 a^4 b^4-\left(a^2+b^2+1\right)^6+540 a^2 b^2 \left(a^2+b^2+1\right)^3\right)\\
    \end{split}
\end{equation}
the two solutions merge at a positive value of $x=x_{ext}$ on the real axis, where 
\begin{equation}\label{xext}
x_{\rm ext}= \frac{1}{6} \left(a^2+b^2+1\right) \left(2 \cos \left(\frac{1}{3}\xi\right)-1\right)
\end{equation}
and  $\xi$ is given by,
\begin{equation}
    \cos (\xi )=  \left(\frac{54 a^2 b^2}{\left(a^2+b^2+1\right)^3}-1\right)
\end{equation}

At $m=m_{\rm ext}$ the black hole is extremal and has nonzero horizon area (and so nonzero entropy) \footnote{While the entropy of the extremal black hole is nonzero at generic values of $a$ and $b$, it vanishes when either $a$ or $b$ vanishes - this follows from the fact that $\phi=\pi$ in this case. In this special case, black holes at extremality are singular; however, they can be approached as a limit of nonsingular (nonzero temperature) black holes: those with $m$ slightly greater than $m_{\rm ext}$.}

At still larger values of $m$, the two (newly minted) positive real roots separate away from each other on the real axis. The black hole turns non-extremal (and remains physical). As $m$ is increased even further, the larger of the two roots increases without bound (tending to 
\begin{equation}\label{xsch}
x \approx  \sqrt{2 m}
\end{equation}
while the smaller of the two roots decreases to zero (while always remaining positive). In this limit, the angular momentum is a small perturbation, and the black hole reduces to an AdS Schwarschild black hole.

\subsubsection{The special case $a=b$}

All expressions presented earlier in this subsection simplify in the special case $a=b$. In this special case, the metric enjoys invariance under $SU(2)_R$, and so can be written in a simple manner in terms of right invariant one-forms. 

\begin{equation}\label{metaequalb}
    ds^2=-\Xi_{a}(1+r^2)(r^2 +a^2) dt^2+\frac{2m}{(r^2 +a^2)\Xi_a^2}\left(dt-a~\sigma_3\right)^2+\frac{(r^2+a^2)dr^2}{\Delta_r}+\frac{(r^2+a^2)}{\Xi_a}\left(\sigma_1^2 +\sigma_2^2 + \sigma_3^2\right)
    \end{equation}
where
\begin{equation}\label{oneforms1}
    \begin{split}
    \sigma_1 &= \frac{1}{2}\left(\sin(\phi) d (2\tilde\theta) - \cos(\phi)\sin(2\tilde\theta)d\psi\right)\\ \sigma_2 &=\frac{1}{2}\left(\cos(\phi) d(2\tilde{\theta}) + \sin(\phi)\sin(2\tilde{\theta})d\psi\right)\\\sigma_3 &=\frac{1}{2}(d\phi - \cos(2\tilde{\theta})d\psi)\\
    \end{split}
\end{equation}
are the usual right invariant oneforms on a (squashed) $S^3$, and 
\begin{equation} \label{phipsi}
    \begin{split}
     \phi&= \phi_1 + \phi_2, ~~~~\psi = \phi_1 - \phi_2\\
    \end{split}
\end{equation}

In this case, the mass and the radius of the extremal black holes are given  by the simple expressions
\begin{equation}\label{m_ext_ab}
    \begin{split}
        m_{\rm ext}&= \frac{1}{128} \left(\sqrt{8 a^2+1}-1\right) \left(\sqrt{8 a^2+1}+3\right)^3\\
    \end{split}
\end{equation}
\begin{equation}\label{rexteq}
\begin{split}
    {(r_+)}_{\rm ext}^2 & = \frac{1}{4}\left( \sqrt{1 +8a^2}-1\right)
\end{split}
\end{equation}
\subsection{Large $r$ behaviors}

While the black hole \eqref{metad} asymptotes to $AdS_5$, this fact is not manifest in the coordinates used in \eqref{metad}, but can be made manifest by changing coordinates from $r, \tilde{\theta}$ to $y, { \theta}$, where $y$ and ${ \theta}$ are defined by the equations
(\cite{Bhattacharyya:2007vs})
\vspace{-0.1cm}

\begin{equation}
\begin{split}
y^2 \sin^2 {\theta} &= \frac{(r^2 + a^2) \sin^2 \tilde\theta}{\Sigma_a},
\\y^2 \cos^2 {\theta} &= \frac{(r^2 + b^2) \cos^2 \tilde \theta}{\Sigma_b}.\\
\end{split}
\end{equation}
\footnote{On adding these two equations we find
$$y^2= \frac{r^2 \left(1 - a^2 \cos^2 {\theta} - b^2 \sin^2 {\theta}\right) + a^2 \sin^2 {\theta} + b^2 \cos^2 {\theta} - a^2 b^2}{\Sigma_a \Sigma_b}, $$}
In these new coordinates, the black hole metric takes the form 
\begin{equation}\label{bhnewcor}
\begin{split}
ds^2 &= - (1 + y^2) dt^2 + \frac{dy^2}{1 + y^2} + y^2 (d{\theta}^2 + \cos^2 {\theta} \, d\psi^2 + \sin^2 {\theta} \, d\phi^2) \\
& + \frac{2m}{y^6 \Delta_{{\theta}}^3} dy^2 + \frac{2m}{y^2 \Delta_{{\theta}}^3} dt^2 \\
& - \frac{4am \sin^2 {\theta}}{y^2 \Delta_{{\theta}}^3} \, dt d\phi - \frac{4b m \cos^2 {\theta}}{y^2 \Delta_{{\theta}}^3} \, dt d\psi \\
& - \frac{2ma^2 \sin^4 {\theta}}{y^2 \Delta_{{\theta}}^3} \, d\phi^2 + \frac{2mb^2 \cos^4 {\theta}}{y^2 \Delta_{{\theta}}^3} \, d\psi^2 \\
& + \frac{2ab m \sin^2 {\theta} \cos^2 {\theta}}{y^2 \Delta_{{\theta}}^3} \, d\psi d\phi,\\
\end{split}
\end{equation}
At larger $y$ this metric \eqref{bhnewcor} reduces to
\begin{equation}
\begin{split}
ds^2 &= - y^2 dt^2 + \frac{dy^2}{y^2} + y^2 (d{\theta}^2 + \cos^2 {\theta} \, d\psi^2 + \sin^2 {\theta} \, d\phi^2).\\
\end{split}
\end{equation}
We see that the metric in these coordinates is manifestly asymptotically $AdS_5$, with the metric on constant $y$ slices being the metric on a round $S^3$ times time.

\subsection{Thermodynamics of the Black Hole}
The energy, angular momenta, entropy, temperature, and the angular speeds of the Kerr AdS$_{5}$ black hole are  given by \cite{Gibbons:2004ai}
\begin{align} \label{thermform}
    \begin{split}&E=\frac{m\pi(2\Xi_a+2\Xi_b-\Xi_a\Xi_b)}{4\Xi_{a}^2\Xi_{b}^2}\\
    &J_a=\frac{2\pi a m}{4\Xi_{a}^2\Xi_{b}}\\
    &J_b=\frac{2\pi b m}{4\Xi_{a}\Xi_{b}^2}\\
    &S=\frac{\pi^2(r_{+}^2+a^2)(r_{+}^2+b^2)}{2\Xi_{a}\Xi_{b}r_{+}}\\
    &T=\frac{r_+}{2\pi}(1+r^2_{+})\left(\frac{1}{r^2_{+}+a^2}+\frac{1}{r^2_{+}+b^2}\right)-\frac{1}{2 \pi r_{+}}\\
    &\omega_a=\frac{a(r_{+}^2+1)}{(r_{+}^2+a^2)}\\
    &\omega_b=\frac{b(r_{+}^2+1)}{(r_{+}^2+b^2)}\\
   & \Xi_{a} = 1-a^2\\
   &\Xi_{b} = 1-b^2
   \end{split}
\end{align}
\footnote{When a black hole is extremal, the LHS of \eqref{horizoneq} has a double root at its horizon. Consequently, the derivative of the LHS of \eqref{horizon} vanishes when evaluated at the horizon radius. As a check on our formulae, we have verified that imposing these conditions sets the RHS of the formula for the temperature (in \eqref{thermform}) to zero.} 
Note that $E$, $J_a$ and $J_b$ diverge in the limit $a\to 1$ or $b\to 1$.

\subsection{Thermodynamics at Extremality}

The formulae \eqref{thermform} can be specialized to extremality 
by plugging \eqref{xext} and \eqref{mextremal} into those formulae. 
As \eqref{xext} and \eqref{mextremal} are rather complicated, the resultant thermodynamical formulae are complicated. All formulae 
simplify, however, if $a$ is taken to be small (at arbitrary values of $b$)
\footnote{We could, of course, also take $b$ to be small at arbitrary values of $a$.}. Working to leading nontrivial order in $a$, we 
find 
\begin{equation}\label{mextrex}
\begin{split}
&(r_{+})_{ext}=\frac{\sqrt{a b}}{(1+b^2)^\frac{1}{4}}\\
&m_{ext}=\frac{b^2}{2}+ab\sqrt{1+b^2}
\end{split}
\end{equation}
and \eqref{thermform} simplify to 

\begin{align}
    &E=-\frac{b^2(b^2-3\pi)}{8(b^2-1)^2}\\
    &J_a=\frac{ab^2\pi}{4(1-b^2)}\\
    &J_b=\frac{b^3\pi}{4(b^2-1)^2}\\
    &S=\frac{\sqrt{ab}b^2\pi^2}{2(1+b^2)^\frac{1}{4}(1-b^2)}
\end{align}
The behaviour of these quantities for small $b$, at leading order, can, of course, be obtained by interchanging $a$ and $b$.

\subsection{Superradiant Instablity}
The Kerr AdS black hole has a superradiant instability either when $\omega_a>1$ or $\omega_b>1$, i.e. either when $r_{+}^2<a$ or $r_{+}^2< b$. Plugging these inequalities into \eqref{horizoneq}, we find that the black hole
has an `$a$' type superradiant instability when 

\begin{equation}\label{supermass}
    m \leq \frac{(1+a)^2(a+b^2)}{2}
\end{equation}
(the condition for a $b$ type instability is obtained by interchanging $a$ and $b$). 

The critical instability mass (lowest mass at which the black hole first suffers from a super radiant instability) always lies above extremality. 
This point can be seen explicitly when $a$ is small. Working to linear order in $a$, the difference between the masses listed in
\eqref{supermass} and \eqref{mextrex} equals 
\begin{equation}\label{diffmass}
\Delta m= a \left( \frac{1}{2} + b^2 - b\sqrt{1+ b^2} \right),  
\end{equation}
a quantity that is everywhere positive for $b\in (0,1)$.

\begin{center}
\begin{figure}[h]
    \centering
\includegraphics[scale=0.5]{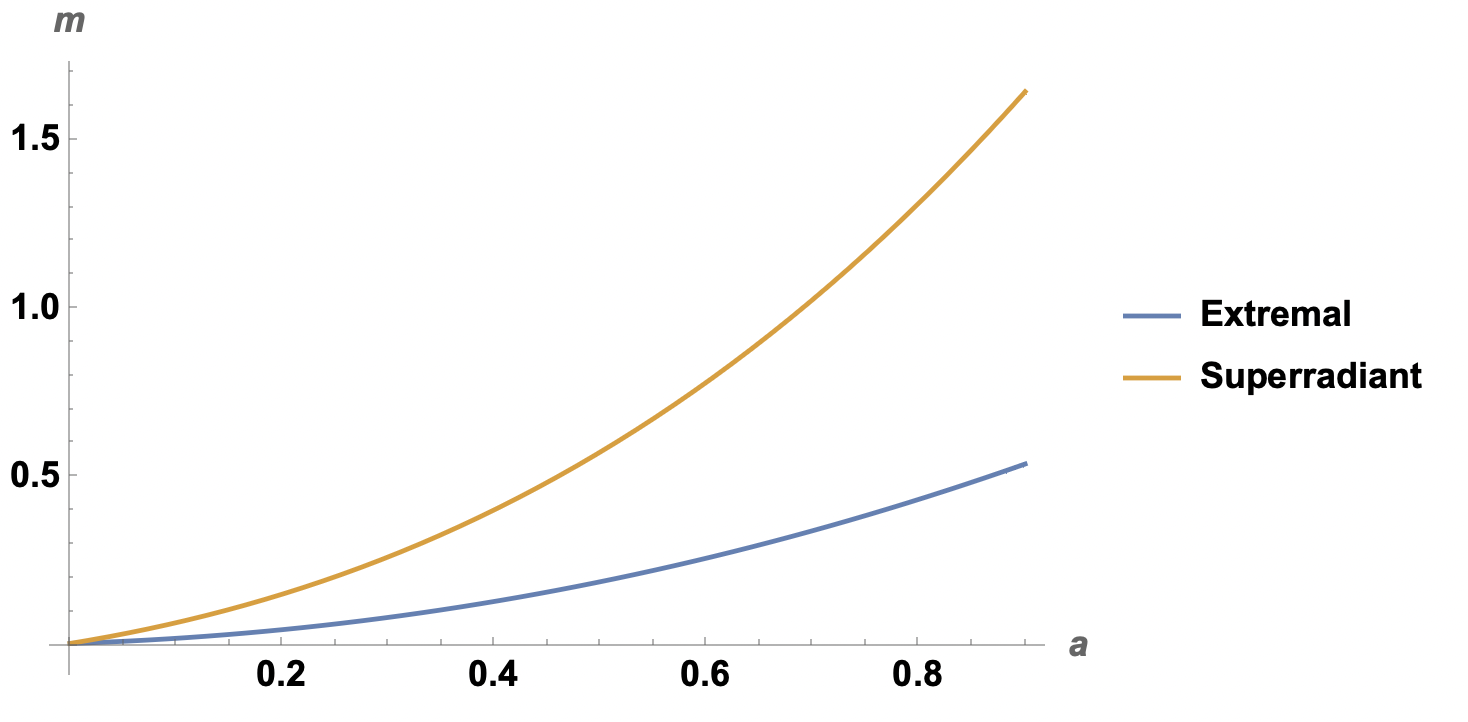}
    \caption{The plot shows that extremal BHs are below the superradiant BHs. While the plots above are presented for $\mathbf{b=0.1}$, the analogous plots at all values of $b \in (0,1)$ are qualitatively similar.}
    \label{extsup}
\end{figure}
\end{center}

While the argument presented above is only accurate at small $a$, the final result is correct at all values of $a$. We can see this by plotting the masses listed on the RHS of \eqref{mextremal} and \eqref{supermass} as a function of $a$ at various fixed values of $b$. We see 
from Fig. \ref{extsup} that the mass at which the black hole becomes super radiant unstable is always larger than the mass at which it becomes extremal. 
\subsection{Charges at boundary of superradiance}

We have seen, above, that black holes lie on the boundary of the superradiant instability when either $r_{+}^2 = a$ or $r_{+}^2 = b$  or when both of these conditions are obeyed. Black holes for which $r_{+}^2 = a$ and $r_{+}^2 = b$ have $\omega_a=\omega_b=1$. These black holes are obtained by setting $a=b$ and by choosing $m$ to saturate the inequality in \eqref{supermass}, i.e. by choosing 
\begin{equation}\label{mnow}
m=\frac{a(1+a)^3}{2}.
\end{equation}
These black holes appear in a one-parameter family (parameterized by $a$ above), are of particular interest to this paper, as they will constitute the central black hole for rank 4 Grey Galaxies. 
In the rest of this subsection, we study these black holes in more detail. 

\subsubsection{Thermodynamical Formulae}

As the radius of the event horizon is given by the simple formula $r_{+}^2=a$, all thermodynamical formulae \eqref{thermform} simplify considerably: we find 
\begin{equation}\label{sameomega}\begin{split}
&T = \frac{1}{2\pi \sqrt{a}}, \\
&E = \frac{a(3+a^2)\pi}{8(1-a)^3}\\
& J = J_{a} = J_b= \frac{\pi  a^2}{(1-a)^3}\\
&S=\frac{\pi ^2 a^{3/2}}{2 (a-1)^2}
\end{split}    
\end{equation}

\subsubsection{Specific Heats}

Along this one parameter curve, the energy as a function of temperature  is given by  
\begin{equation}\label{Et}
 E= -\frac{\pi+48 \pi ^5 T^4}{8 \left(1-4 \pi ^2 T^2\right)^3}, ~~~~   \frac{1}{2 \pi} \leq T < \infty
\end{equation}

\footnote{The limit $T= \frac{1}{2\pi}$ corresponds the big black hole limit $a=1$, while the 
limit $T=\infty$ corresponds to the small black hole limit $a=0$.} In the limit 
$T \rightarrow \infty$ \eqref{Et} simplifies to  
\begin{equation}\label{Etsimp}
    E \approx \frac{3}{32 \pi T^2}
\end{equation}

We can calculate the specific heat \footnote{Note, of course, that this specific heat is calculated at 
the constant values $\omega_a=\omega_b=1$. In other words, it is a specific heat evaluated at constant $\omega$ rather than at constant $J$. }
of these black holes  by differentiating \eqref{Et} w.r.t temperature; we find 
\begin{equation}
    c=\left( \frac{dE}{dT} \right)\Big|_{\omega_a=\omega_b=1} 
    = -\frac{3 \pi ^3 T \left(4 \pi ^2 T^2+1\right)^2}{\left(1-4 \pi ^2 T^2\right)^4} 
\end{equation}
    We note that the specific heat is always negative. Note that the specific heat 
goes to zero in the small black hole limit ($T \rightarrow \infty$) but diverges in 
the large black hole limit $T \rightarrow \frac{1}{2 \pi}$. Quantitatively, 
in the limit $T \rightarrow \frac{1}{2\pi}$, 
\begin{equation}
    c \approx -\frac{3}{128 \pi ^2 \left(T-\frac{1}{2 \pi }\right)^4}
\end{equation}
while in the large $T$ limit, 
\begin{equation}
    c \approx -\frac{3}{16\pi T^3}
\end{equation}

\subsubsection{An inequality for $E$ as a function of $J$}

We can, in principle, use the third equation in \eqref{sameomega} to solve for $a$ in terms of $J$, and plug this solution into the second of \eqref{sameomega} to obtain 
$E=E_{\omega}(J)$. This gives us $E$ as a function of $J$ along this curve. Notice that 
$E_{\omega}(0)= 0$. By differentiating the second and third of \eqref{sameomega}, and dividing the results, we find an exact parametric expression for the gradient of the function $E_{\omega}(J)$
\begin{equation}\label{derom}
\frac{dE_{\omega}(J)}{d J} = \frac{3 (a+1)^2}{2 a (a+2)} 
\end{equation} 
Notice that $J(a)$ varies from $0$ to $\infty$ as $a$ varies from $0$ to unity. It is easy
to convince oneself that the RHS of \eqref{derom} is a monotonically decreasing function 
in the range $a\in (0,1)$. The slope \eqref{derom} blows up at $a=0$ ($J=0$), tends to 
$2$ at $a=1$ ($J=\infty$), and is greater than 2 at every intermediate value. It follows in particular that 
\begin{equation}\label{ej}
E_{\omega}(J) = \int_0^J \frac{dE_{\omega}(J)}{d J} > 2 J
\end{equation}
so that, everywhere,  $E_{\omega}(J)$ obeys the unitarity bound $E> J_1+J_2$
(recall that in the current context $J_1=J_2=J$ so the unitarity bound simplifies to \eqref{ej}).

\subsection{Boundary Stress Tensor for Black Hole} \label{bhstr}

\subsubsection{Components of the boundary stress tensor}
The boundary stress-tensor of Kerr-$AdS_5$ black holes is easily read off 
from the form of the metric presented in \eqref{bhnewcor}, and is given by \cite{Bhattacharyya:2007vs} : 
\begin{equation} \label{bcs1} 
    \begin{split}
        T^{tt} &= \frac{m}{8\pi G_{5}}(4 \gamma^6-\gamma^4)\\T^{\phi_1\phi_1} &= \frac{m}{8\pi G_5}\gamma^4\left ( 4\gamma^2 a^2 + \frac{1}{\sin^2({\theta})}\right)\\ T^{\phi_2\phi_2} &= \frac{m}{8\pi G_5}\gamma^4\left ( 4\gamma^2 b^2 + \frac{1}{\cos^2({\theta})}\right)\\T^{t\phi_1} &= \frac{4m}{8\pi G_5} a\gamma^6, ~~T^{t\phi_2} = \frac{4m}{8\pi G_5} b\gamma^6\\T^{\phi_1\phi_2}&=\frac{4m}{8\pi G_5} ab\gamma^6,~~T^{{\theta}{\theta}} = \frac{m}{8\pi G_5}\gamma^4\\
    \end{split}
\end{equation}
where $\gamma^{-2} = 1-a^2\sin^2{\theta}-b^2\cos^2{\theta}$. The boundary stress tensor above can be rewritten in terms of $N$ by setting 
$G_5= \frac{\pi}{2 N^2}$

\subsubsection{Fluid nature of the boundary stress tensor}

Notice that the boundary stress tensor \eqref{bcs1} can be succintly rewritten \cite{Bhattacharyya:2007vs, Bhattacharyya:2008mz} in the `fluid form'
\begin{equation}\label{fst}
\begin{split}
T^{\mu \nu}&= \frac{N^2 m}{4 \pi^2} \gamma^4(\theta)\left( 4 u^{\mu}_{~bh} u^{\nu}_{~bh} + g^{\mu\nu} \right)\\
u^{\mu}_{~bh}&= \gamma \left( \partial_t - a \partial_{\phi_1} - b \partial_{\phi_2} \right), ~~~  \gamma^{-2} = 1-a^2\sin^2 {\theta}-b^2\cos^2{\theta}\\ 
\end{split}
\end{equation}
where $g^{\mu\nu}$ is the inverse metric on the unit 3 sphere  (see \eqref{sthree}). Note that $u^\mu_{bh}$ is a velocity vector field, in the sense that $u^\mu_{bh} u^\nu_{bh} g_{\mu \nu}=-1.$ \eqref{fst}  is the stress tensor of an equilibrated perfect fluid with fluid velocity $u_{bh}^\mu$.  

For large black holes (i.e. when $r_+ \gg 1$)
\footnote{For all allowed values of 
$a$ and $b$ it follows from \eqref{horizoneq} that this inequality is equivalent $m \gg 1$.} we see from \eqref{thermform} that 
$\omega_a \approx a$ and $\omega_b \approx b$. In this case the 
effective velocity field $u_{bh}^\mu$ reduces to 
\begin{equation}\label{simpv}
u_{bh}^{\mu} = \gamma \left( \partial_t - \omega_a \partial_{\phi_1} - \omega_b \partial_{\phi_2}
\right), ~~~  \gamma^{-2} = 1-\omega_a^2\sin^2{\theta}-\omega_b^2\cos^2{\theta}
\end{equation}
i.e. to the velocity field of a fluid rotating with angular velocities 
$\omega_a$ and $\omega_b$ on the unit sphere. As  
$r_+$ retreats to smaller values,  $a< \omega_a$. It follows that 
the effective angular velocity of the `black hole fluid' is smaller than $\omega_a$ and $\omega_b$ when the central black hole is small, but tends to these values in the limit that the black hole radius becomes large. 

The fluid discussed in this section is that of the fluid gravity correspondence \cite{Bhattacharyya:2007vs, Bhattacharyya_2008, Bhattacharyya:2008mz, Bhattacharyya:2008xc}

\subsection{Energy Density as a function of ${\theta}$}

\begin{figure}[h]\label{bdryst}
\centering
\includegraphics[scale=0.80]
{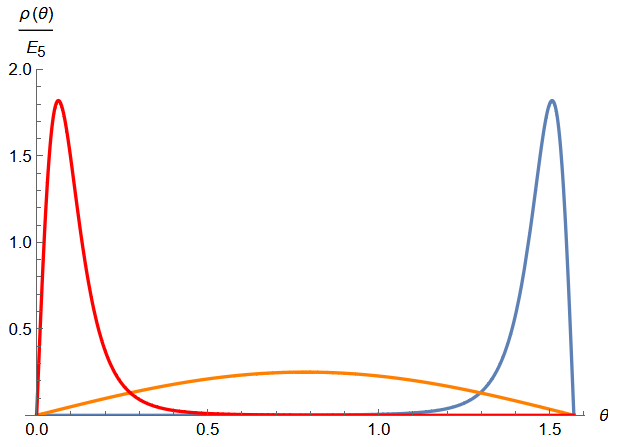}
\caption{Here the red curve is for $b \rightarrow 0.99,a\rightarrow0.01$ and blue  curve is for $ a\rightarrow0.99,b\rightarrow0.01$ , finally the orange curve is for $a\rightarrow0.99,b\rightarrow0.99$.}
\end{figure}

The energy density of the black hole solution is simply equal to the time-time component of the boundary stress tensor $T^{tt}$. Integrating this over $\phi_1$ and $\phi_2$ (with the usual measure factor on $S^3$) yields the 
$\rho({\theta})$, the energy per unit ${\theta}$, normalized so that 
$E= \int d{\theta} \rho({\theta})$ ($E$ is the full black hole energy). We find 
\begin{equation}
    \begin{split}
        E_5 &=\int d\Omega_3 T_{0}^{~0}\\=& \frac{m}{8\pi G_{5}}\int_{0}^{2\pi}d\phi_1\int_{0}^{2\pi}d\phi_2\int_{0}^{\frac{\pi}{2}}d{\theta}\sin({\theta})\cos({\theta})(4 \gamma^6-\gamma^4)\\
        \equiv & \int \rho({\theta}) d{\theta}\\
        =& \frac{m\pi(2-2a^2+2-2b^2-(1-a^2)(1-b^2))}{4G_5 (1-a^2)^2(1-b^2)^2}\\
    \end{split}
\end{equation}
where 
\begin{equation}\label{enden}
\rho({\theta})\ =\frac{m \pi}{2 G_5 }\frac{\sin({\theta})\cos({\theta})(3+a^{2}\sin^{2}({\theta})+b^{2}\cos^{2}({\theta}))}{(1-a^{2}\sin^{2}({\theta})-b^{2}\cos^{2}({\theta}))^{3}}
\end{equation}
In Fig \ref{bdryst}, we plot $\frac{\rho({\theta})}{E}$ versus ${\theta}$ for 
three different values of $a$ and $b$, namely $(a=0.01,b=0.99)$, $(a=0.99,b=0.01)$ and $( a=0.99,b= 0.99)$.

Notice that the energy density is uniformly distributed on the sphere when $a=b$ (in fact, in this case, $T^{tt}$ is exactly spherically symmetric; a result that follows from the fact that this choice of parameter preserves $SU(2)_R$). When $b \gg a$ or $a\gg b$,  the energy density clumps around $0$ and $\frac{\pi}{2}$ respectively. \footnote{This happens because the fluid tries to move as far as possible from the axis of rapider rotation (as a consequence of its larger centrifugal force).}

\section{The Phase diagram} \label{pd}
\begin{figure}[h]
    \centering
    \includegraphics[width=0.6\textwidth]{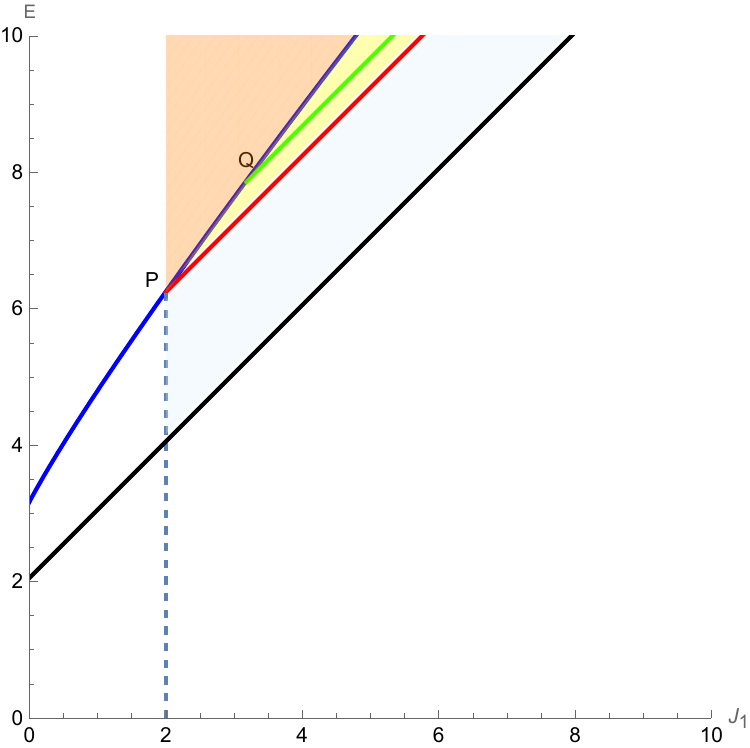}
    \caption{A constant $J_2$ ($J_2=2$) slice of the phase diagram. 
    We have plotted $E$ on the $y$ axis and $J_1$ on the $x$ axis. The black line is the unitarity bound $E=2+J_1$.  The blue curve is the 
    line $\omega_1=1$. When $J_1 >2$, this curve represents a phase boundary. Above this curve, the dominant phase is the usual vacuum black hole. Below this curve (shaded yellow region) we have a rank 
    2 Grey Galaxy. All points on the green 45-degree line (in this phase) have the same entropy, given by the entropy of the vacuum $\omega_1=1$ black hole at the lower end of the green line. The rank 2 Grey Galaxy phase ends on the red 45-degree line, which meets the $\omega_1=1$ curve at $J_1=2$. Below the red curve (shaded blue region) the dominant phase is the rank four Grey Galaxy all the way down to the unitarity bound. We have not attempted to sketch the phase diagram for $J_1<2$; part of the phase diagram is best visualized in constant $J_2$ slices. 
    }
    \label{EvsJ1}
\end{figure}

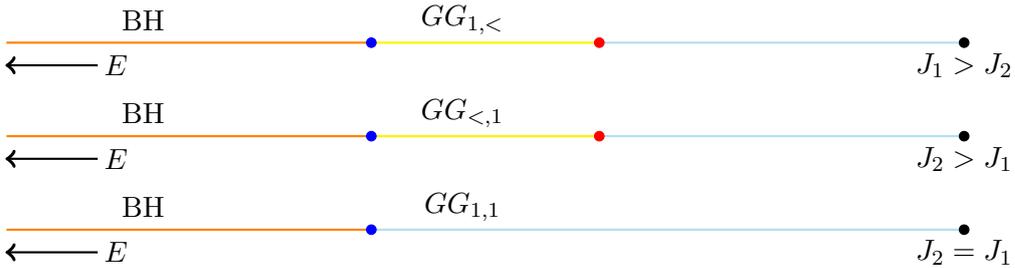
\begin{figure}
\begin{tikzpicture}[scale=0.6]
\draw[orange,thick] (1,0) -- (9,0);
\draw[yellow,thick] (9,0) -- (14,0);
\draw[LightBlue2,thick] (14,0) -- (22,0);
    \filldraw[blue] (9,0) circle[radius=3pt];
    
    \filldraw[red] (14,0) circle[radius=3pt];
    
    \filldraw[black] (22,0) circle[radius=3pt];
                \draw (4,0.5) node{$\rm{BH}$};
                \draw (11,0.5) node{$GG_{1,<}$};
                \draw (22,-0.5) node{$J_1>J_2$};
              \draw
[
postaction={decorate,decoration={markings , 
mark=at position .10 with {\arrow[black,line width=0.5mm]{<};}}}
][black, thick] (1,-0.5)--(3,-0.5);
\draw (3.4,-0.5) node{$E$};
\end{tikzpicture}

\begin{tikzpicture}[scale=0.6]
\draw[orange,thick] (1,0) -- (9,0);
\draw[yellow,thick] (9,0) -- (14,0);
\draw[LightBlue2,thick] (14,0) -- (22,0);
    \filldraw[blue] (9,0) circle[radius=3pt];
    
    \filldraw[red] (14,0) circle[radius=3pt];
    
    \filldraw[black] (22,0) circle[radius=3pt];
                \draw (4,0.5) node{$\rm{BH}$};
                \draw (11,0.5) node{$GG_{<,1}$};
                \draw (22,-0.5) node{$J_2>J_1$};
              \draw
[
postaction={decorate,decoration={markings , 
mark=at position .10 with {\arrow[black,line width=0.5mm]{<};}}}
][black, thick] (1,-0.5)--(3,-0.5);
\draw (3.4,-0.5) node{$E$};
\end{tikzpicture}

\begin{tikzpicture}[scale=0.6]
\draw[orange,thick] (1,0) -- (9,0);

\draw[LightBlue2,thick] (9,0) -- (22,0);
\filldraw[blue] (9,0) circle[radius=3pt];

    \filldraw[black] (22,0) circle[radius=3pt];
                \draw (4,0.5) node{$\rm{BH}$};
                \draw (11,0.5) node{$GG_{1,1}$};
                \draw (22,-0.5) node{$J_2=J_1$};
              \draw
[
postaction={decorate,decoration={markings , 
mark=at position .10 with {\arrow[black,line width=0.5mm]{<};}}}
][black, thick] (1,-0.5)--(3,-0.5);
\draw (3.4,-0.5) node{$E$};
\end{tikzpicture}
\caption{Here we plot the line diagram, which shows the transitions of the different phases as we lower the energy $E$ (Keeping the $J_1$ and $J_2$ fixed) starting from the vacuum black hole phase. These plots are presented for three different cases; namely $J_1> J_2$, $J_2>J_1$, and $J_1 = J_2$.}
\label{figline}
\end{figure}

In this section, we determine the phase diagram of our system as a function of the total energy $E$, the total angular momentum $J_1$, and the total angular momentum $J_2$, working under the assumption that the only relevant phases are vacuum black holes and Grey Galaxies. 
\footnote{When this assumption fails - e.g. at  low energies in $AdS_5 \times S^5$ (see the introduction)- only a part of the phase diagram presented in this section is physical.}

Consider the three-dimensional space parameterized by $E$, $J_1$ and $J_2$. When visualizing this space we view the energy axis as lying on the $z$ axis and refer to points with larger values of $E$ as 
lying `above' points with smaller values of $E$.

Our charge space hosts three important two-dimensional surfaces. These are 
\begin{itemize}
\item The unitarity plane $E=J_1+J_2$. The unitarity bound tells us that there are no states below this plane. 
\item The surface $S_1$ on which Kerr Black holes have $\omega_1=1$
but $\omega_2\leq 1$ (see \eqref{thermform} for a concrete expression). 
\item The surface $S_2$ on which Kerr Black holes have $\omega_2=1$ but $\omega_1 \leq 1$ (see \eqref{thermform} for a concrete expression). 
\end{itemize}

The surfaces $S_1$ and $S_2$ intersect on the line $C_{12}$. Along this line, $J_1=J_2=J$ and the energy is given as a function of $J$
by the function $E_{\omega} (J)$ defined above \eqref{derom}).

Our system has four phases. The first of these is the `vacuum' black hole phase, in which the bulk is described by Kerr AdS black holes. We call this the $(BH)$ phase. This $(BH)$ phase is stable only if $\omega_1<1$ and $\omega_2<1$, i.e. this phase exists only `above' the higher of the two sheets $S_1$ and $S_2$. 

The second phase is a Grey Galaxy (GG) with a central black hole that has  $\omega_1=1$ and $\omega_2<1$, in equilibrium with a gas that has $E=J_1$. We call this the $(GG)_{1, <}$ phase. This phase only exists when $J_1>J_2$ 
\footnote{When $J_2 >J_1$, $\omega_2>\omega_1$, so it is impossible to find $\omega_1=1$ and $\omega_2<1$.}. Points in this phase are obtained by starting on any point on $S_1$ with $J_1>J_2$, and then shooting a ray upwards at 45 degrees in the $E-J_1$ plane. Note that this ray is shot at constant $J_2$. 
This is depicted in Fig \ref{EvsJ1}, where we present a plot of the $E-J_1$ plane at $J_2=2$. The blue curve in Fig \ref{EvsJ1} represents the intersection of $S_1$ with this plane. Note that the slope of this curve is always greater than 45 degrees, and asymptotes to 45 degrees at large $J_1$ (this can be shown to be true at all values of $J_2$). It follows that 45 degree lines shot out from $S_1$ - like the green line in Fig \ref{EvsJ1}-
never intersect the surface $S_1$. \footnote{The diagram in Fig \ref{EvsJ1} is qualitatively similar to Fig 1. of \cite{Kim:2023sig}, which plots $E$ vs $J$ for black holes in $AdS_4$. The one qualitative difference is that the unitarity curve - the black line in Fig \ref{EvsJ1} is offset from the origin (the offset is by $J_2$, which has no analogue in $AdS_4$.} 
It follows, therefore, that no point in the space $(E, J_1, J_2)$ lies simultaneously in 
the $(BH)$ phase and the $GG_{1,<}$ phase. 
The point $P$ in Fig. \ref{EvsJ1} marks the point on $S_1$ at which $J_1=J_2$ (in the case of Fig \ref{EvsJ1}, $J_2=2)$. This point lies on the boundary of the surface $S_1$. Note that the 45-degree lines shot out from $P$ - the line drawn in red in Fig  \ref{EvsJ1}, - gives the lowest points (i.e. the points with the smallest energy for any given $J_1$ and $J_2$) that are accessed in this phase. \footnote{Note also that $P$ lies on the curve $C_{12.}$} The union of these (red) lines for all values of $J_2$ makes us a second surface
which we denote by $B_1$. The set of points in  $(E, J_1, J_2)$ space,  that make up the $(GG)_{1,<}$ phase are those bounded from above by the surface $S_1$, and from below by the surface $B_1$ (this is the region shaded yellow in Fig \ref{EvsJ1}). As is clear from Fig. \ref{EvsJ1}, there is a unique ray that reaches every point in this region.

The third phase - the $(GG)_{<, 1}$ phase - is a GG with a central black hole that has  $\omega_2=1$ and $\omega_1<1$, in equilibrium with a gas that 
has $E=J_2$. This phase only exists when $J_2>J_1$, and points in this phase are obtained by starting on any point on $S_2$ and then shooting a ray upwards at 45 degrees in the $E-J_2$ plane at fixed $J_1$. The set of points that lie in this phase is characterized as in the previous paragraph. These points are bounded from above by $S_2$ and from below by $B_2$ 
($B_2$ is defined in a manner entirely analogous to $B_1$, but with $1 \rightarrow 2$). 

The fourth phase - The $(GG)_{1,1}$ phase - consists of a  GG with a central black hole that has  $\omega_1=\omega_2=1$, in equilibrium with a gas that has $E=J_1+J_2$. Points in this phase are obtained starting on the curve $C_{12}$ and shooting out rays that obey $\Delta E= \Delta J_1 + \Delta J_2$, in every positive direction \footnote{By a positive direction we mean a direction in which $J_1$ and $J_2$ both increase as $E$ increases.} in the $J_1J_2$ plane.

We will now demonstrate that points that lie in the fourth phase always lie below both $B_1$ and $B_2$, 
and, moreover, that every point below $B_1$ and $B_2$
(but above the unitarity plane) hosts a unique $GG$ in the $(GG)_{1,1}$ phase.

Let us consider some given values of $J_1$ and $J_2$. For definiteness, we suppose that 
$J_1=\alpha$ and  $J_2=\beta$ with $\alpha>\beta$ (the other case can be discussed analogously). Since $J_1>J_2$, there exists a (unique) point $O_1(\alpha, \beta)$ on the 
sheet  $B_1$ with the same values of $J_1$ and $J_2$, namely with $J_1=\alpha$ and $J_2=\beta.$ Let us denote the energy of this point by $E_{B_1}(\alpha, \beta)$. 
It follows that the point $O_1(\alpha, \beta)$ has coordinates 
$$(J_1, J_2, E)= (\alpha, \beta, E_{B_1}(\alpha, \beta)).$$

By definition, the point $(\alpha, \beta, E_{B_1}(\alpha, \beta)$ is connected - by a 45 degree line in the $E-J_1$ plane - to some point $P_{12}(\alpha, \beta)$ on $C_{12}$ (this is the point $P$ in Fig \ref{EvsJ1}). Since this line propagates at constant $J_2$, it follows that $J_2=\beta$, and the coordinates $(J_1, J_2, E)$ of $P_{12}(\alpha, \beta)$ are 
$$(\beta, \beta, E_{\omega}(\beta)).$$

Since the points $O_1(\alpha, \beta)$, and the points 
$P_{12}(\alpha, \beta)$ are connected by a 45 degree line in the $E-J_1$ plane, it follows that 
\begin{equation}\label{fofiv}
E_{\omega}(\beta) - E_{B_1}(\alpha, \beta)= \beta - \alpha
\end{equation} 

Let us now search for other points in this phase with $J_1=\alpha$ and $J_2=\beta$. Such a point must lie at the endpoints of positive rays shot out from $C_{12}$. As 
the rays must be positive, it follows that it must originate at points on $C_{12}$ with $J< \beta$. \footnote{Else the ray would have to proceed in a direction that decreases $J_2$, which violates its positivity.}. Let us study a point that originates at 
$J=\beta - x$, i.e. from the point $(\beta-x, \beta-x, E_{\omega}(\beta-x)).$ This ray has to end up at $(J_1, J_2)= (\alpha, \beta)$. Since the ray obeys $\Delta E = \Delta J_1+ \Delta J_2$, it follows that it lands up at the point 
$$\left(\alpha, \beta, E_{\omega}(\beta-x) + \left(\alpha - (\beta-x) \right)  + \left( \beta -(\beta-x) \right) 
\right) =\left(\alpha, \beta, E_{\omega}(\beta-x) + \alpha -\beta +2x \right)$$
Using \eqref{fofiv}, the coordinates of this point simplify to 
$$\left(\alpha, \beta, E_{B_1}(\alpha, \beta) + E_{\omega}(\beta-x) -E_{\omega}(\beta) +2x \right)$$

Now we have already explained (see around \eqref{ej}) that the curve $E_{\omega}(y)$ is monotonic with a slope everywhere greater than two. It follows, as a consequence, that the quantity 
$E_{B_1}(\alpha, \beta) + E_{\omega}(\beta-x) -E_{\omega}(\beta) +2x$
starts out equal to $E_{B_1}(\alpha, \beta)$ when $x=0$, 
and then decreases monotonically, (as $x$ is increased) until it equals $\alpha+\beta$ at $x=\beta$ (this is the largest allowed value of $x$). We have just proved, in other words, that every point between $B_1$ and the unitarity plane represents a unique $(GG)_{1,1}$ black hole.  

The structure of the phase diagram (in the microcanonical ensemble) is now clear. This diagram is best understood as follows. Let us sit at fixed values of $J_1$ and $J_2$ and examine which phase we occupy as we vary $E$ (see Fig. \ref{figline}). At large values of $E$ we are always in the $(BH)$ phase. 

Let us first assume that $J_1>J_2$. In this case, as we lower energy, we eventually hit the sheet $S_1$, and transit to the $(GG)_{1, <}$ phase 
(see the first of Fig. \ref{figline}). Upon further lowering energy we hit the sheet $B_1$ and transit to the $(GG)_{1, 1}$ phase. Finally, further lowering the energy takes us to the unitarity sheet. \footnote{The size of the energy interval that lies in the $(GG)_{1, <}$ depends on $J_1$ and $J_2$. In particular, this interval goes to zero when $J_1 -J_2 \rightarrow 0$.} The case $J_2>J_1$ is similar: the discussion of the previous paragraph applies unchanged $1 \leftrightarrow 2$ (see the second of Fig. \ref{figline}).

Finally, let us consider the case $J_1=J_2$. In this case, as we lower energy, we eventually hit the line $C_{12}$ and transit directly into the $GG_{1,1}$ phase. Further lowering takes us to the unitarity sheet 
(see the last of Fig. \ref{figline}).

We emphasize that the discussion above is completely quantitative.
It is, for instance, a simple matter to produce a \texttt{Mathematica} file that allows the reader to input any values of the energy and two angular momenta, and outputs the phase that these charges lie in, the energy and angular momenta of the seed black hole (at those values of total charges), and the entropy of this seed black hole. We have written such a file, and would be happy to share the same upon request.


As explained in the introduction, while Grey Galaxies of rank 2 are qualitatively similar to Grey Galaxies in $AdS_4$, Grey Galaxies of rank 4 have several new features. Throughout the rest of this paper, we focus on Grey Galaxies of rank 4. Over the next three sections, we will build towards a (conjectured) formula for the boundary stress tensor of such Grey Galaxies. 

\section{Gas thermodynamics in the canonical ensemble}\label{pf}

In thermal $AdS$ (and, effectively, in the gas part of a Grey Galaxy phase), the bulk partition function is a product over partition functions $Z_{\Delta, J_L, J_R}$ for each field theory single trace primary operator (here $\Delta$, $J_L$, and $J_R$ denote the scaling dimension, $SU(2)_L$ representation and $SU(2)_R$ representation of the primary). 
In this brief section, we determine the multiparticle partition function of 
\begin{itemize}
\item The bulk field dual to a scalar primary operator of dimension $\Delta$ and all its descendants. 
\item All bulk fields arising out of the 10-dimensional dilaton in IIB supergravity.
\item All bulk fields of IIB supergravity on $AdS_5\times S^5$
\end{itemize}
In each case, we work in the limit that $1-\omega_1$ and $1-\omega_2$ are both small (consequently the results of this subsection will be of use when
studying Grey Galaxies of rank 4). 

As we explain below, a striking feature of our results is that all partition functions we study take the `fluid form' even though we are not necessarily working at high temperatures.  

\subsection{Partition function for a single scalar in $AdS_5$}

Consider a primary $O$ with $J_L=J_R=0$. Its descendants take the form 
\begin{equation}\label{desc}
(\partial^2)^n \partial_{\mu_1} \ldots \partial_{\mu_J} O
\end{equation}
where the derivatives  $\partial_{\mu_1} \ldots \partial_{\mu_J}$ 
are understood to act in a `trace removed' manner. The derivatives 
$\partial_{\mu_1} \ldots \partial_{\mu_J}$ generate a spin 
$(\frac{J}{2}, \frac{J}{2})$ representation of $SU(2)_L \times SU(2)_R$. Let $m_L$ and $m_R$ represent the quantum numbers of 
$J_L^z$ and $J_R^z$.
Clearly $m_L$ and $m_R$ each run over the range 
$(-J/2, J/2)$. It follows that the full partition function 
$$ {\rm Tr} e^{- \beta H + \beta(\omega_1+ \omega_2) J^z_L + 
\beta (\omega_1-\omega_2) J^z_R}$$
\footnote{Note that $J_L$ and $J_R$ are related to the `orthogonal two-plane rotation quantum numbers' $J_1$ and $J_2$  via
\begin{equation*}
    \begin{split}
        J_L= \frac{J_1 + J_2}{2}, \quad J_R = \frac{J_1 - J_2}{2}, \quad 
    \end{split}
\end{equation*}
Consequently, the partition function listed above can also be written as 
\begin{equation*}\label{so4qnum}
\ln Z=-\sum_{n=0}^{\infty}\sum_{J=0}^{\infty}\sum_{|m_1| +|m_2|\leq J}\ln (1-e^{-\beta(\Delta + 2n + J) + \beta\omega_1 m_1 + \beta \omega_2 m_2})
\end{equation*}
}
(obtained by multiparticling over operators in the given $SO(4,2)$ multiplet) is given by 
\begin{equation} \label{gaspf}
\ln Z(\beta,\omega_1,\omega_2)=-\sum_{n=0}^{\infty}\sum_{J=0}^{\infty}\sum_{m_L=-J/2}^{J/2}\sum_{m_R=-J/2}^{J/2}\ln\left(1-e^{-\beta(\Delta + 2n + J) + \beta\omega_1(m_L +m_R) + \beta \omega_2(m_L -m_R)}\right)
\end{equation}

The key point here is that the partition function in \eqref{gaspf}
diverges when either $\omega_1$ or $\omega_2$ tend to unity.
\footnote{In the case that $\omega_1$ is parametrically near to unity, but $\omega_2<$ unity (or vice versa), the divergence has its origins, respectively, in modes for which $n$, $\frac{J}{2}-m_L$ 
and $\frac{J}{2}-m_R$ are kept fixed at values of order unity, while $J$ runs over very large values. In the case that $\omega_1$ and $\omega_2$ are both parametrically near to unity (the situation of main interest for this section), the divergence has its origin in modes for which $n$, $\frac{J}{2}-m_L$ are fixed, while $J$ and $m_R$ run over very large values.} 
This point may be made explicit by expanding the logarithm in 
\eqref{gaspf} in a Taylor series and then performing the summations over $n$, $m_L$, $m_R$ and $J$ (see Appendix \ref{partapp}). Specializing to the case in which $\omega_1$ and $\omega_2$ are both very near unity and working to leading order both in $1-\omega_1$ and in $1-\omega_2$, we obtain 
\begin{align} \label{zexact}
     \ln Z(\beta,\omega_1,\omega_2)&\approx\sum_{q=1}^{\infty}\frac{1}{q}\Big[-\frac{e^{-q\beta  (\Delta -6) }}{\left(e^{2 \beta  q}-1\right)^4}+\frac{e^{-q\beta  (\Delta -2) }}{4 \beta ^2 q^2 \left(1-\omega _1\right) \left(1-\omega _2\right) \sinh ^2(\beta  q)}\Big]\nonumber\\
     &\approx \frac{h_\Delta(\beta)}{\left(1-\omega _1\right) \left(1-\omega _2\right)}\\
     h_{\Delta}(\beta)&= \sum_{q=1}^{\infty} \frac{e^{-q \beta  (\Delta -2) }}{4 \beta ^2 q^3  \sinh ^2(\beta  q)}
 \end{align}
 While the first term in the first line of \eqref{zexact} is of order unity as $\omega_i \to 1$, the second term (again on the first line of \eqref{zexact}) diverges in this limit. When either $\omega_1$ or $\omega_2$ (or both) is near to unity, therefore, the gas partition function is well approximated by the second term, i.e. by the second line of 
 \eqref{zexact}. 

At leading order in $1-\omega_1$ and $1-\omega_2$, the second line of 
\eqref{zexact} may equivalently be rewritten as 
\begin{equation} \label{zexactn}
  \ln Z \approx \frac{4h_\Delta(\beta)}{\left(1-\omega _1^2\right) \left(1-\omega _2^2\right)}
\end{equation}
\eqref{zexactn} takes precisely the form of the partition function of a `conformal' fluid (with the rotational chemical potentials $\omega_{1}$ and $\omega_{2}$) on $S^3$ \cite{Bhattacharyya:2007vs}, with  $4 h_\Delta(\beta)$ playing the role of the `zero angular velocity partition function' of the same fluid. 
 
The thermodynamical charges that follow from \eqref{zexactn} also, of course, take the fluid form \cite{Bhattacharyya_2008} and are explicitly given by 
\begin{equation} \label{thermo}
\begin{split} 
&E= \frac{8h_{\Delta}(\beta)}{\beta} \left(\frac{1}{(1-\omega _2^2)^2(1-\omega_1^2)}+\frac{1}{(1-\omega _1^2)^2(1-\omega_2^2)}\right)
\\
&J_1=\frac{1}{\beta}\frac{8h_\Delta(\beta)}{(1-\omega _1^2)^2(1-\omega_2^2)} \\
&J_2= \frac{1}{\beta}\frac{8h_\Delta(\beta)}{(1-\omega _2^2)^2(1-\omega_1^2)}  \\
& S=\frac{\left(1- \beta \partial_\beta \right) 4h_\Delta(\beta)}{\left(1-\omega _1^2\right) \left(1-\omega _2^2\right)}
\end{split}
\end{equation}
Note that while all the charges diverge when $\omega_i \to 1$ ($i=1,2$), \footnote{We assume here that ${\cal O}(1-\omega_1)={\cal O}(1-\omega_2)={\cal O}(1-\omega)$).} the energy and two angular momenta diverge like $\frac{1}{(1-\omega)^3}$ and the entropy diverges like $\frac{1}{(1-\omega)^2}$.
In both these cases, the entropy is subdominant compared to the energy, a point that will prove physically significant below. 

\subsection{Partition function for the $10d$  dilaton}
The partition function for all scalars arising from the $10d$ dilaton can be computed from the partition function in $AdS_5$ evaluated in \eqref{zexact}, by noting that the KK reduction of the 10d dilaton gives rise to operators of dimension $\Delta = n+4$ with $n=0, 1 \ldots \infty$. 
The operators of dimension $4+n$ transform as completely symmetric traceless tensors of the R symmetry $SO(6)$. The dimension of this irreducible representation is $d_n$ where
\begin{align}
    d_n=\frac{(2n+4)(n+1)(n+2)(n+3)}{24}.
\end{align}
As a consequence, the partition function for all KK modes arising out of the dilaton is given by 
\begin{align}\label{part10dil}
    \ln Z&\approx \sum_{n=0}^{\infty}d_n\frac{4h_{\Delta}(\beta)}{\left(1-\omega _1^2\right) \left(1-\omega _2^2\right)}\\
     h_{\Delta}(\beta)&= \sum_{q=1}^{\infty} \frac{e^{-q \beta  (\Delta-2) }}{4 \beta ^2 q^3  \sinh ^2(\beta  q)},\quad \Delta=n+4,
\end{align}
Computing the sum over $n$ we find
\begin{align}\label{hphi}
\ln Z&=\frac{1}{(1-\omega_1^2)(1-\omega_2^2)}\sum_{q=1}^{\infty}\frac{4\text{csch}^6\left(\frac{\beta  q}{2}\right) \text{csch}(\beta  q)}{128 \beta ^2 q^3} \nonumber\\
&=\frac{4h_{\phi}(\beta)}{(1-\omega_1^2)(1-\omega_2^2)}.
\end{align}
All thermodynamical formulae for this case are given by 
\eqref{thermo} with the replacement $h_{\Delta}(\beta) \rightarrow h_{\phi}(\beta)$. 

\subsection{The partition function for II Sugra on $AdS_5 \times S^5$ } 

In the previous subsection, we worked out the gas partition function of all modes that arose out of the Kaluza Klein reduction of the dilaton in $AdS_5 \times S^5$. It is not too difficult to repeat this exercise, for all fields of IIB Sugra 
on $AdS_5 \times S^5$, i.e. work out the multi-particle gas partition function over the full spectrum of small fluctuations on $AdS_5 \times S^5$ (at arbitrary values of $\omega_i$) and then specialize the final result to $\omega_1 \approx 1$ and $
\omega_2 \approx 1$. In this subsection, we present only our final answer (the interested reader will find details in Appendix \ref{fullgas}). Once again we find a result of the fluid dynamical form 
\begin{equation}\label{fflnz}
\ln Z= \frac{4h_{YM}(\beta)}{(1-\omega_1^2)(1-\omega_2^2)}.
\end{equation}
where 
\begin{equation}\label{hym}
h_{YM}(\beta) = h^B_{YM}(\beta) + h^F_{YM}(\beta)
\end{equation}
$h^B_{YM}(\beta)$ and $h^F_{YM}(\beta)$ (which, respectively, denote the bosonic and the fermionic contributions 
$h_{YM}$) are given by 

\begin{equation}\label{hbeta}
    \begin{split}
        h^B_{YM}(\beta) &=\sum_{n=1}^{\infty}\frac{(22 \cosh (\beta  n)+17 \cosh (2 \beta  n)+6 \cosh (3 \beta  n)+\cosh (4 \beta  n)+18) \text{csch}^7\left(\frac{\beta  n}{2}\right) \text{sech}\left(\frac{\beta  n}{2}\right)}{128 \beta ^2 n^3}\\ h^F_{YM}(\beta)&=\sum_{n=1}^{\infty} \frac{(-1)^{n+1}\cosh (\beta  n) (\cosh (\beta  n)+\cosh (2 \beta  n)+2) \text{csch}^7\left(\frac{\beta  n}{2}\right)}{8 \beta ^2 n^3}\\
    \end{split}
\end{equation}
Once again, all thermodynamical formulae for this case are given by \eqref{thermo} with the replacement $h_{\Delta}(\beta) \rightarrow h_{YM}(\beta)$ (see \eqref{hym}).
\section{Bulk Gas Stress tensor}\label{section5}

\subsection{The problem addressed}

Stationary black holes in $AdS_5$ always live in equilibrium with surrounding thermal gas. When the black hole angular velocities are less than (and well separated from) unity, the bulk gas carries energy of order unity, and so its backreaction on the bulk metric is of order $G_N$ \footnote{Here $G_N$ is Newton's constant.}$\sim \frac{1}{N^2}$ and so is negligible in the large $N$ limit. 
As we have explained in section \ref{pf}, however, this is no longer the case when one or both of $\omega_i$ approach unity.
In particular, when we set 
\begin{equation}\label{scom} \begin{split}
&1- \omega_1 =\frac{\alpha_1}{N^\frac{2}{3}}\\
& 1- \omega_2 =\frac{\alpha_2}{N^\frac{2}{3}}\\
\end{split}
\end{equation}
($\alpha_1$ and $\alpha_2$ are numbers of order unity)
then the energy and angular momentum in the bulk gas are both 
of order $N^2$. In this situation the backreaction of the 
bulk gas cannot be totally ignored. As in \cite{Kim:2023sig},
we will find that this energy is distributed over a very large region (upto a radial coordinate of order $N^{\frac{1}{3}}$). As a consequence, the energy density of the bulk gas is parametrically suppressed in $N$. For this reason, the backreaction of the bulk gas is accurately captured by a linear order. In leading order, moreover, the total backreaction is (effectively) a superposition of the backreaction from the black hole and the back reaction of the gas (see \cite{Kim:2023sig} for a detailed discussion of these points which carry over unchanged to this paper). 

In this section (see subsection \ref{absagg} for a discussion) we compute the bulk stress tensor for a Grey Galaxy built out of a single 10d bulk scalar field (`the dilaton') in $AdS_5 \times S^5$. As explained in the introduction, we postpone a discussion of the generalization to more realistic bulk matter to section \ref{cfbtst}.

\subsection{Computation in 10d Euclidean Space}

\subsubsection{Setting up the computation}\label{sec10d}

We wish to compute the expectation value of the bulk stress tensor for the dilaton in the thermal ensemble defined by 
\begin{equation}\label{pfform}
    \mathcal{Z}=\rm{Tr}\left[e^{-\beta( H-\omega_1 J_1-\omega_2 J_2)}\right]
\end{equation}
\eqref{pfform} is computed by a Euclidean path integral with  coordinate identifications given by
\begin{align} \label{bcs}
    (\tau, \phi_1,\phi_2,\Omega_i)\sim(\tau+\beta,\phi_1-i \beta\omega_1,\phi_2-i\beta\omega_2,\Omega_i)
\end{align}
\footnote{Let us define $\omega_i=i {\tilde \omega_i}$. When rewritten in terms of ${\tilde \omega_i}$, the Boltzmann factor can be rewritten as $e^{-\beta H} e^{ \left(i \beta \omega_1 J_1+i\beta \omega_2 J_2\right)}$. Now the operator $e^{i \alpha_iJ_i}$ affects a rotation by angle $\alpha_i$, and the path integral identifies the fields at $\tau=0$ with the appropriately rotated fields at 
$\tau=\beta$. On substituting ${\tilde \omega}_i = i \omega_i$ we recover the identification \eqref{bcs}.} 

 In this section we proceed with the computation roughly along the lines of the analysis presented in section 4.3 of \cite{Kim:2023sig} with one difference that we will highlight below. 
\footnote{As in 4.3 of \cite{Kim:2023sig}, it is sufficient to work with thermal $AdS_5$ rather than in the background of a black hole, as the dominant contribution from the gas comes from very large values of $r \sim N^{\frac{1}{3}}$, where these two spaces are essentially identical.} 
We find the one-point function of the stress tensor by evaluating the two-point function of the bulk scalar at separated points, taking the necessary derivatives 
(see \eqref{tform} below), and taking the coincident limit. After performing a temperature independent renormalization 
of this result, we find an unambiguous finite result for the (temperature dependent) part of the stress tensor.
 
The computation performed in this subsection differs qualitatively 
from that performed in section 4.3 of \cite{Kim:2023sig} in one respect. Instead of Kaluza Klein decomposing the 10-dimensional dilaton into its various distinct 5-dimensional fields, determining the stress tensor field by field, and
then summing these stress tensors, we 
work directly with the 10-dimensional dilaton, and so directly find the answer for the stress tensor after the summation. Remarkably enough the answer we find 
for the stress tensor from this (effectively summed up) 10-dimensional procedure is considerably simpler than the 
contribution from any one of its 5-dimensional KK modes. \footnote{As a consistency check of the results of 
 this subsection, however, in Appendix \ref{summing} we obtain the final result for the bulk stress tensor (\eqref{mats}) using the more tedious method (i.e. by summing the contributions from each of the (infinitely many) 5d fields).  }

Let us now proceed with the computation. 10-dimensional dilaton is a minimally coupled scalar field, and bulk stress tensor of the dilaton field is given by the simple formula
\begin{equation}\label{tform1}
T_{\mu\nu}=\partial_{\mu}\Phi\partial_{\nu}\Phi - \frac{g_{\mu\nu}}{2}(\partial\Phi)^2
\end{equation}
The expectation value of $T_{\mu\nu}
$ in the ensemble \eqref{pfform}
 is given by 
\begin{equation}\label{tform}
\langle T_{\mu\nu} \rangle = \lim_{x_1 \rightarrow x_2}  \left[ \partial^{x_1}_{\mu} \partial^{x_2}_{\nu} {\tilde G}(x_1-x_2)  - \frac{g_{\mu\nu}}{2}\left(   \partial^{\alpha, x_1} \partial^{x_2}_{\alpha} {\tilde G}(x_1-x_2)\right) \right]
\end{equation}
where ${\tilde G}(x_1-x_2)$ is the propagator on $AdS_5 \times S^5$ with the identifications \eqref{bcs}. Using the method of images, it follows that  ${\tilde G}(x_1-x_2)$ is given by a sum over propagators on ordinary (non-thermal) Euclidean $AdS_5 \times S^5$ via the method of images: 
\begin{equation}\label{tg}
{\tilde G}(x_1, x_2) = \sum_{q=-\infty}^\infty G\left(x_1, R^{q}(x_2) \right)
\end{equation} 
where the symbol $R^{q}$ denotes the combined action of a translation in global time by $\beta$, a rotation by angle $-q i \omega_1 \beta$ in the first two planes and 
the angle $-q i \omega_2\beta$ in the second two plane. The propagator 
$G\left(x_1, x_2\right)$, in turn,  
is given \cite{Dorn:2003au, Dai:2009zg} by the remarkably simple formula 
\begin{equation}\label{propmt}
 G\left(x_1,x_2 \right)=  \frac{\Gamma[4]}{4\pi^5}\left( \frac{1}{u + v}\right)^{4}\nonumber
\end{equation}

where $v$ is the chordal distance between the two points on $S^5$,  and $u$ is the chordal distance between the two points on $AdS_5$ (see Appendix \ref{propagator} for a review of the elegant derivation \cite{Dorn:2003au, Dai:2009zg} and for a definition of these chordal distances). As we will see below, all terms in the stress tensor that involve derivatives in the $S^5$ directions will turn out to be subleading, and so can be ignored. For our purposes, consequently, $v$ in \eqref{propmt} can simply be set to zero (recall that all images lie at the same point on $S^5$). On the other hand 
our image points are separated in $AdS_5$, 
so we need the expression for the chordal distance  $u$.  Working in global coordinates (the Euclidean continuation of the coordinates used in \eqref{metric} below), it is not difficult to verify that the chordal distance (defined in \eqref{udef} the Appendix) between points with coordinates $(r, \theta, \phi_1, \phi_2, \tau)$ and $ (r', \theta', \phi'_1, \phi'_2, \tau')$ is given by
\begin{align}\label{chordfull}
	u&=-2+ 2\Big[\sqrt{(1+r^2)(1+r'^2)}\cosh(\tau^E-{\tau'}^E)\nonumber\\
 &~~~~~~~~~~~~~~~~~~~~~~~~~~~~-r r'\left(\sin\theta\sin\theta'\cos\left(\phi_1-\phi_1'\right)
 +\cos\theta\cos\theta'\cos\left(\phi_2-\phi_2'\right)\right)\Big]
\end{align}

The `method of images',  \eqref{tg},  instructs us to evaluate the propagator between points related so that $\tau'-\tau=q\beta$, $\phi_1-\phi_1'=-i q\beta\omega_1$, $\phi_2-\phi_2'=-i q\beta\omega_2$, $r=r'=r, \theta=\theta' =\theta$ \footnote{In computing the stress tensor we need to take derivatives w.r.t. the coordinates 
$1$ and $2$. For this purpose we should, really, leave $r_1$ and $r_2$ as distinct until after we have taken all derivatives. As in \cite{Kim:2023sig}, however, all derivatives w.r.t. either 
$r$ or $\theta$ turn out to be subleading: the leading order stress tensor is obtained from derivatives in only the time, $\phi_1$ and $\phi_2$ directions. For this reason, we simply set $r=r'=r$
and $\theta=\theta'=\theta$ right at the beginning of the computation.} 
and $\Omega_{i}=\Omega_{'i}$ for the $q^{\rm th}$ image. It follows that the chordal distance between a point and its $q^{th}$ image (in $AdS_5$) is given by 
\begin{align}\label{chord}
	u_q&=-2+ 2\Big[  (1+r^2) \cosh(q\beta + \tau'-\tau)\nonumber\\
 &-r^2\left(\sin^2{\theta}\cos{\left(\phi_1-\phi_1'-i q\beta \omega_1\right)}+\cos^2{\theta}\cos{\left(\phi_2-\phi_2'-iq\beta\omega_2\right)}\right)\Big]\
\end{align}
\footnote{As we have mentioned above, the chordal distance between images on the sphere simply vanishes. }
(In the formula \eqref{chord} above we will set 
\begin{equation}\label{tauph}
\tau'=\tau, ~~~ \phi_1'=\phi_1, ~~~
\phi_2'=\phi_2
\end{equation}
after taking all relevant derivatives).

\subsubsection{The large $r$ scaling limit}

We wish to evaluate the bulk stress tensor at very large values of $r$. As we have explained around \eqref{tform}, the stress tensor is given by evaluating various derivatives of the Greens function \eqref{tg} evaluated between the image points described above. At large values of $r$, the quantity $u_q$ above generically becomes very large so the stress tensor (at large $r$) generically becomes very small. This generic expectation fails (and so we receive a substantial contribution to the bulk stress tensor) only if the bulk point and its image are separated in a nearly lightlike manner. This is possible (even though we are working in Euclidean spacetime) because the angular separation between images is imaginary (in contrast to the temporal separation, which, in Euclidean spacetime, is always real). If these two separations are nearly equal and opposite, points and their images can be effectively nearly lightlike separated.

In global $AdS_5$, the length of the time circle is $\sqrt{1+r^2} \beta$. The lengths of the two angular circles are $\beta r \omega_1 \sin \theta $ and $\beta r \omega_2 \cos \theta$. When $\omega_1$ and $\omega_2$ are both (substantially, i.e. 
by order unity) less than one, the effective length of the angular circle, 
which equals $r\beta\sqrt{( \omega_1^2 \sin^2 \theta+ \omega_2^2 \cos^2 \theta)}$, is always substantially (i.e. by an order one fraction) less than the length of the time circle, and the bulk stress tensor is very small at large $r$. When $\omega_1$ is very near to unity, but $\omega_2$ is substantially smaller than unity, the angular circle approaches the size of the time circle (at large $r$) in a small range of $\theta$ values around $\frac{\pi}{2}$. In this case, the stress tensor is substantial upto $r \sim \frac{1}{\sqrt{1-\omega_1}}$. In this case, we find a two-spatial dimensional disk of matter around 
$\theta= \frac{\pi}{2}$. 
Similar remarks apply to the case $\omega_2$ near unity, and $\omega_1$ substantially less than unity. In this section, we focus attention on the case 
that $\omega_1$ and $\omega_2$ are both of order unity. In this case, we find a substantial contribution to the bulk stress tensor at every value of $\theta$, 
upto radial coordinates of order 
$$ r \sim\frac{1}{\sqrt{(1-\omega_1\sin^2{(\theta})-\omega_2\cos^2{(\theta)})}}$$
For the reasons mentioned above, the bulk stress tensor of this subsection is non negligible when $x^2$ defined by 
\begin{align}\label{reo1o2}
  r^2(1-\omega_1\sin^2{(\theta})-\omega_2\cos^2{(\theta)})=x^2
\end{align}
remains of unity or smaller (even though $r$ is taken to be very large). 

To first order in $1-\omega_1$ and $1-\omega_2$, 
\begin{equation}\label{gmaaa}
2(1-\omega_1\sin^2{(\theta)}-\omega_{2}\cos^2{(\theta})) = 1-\omega_1^2 \sin^2{(\theta)}-\omega_{2}^2\cos^2{(\theta)}=\frac{1}{\gamma^2(\theta)},
\end{equation} 
where $\gamma(\theta)$ was defined in \eqref{gammafac}.
Our definition of the rescaled coordinate $x$ may, therefore, 
be rewritten as
\begin{equation}\label{defx}
x^2 =  \frac{r^2}{2\gamma^2(\theta)}.
\end{equation}

In this scaling limit, the formula for the chordal distance $u_q$ (with the replacement \eqref{tauph})
simplifies to 
\begin{align}\label{cho1o2}
   u_q&= 2 \left(\cosh (\beta  q)+\beta  q r^2 \sinh (\beta  q) \left(1-\omega _1 \sin ^2(\theta )-\omega _2 \cos ^2(\theta )\right)-1\right)\nonumber\\
   &=2 \left(\cosh (\beta  q)+\beta  q x^2 \sinh (\beta  q)-1\right)
\end{align}
(in going from the first to the second line of \eqref{cho1o2} we have used 
\eqref{reo1o2}). 
Similarly, we find
\begin{equation}
\begin{split}
&\partial_{\tau} u_q= -\partial_{\tau'} u_q=-2\left(\left(r^2+1\right) \sinh (\beta  q)\right) \\
&\partial_{\phi_1} u_q= -\partial_{\phi_1'} u_q=2i r^2 \sin ^2(\theta) \sinh \left(\beta  q \omega _1\right) \\
&\partial_{\phi_2} u_q= -\partial_{\phi_2'} u_q=2i r^2 \sin ^2(\theta) \sinh \left(\beta  q \omega _2\right)
\end{split}
\end{equation}

\subsubsection{Simplification from Chirality}\label{chir}

As in \cite{Kim:2023sig} the bulk gas is effectively chiral (it moves, at almost the speed of light, in the direction of the vector field \footnote{As written, the vector field $w^{\mu}\partial_{\mu}$ is
not a velocity vector field as $w^2 \neq -1$. The velocity vector field of the gas in the bulk equals 
\begin{equation}\label{gvno}
\gamma(x, \theta)\left( \partial_t -\omega_1 \partial_{\phi_1} - \omega_2 \partial_{\phi_2} \right), ~~~~\gamma(x, \theta) 
= \frac{1}{\sqrt{1+ \frac{r^2}{\gamma^2(\theta)}}}= \frac{1}{\sqrt{1+2 x^2}}  
\end{equation} 
In the limit  $\omega_1 \to 1$ and $\omega_2 \to 1$, this vector field is clearly proportional to $w^\mu \partial_\mu$, with the $x$ dependent proportionality factor displayed above. 
}
\begin{equation}\label{fv} 
w^\mu \partial_\mu =  \left( \partial_t -\partial_{\phi_1} -\partial_{\phi_2} \right) 
\end{equation} 
For this reason (as we will see below) 
at leading order 
\begin{equation}\label{dirder}
g^{\mu_1 \nu_1} g^{\mu_2 \nu_2} \partial_{\mu_1}\partial_{\mu_2}\langle\Phi(x_1)\Phi(x_2)\rangle_q \propto u^{\nu_1}
u^{\nu_2}
\end{equation}
(here $q$ denotes the contribution from the $q^{th}$ and $-q^{th}$ image: as in Sec 4.3 of \cite{Kim:2023sig} we have renormalized our answer by dropping the (temperature and $\omega_i$ independent) contribution from $q=0$). As $u^\mu$ is an almost lightlike vector, the second term 
in \eqref{tform} is of subleading order, and we find 
\begin{align}
T_{\mu_1\mu_2}=-2\sum_{q=1}^{\infty} K^q_{\mu_1\mu_2}
\end{align}

\begin{equation}\label{olmju1}
\begin{split}
K^q_{\mu_1\mu_2}&=\partial_{\mu_1}\partial_{\mu_2}\langle\Phi(x_1)\Phi(x_2)\rangle_q\nonumber\\
&=\partial_{u}^2G(u,v)\frac{\partial u_q}{\partial\mu_1} \frac{\partial u_q}{\partial \mu_2}+ \partial_u G(u,v) \frac{\partial^2 u_q}{\partial \mu_1\partial \mu_2}
\end{split}
\end{equation}

(After all derivatives are taken, $u_q$ is evaluated on the configuration $\eqref{tauph}$ )

The expression on the RHS of \eqref{olmju1}
has two kinds of terms: those proportional to a second derivative of $u_q$ and those proportional to the square of the first derivatives of $u_q$. It is not difficult to verify that terms of the first sort are subleading (in the large $r$ limit) to terms of the second sort. 
\footnote{
For instance, 
\begin{equation}\label{deru}
    \begin{split}
           \frac{\partial u}{\partial\tau^E}\frac{\partial u}{\partial\tau^{\prime E}}&=-4 r^4 \sinh^2{(q\beta)}\nonumber\\
           &=-16\gamma^4(\theta)x^4\sinh^2(q\beta)\\
           \frac{\partial^2 u}{\partial \tau^E \partial {\tau'}^E}&=-2 r^2 \cosh{(q\beta)}\nonumber\\
           &=-4\gamma^2(\theta)x^2\cosh(q\beta)
    \end{split}
\end{equation}
Note that while the RHS of the first line of \eqref{deru} is of order $r^4$, the second line is of order $r^2$, illustrating the fact that terms of the schematic form $\partial^2 u$ are subleading (and so can be ignored) compared to terms of the schematic form $(\partial u)^2$. }

\subsubsection{Final Result at leading order}\label{stressf}

From the discussion of the previous subsection, we see that $K^q_{\tau\tau}$ simplifies, at leading order, to 
\begin{equation}
    K^q_{\tau\tau}=\partial_{u}^2G(u,v)\frac{\partial u_q}{\partial\tau} \frac{\partial u_q}{\partial \tau'}|_{u=u_q}
\end{equation}
Putting everything together, in leading order we find 
\begin{align}
    K^q_{\tau\tau}&=-16\gamma^4(\theta)x^4\sinh^2(q\beta)\partial^2_{u}G(u,v)|_{v\rightarrow 0}\nonumber\\
    &=-\frac{ 4\gamma^4(\theta)x^4 \sinh ^2(\beta  q)\Gamma[6]}{\pi ^5 u_q^6}
\end{align}
\begin{align}
    \langle T_{00}\rangle&=\sum_{q=1}^{\infty}\frac{8\gamma^4(\theta)x^4\sinh^2(q\beta)\Gamma[6] }{\Gamma[2]\pi^5 u_q^6}
\end{align}
 
Other components of the stress tensor are computed in a similar manner. At leading order we find 
\begin{align}\label{stress10}
    \begin{split}
         \langle  T_{0\phi_1}\rangle&=\sum_{q=1}^{\infty}\frac{4\gamma^4(\theta)x^4\sinh^2(q\beta)\sin^2\theta\Gamma[6]}{\pi ^5 u_q^6}\\
           \langle  T_{0\phi_2}\rangle&=\sum_{q=1}^{\infty}\frac{4\gamma^4(\theta)x^4\sinh^2(q\beta)\cos^2\theta\Gamma[6]}{\pi ^5 u_q^6}\\
        \langle  T_{\phi_1\phi_1}\rangle&=\sum_{q=1}^{\infty}\frac{4\gamma^4(\theta)x^4\sinh^2(q\beta)\sin^4\theta\Gamma[6]}{\pi ^5 u_q^6}\\
        \langle T_{\phi_2\phi_2}\rangle&=\sum_{q=1}^{\infty}\frac{4\gamma^4(\theta)x^4\sinh^2(q\beta)\cos^4\theta\Gamma[6]}{\pi ^5 u_q^6}\\
       \langle T_{\phi_1\phi_2}\rangle&=\sum_{q=1}^{\infty}\frac{4\gamma^4(\theta)x^4\sinh^2(q\beta)\sin^2\theta\cos^2\theta\Gamma[6]}{\pi ^5 u_q^6} \\
       \langle T_{\theta i}\rangle&=0, \quad\quad i\in \{\tau_E, \theta,\phi_1,\phi_2\}\\
       \langle T_{ri}\rangle &=0\\
       \langle T_{\alpha\beta}\rangle&=0\\
       \langle T_{\mu\alpha}\rangle&=0, \quad y_{\alpha},y_{\beta}\in S^5,\quad x_{\mu}\in AdS_5
    \end{split}
\end{align}
In the matrix form,

\begin{equation}\label{mats}
   \langle T_{\mu\nu}\rangle= 4\mathcal{F}(x)\gamma^4(\theta)\left(
\begin{array}{ccccc}
 1 & 0 & 0 & \sin ^2(\theta ) & \cos ^2(\theta ) \\
 0 & 0 & 0 & 0 & 0 \\
 0 & 0 & 0 & 0 & 0 \\
 \sin ^2(\theta ) & 0 & 0 & \sin ^4(\theta ) & \sin ^2(\theta ) \cos ^2(\theta ) \\
 \cos ^2(\theta ) & 0 & 0 & \sin ^2(\theta ) \cos ^2(\theta ) & \cos ^4(\theta ) \\
\end{array}
\right)
\end{equation}
here, $\mu ,~\nu \in \{t, r,\theta, \phi_1,\phi_2 \}$ and 
\begin{equation}\label{fxdef}
    \begin{split}
         \mathcal{F}(x)=&\sum_{q=1}^{\infty}\frac{2x^4\Gamma[6]\sinh^2{\beta q}}{\pi ^5 u_q^6} \\=&\sum_{q=1}^{\infty}\frac{x^4\Gamma[6]\sinh^2{\beta q}}{2^5\pi ^5 ( \left(\cosh (\beta  q)+\beta  q x^2 \sinh (\beta  q)-1\right))^6}\\ 
    \end{split}
\end{equation}
The final form of the bulk stress tensor can be succinctly rewritten as 
\begin{equation}\label{bst}
T_{\mu\nu}=4\mathcal{F}(x)\gamma^4(\theta) w_\mu w_\nu
\end{equation}
where $w_\mu$ is defined as the velocity vector field \eqref{fv}, lowered using the boundary metric\footnote{Note, as a consequence of our definition (lowering with the boundary rather than the bulk metric), that the bulk one-form field corresponding to the fluid velocity $w^\mu \partial_\mu$ does not equal $\omega_\mu dx^\mu$. To leading order, instead, this bulk oneform field 
equals $r^2 \omega_\mu dx^\mu$.}
$ds^2=-dt^2 + d \Omega_3^2$. Note that the  
stress tensor \eqref{bst} takes the form \eqref{dirder} as foreshadowed above. 
\footnote{Using \eqref{gvno} this stress tensor can be rewritten in terms of the local fluid velocity. This rewriting is not very useful, however as $T_{\mu\nu}$ in \eqref{bst} is more simply written in terms of $\gamma(\theta)$ rather than $\gamma(x, \theta)$ in \eqref{gvno}. In other words \eqref{bst} does not 
admit a particularly natural rewriting in terms of the bulk local fluid velocity.}

We will find it useful below to have explicit expressions for $\mathcal{F}(x)$ at large and small values of $x$. At small values of $x$ we find 
\begin{equation} \label{cfsmallx}
\begin{split}
&\mathcal{F}(x)=x^4 D  + {\cal O}(x^6) \\
& D= \left( \frac{\Gamma[6]}{2^6\pi ^5}
\sum_{q=1}^{\infty}\frac{1}{2\sinh^2{\beta q} (\cosh (\beta  q)-1)^6}\right)\\
\end{split}
\end{equation}
whereas at large values of $x$
\begin{equation} \label{cflargex}
\begin{split}
&\mathcal{F}(x)=
\frac{B}{x^8}+ {\cal O}\left(\frac{1}{x^{10}}\right) \\
& B= \frac{15  \Gamma[6]}{ 4\pi ^5 \beta ^6}
\sum_{q=1}^{\infty}\frac{ \text{csch}^4(\beta  q)}{ q^6 }
\end{split} 
\end{equation}

\subsection{The improved bulk stress tensor and bulk fluid dynamics}\label{impbulk}

\subsubsection{The advantage of `Improving' the bulk stress tensor}

Consider the leading order bulk stress tensor $T_{\mu\nu}$ (see \eqref{bst}). It is easy to verify that $T_{\mu\nu} T^{\mu\nu} \sim \frac{\gamma^8(\theta)}{r^4} \sim \gamma^4$. \footnote{As usual,  we take $r$ to scale like $\gamma(\theta)$). } As the natural invariant formed out of the leading order stress tensor is itself of order $\gamma^4$, we say that the stress tensor is of order $\gamma^2$. 

The stress tensor computed \eqref{bst} is, of course, only the first term in an expansion of the true bulk stress tensor  \eqref{tform} in a power series expansion in $1/\gamma^2(\theta)$. More precisely, the true bulk stress tensor \eqref{tform} can be expanded as
\begin{equation}\label{sttherm}
\langle T_{\mu \nu} \rangle = \left( 4\mathcal{F}(x)\gamma^4(\theta) w_\mu w_\nu \right) +T^{(1)}_{\mu\nu} + T^{(2)}_{\mu \nu} + \ldots
\end{equation}
where $T^{(1)}_{\mu\nu}$ is of order unity,  $T^{(2)}_{\mu\nu}$ is of order $\frac{1}{\gamma^2(\theta)}$, etc. 

Of course the true stress tensor (the LHS of \eqref{sttherm}) is conserved, so that 
\begin{equation}\label{ftenscons}
\nabla^\mu \langle T_{\mu \nu} \rangle =0.
\end{equation}
By explicit calculation we find that  
\begin{equation}\label{bstin}
\nabla^\mu \left( 4\mathcal{F}(x)\gamma^4(\theta) w_\mu w_\nu \right)= V_\nu, ~~~~~V_\mu V^\mu \sim {\cal O}(1)
\end{equation}
In other words, the leading order stress tensor is not, by itself conserved, but the violation of 
conservation happens only at order unity (recall that the leading stress tensor itself was of order 
$\gamma^2(\theta)$ ), and so is of the form that can be cancelled by subleading corrections to \eqref{bst}
\footnote{Explicitly, this works as follows.
Plugging \eqref{bstin} and \eqref{sttherm} into \eqref{ftenscons}, we see that it must be that 
\begin{equation}\label{bstinss}
\nabla^\mu T^{(1)}_{\mu \nu}= V_\nu + {\cal O}\left(\frac{1}{\gamma^2(\theta)}\right)
\end{equation}}

While the non conservation of the leading order stress tensor \eqref{bstin} is thus not a problem in principle, it is an irritation in practice \footnote{Because the linearized Einstein equations are ill posed when the stress tensor on the RHS is not conserved.}. For this reason, we will pause, in the rest of the subsection, to `improve' the stress tensor \eqref{bst}, i.e, to find a correction term to the bulk stress tensor, that is itself of order 
$\gamma^0(\theta)$ (and so does not modify the 
stress tensor at leading order), but is chosen in
a manner that ensures that the resultant bulk stress 
tensor is exactly conserved \footnote{A similar maneuver was executed in a different context in  \cite{Dandekar:2017aiv} (the context there was the investigation of the membrane like behaviour of black holes at large $D$). }. In fact such an improvement term turns out to be surprisingly easy to find. 
\footnote{In a different but somewhat similar context (namely the study of black hole dynamics at large $D$ along the lines developed in \cite{Bhattacharyya:2015dva, Bhattacharyya:2015fdk, Dandekar:2016fvw, Bhattacharyya:2016nhn}), a conceptually similar all order improvement to the stress tensor of the large D black hole `membrane' was implemented in \cite{Dandekar:2017aiv}.}

\subsubsection{The improved stress tensor from a fluid inspired ansatz}\label{impbst}

Inspired by the analyses of 
\cite{Bhattacharyya:2007vs, Banerjee:2012iz, Bhattacharyya:2008mz, Jensen:2012jh}, we search for an improved version \eqref{bst} that takes the 
`perfect fluid' form
\begin{equation}\label{stform}
    T^{\mu\nu}= T_1(r,\theta)w^{\mu}w^{\nu} + T_2(r,\theta) g^{\mu\nu},
\end{equation}
where $w^{\mu}$ is the vector field \eqref{fv} (recall that $w^\mu$ is proportional to the bulk fluid velocity)

It is not difficult to verify that the conservation condition $\nabla_{(b)}^{\mu}T_{\mu\nu}=0$ \footnote{The subscript $(b)$ tells us that the covariant derivative is taken in the background of vacuum $AdS_5$.} is obeyed provided the coefficient functions  $T_i(r,\theta)$  obey the following  differential equations:
\begin{equation}\label{consequ}
\begin{split}
        &\frac{\partial T_2(r,\theta )}{\partial r}+\frac{r T_1 (r,\theta )}{\gamma (\theta )^2}=0, \\
        &\frac{\partial T_2(r,\theta )}{\partial \theta}-\frac{r^2 \gamma '(\theta ) T_1 (r,\theta )}{\gamma (\theta )^3}=0.
    \end{split}
\end{equation}
 
\footnote{The two equations listed above correspond, respectively, to the $r$ and $\theta$ components of the conservation condition.}

Imposing the consistency condition that arises from the commutativity of partial derivatives acting on $T_2$,  yields the following differential equation for
the function $T_1$: 
\begin{equation}\label{toneconst}
    r \gamma '(\theta ) \frac{\partial T_1(r,\theta )}{\partial r}+\gamma (\theta ) \frac{\partial T_1(r,\theta )}{\partial \theta}=0.
\end{equation}
This equation is easily solved using the method of characteristics; the most general solution is given by 
\begin{equation}
T_1(r,\theta)=f\left(\frac{r}{\gamma(\theta)}\right).
\end{equation} for some function $f$. The term proportional to $w^\mu w^\nu$ in \eqref{stform}  matches \eqref{bst} at leading order, provided 
we choose 
\begin{equation}
    f\left(\frac{r}{\gamma(\theta)}\right)= 4\mathcal{F}\left(\frac{r}{\sqrt{2}\gamma(\theta)}\right)\frac{\gamma^4(\theta)}{r^4}.
\end{equation}
\footnote{Lowering both indices w.r.t. the bulk spacetime metric, turns $w^\mu w^\nu$ into $r^4 w_\mu w_\nu$ (recall we have defined $w_\mu$ as $w^\mu$ lowered with the boundary metric), matching 
\eqref{bst}.}

With the function $T_1$ in hand, we can now use 
\eqref{consequ} to evaluate the function $T_2$ by 
simple integration. We obtain
\begin{equation}
     T_2(r,\theta)= P_{\infty}+\int_{x}^{\infty} dz~\frac{2\mathcal{F}(z)}{z^3},
\end{equation}
where the variable $x$ was defined in \eqref{defx}: 
$x=\frac{r}{\sqrt{2}\gamma(\theta)}$ and $P_{\infty}$ is an integration constant. Since the bulk stress tensor must vanish at $x=\infty$, we conclude that $P_\infty=0$. Note that the function $T_2$ is of order unity (in $\gamma(\theta)$ counting) as anticipated above. 

In summary, our final expression for the improved bulk stress tensor in  $AdS_5$ takes the form 
\begin{equation}\label{frst}
\begin{split}
    T^{\mu\nu}(r,\theta) &=\frac{\mathcal{F}\left(x\right)}{x^4}w^{\mu}w^{\nu} + \left(\int_{x}^{\infty} dz~\frac{2\mathcal{F}(z)}{z^3}\right)g^{\mu\nu} \\
&=\frac{\mathcal{F}\left(x\right)(1+2x^2)}{x^4}u^{\mu}u^{\nu} + \left(\int_{x}^{\infty} dz~\frac{2\mathcal{F}(z)}{z^3}\right)g^{\mu\nu}.
    \end{split}
\end{equation}
where, in the second expression, we have presented the stress tensor in terms of the fluid velocity $u^{\mu} = \frac{1}{\sqrt{1+2x^2}}w^{\mu}$. $u^\mu$
is a vector field proportional to $w^\mu$, but normalized (using the bulk metric) so that 
$u^{\mu}u_{\mu}=-1$. The stress tensor 
\eqref{frst} obeys the conservation equation 
\eqref{ftenscons} (in the undeformed background $AdS_5$ spacetime) without any approximation.

\subsubsection{The stress tensor of a perfect bulk fluid}\label{bulkstrten}

In the previous subsubsection we used 
fluid dynamics as inspiration to propose an ansatz for the form of the improved bulk stress tensor, and then proceeded to use conservation to determine the detailed form of this improved stress tensor. In this subsubsection \footnote{The material from previous subsubsection and Appendix \ref{eqflu} are not used anywhere else in this paper. This subsubsection and the related Appendix can be thought of as tantalizing observations that should (hopefully) find 
explanations in future work.} and 
Appendix \ref{eqflu}, we explain that the relationship of our bulk system to the fluid description is more than a useful trick. We demonstrate, infact, that the partition function of our bulk gas is the integral of a bulk local quantity, namely 
\begin{equation} \label{pfint}
    W = \ln Z = \int d^p x \sqrt{g_p}\, \frac{P\big(T(x)\big)}{T(x)}\, 
\end{equation}
 where
\begin{equation}\label{txpt}
\begin{split}
    P(x) &= \int_x^\infty dz\, \frac{2\mathcal{F}(z)}{z^3} \\
    T(x)&=\frac{T_0}{\sqrt{1+2x^2}}
\end{split}
\end{equation} 
Moreover the stress tensor presented in \eqref{frst} also agrees exactly with the stress tensor obtained by varying \eqref{pfint} w.r.t. the metric (see Appendix \ref{eqflu} and \cite{Banerjee:2012iz} for precisely what this means). 

The reason these observations are significant goes as follows. In their studies of the structure of hydrodynamics in equilibrium, the authors of \cite{Banerjee:2012iz} studied the structure of the partition function of any quantum field theory (like the one loop bulk gas studied in this section) on a Euclidean manifold with compactified `time' along an appropriate killing direction. In that context they concluded that the partition function of any field theory, at leading order in the high `temperature' limit, takes the form \eqref{pfint}, with the function $P(T)$ being identified with the thermodynamical pressure as a function of temperature of the system. 
We see from \eqref{pfint} that our bulk gas does both more -  and less - than what is expected from this point of view. 

The bulk gas does more than what is expected in the following sense: the partition function of our bulk fluid takes the form \eqref{pfint} even though the local temperature of the gas is not large (in, for instance, units of bulk curvature or the masses of bulk fields). 
Recall that the `thermal' compactification circumference - of the angular velocity twisted Euclidean circle - in our context equals $1/T(x)$ 
(see \eqref{txpt}) and does not go to zero in the scaling limit studied in this paper.

The sense in which the bulk gas does less than what is expected of it is the following: the function $P(T)$ \footnote{In \eqref{txpt}, $P$ is presented as a function of the coordinate $x$. It can be converted into a function of $T$ by solving for $x$ in terms of $T$ from the second of \eqref{txpt}.} that appears in \eqref{txpt}
does not appear to have a simple thermodynamical significance. 
As far as we can tell, it is not possible to infer the function $P(T)$ in \eqref{txpt} from the thermodynamical properties of the bulk gas in flat space (or a flat box). 

We do not clearly understand why the bulk partition function of our thermal gas is so simple, even though the gas is not necessarily at high temperature, or the 
detailed origin of the function $P(T)$ in \eqref{txpt}.
(However, these two points seem very related: in the case that the bulk gas is taken to be a 5 dimensional massless field, and $T_0$ is taken to be large, it is not difficult to check that the pressure function 
in \eqref{txpt} reduces to the expected thermodynamical pressure of this gas). We leave further analysis of this very interesting (but somewhat confusing) point to future work.  
The authors of \cite{Banerjee:2012iz} also computed the stress tensor that emerges out of the partition function \eqref{pfint}, and found that it takes the form 
\begin{equation}\label{eq:perfect_fluid_tensorint}
 \begin{split}   
 T_{\mu\nu} &= \left(\epsilon \left(T(x) \right)+ {P}\left(T(x) \right) \right)  u_\mu u_\nu + {P}\left(T(x) \right) g_{\mu\nu} \\
  \epsilon(T) &=  -P + T 
  \frac{\partial P}{\partial T}\\
\end{split}
\end{equation}
Using \eqref{txpt} (and the identity $\frac{\partial P}{\partial T}= \frac{\left(\frac{\partial P}{\partial x}\right)}{\left(\frac{\partial T}{\partial x}\right)}$), it is easy to specialize \eqref{eq:perfect_fluid_tensorint} to the current situation. 
In \eqref{eqflu} we demonstrate that the result agrees perfectly with 
\eqref{frst}.

\section{Back-reactions on the metric}\label{section6}

In this section, we compute the response (the backreaction) of the bulk metric to the bulk stress tensor \eqref{bst} computed in the previous section. 
We continue to work within the model of the previous section (the bulk matter is given by a single massless 10d scalar field). We discuss the generalization to a more realistic bulk matter content in the next subsection. 

\subsection{$AdS_5$ metric in terms of 
right invariant one forms} 

The metric on a unit $S^3$ 
\begin{equation}\label{sthree}
ds^2=  d\theta^2 + \sin(\theta)^2d\phi_1^2 + \cos(\theta)^2d\phi_2^2
\end{equation}
\footnote{Here $\phi_1$ and $\phi_2$ represent 
rotations in the two orthogonal two planes 
in the embedding $R^4$}
can be rewritten as
\begin{equation}\label{newsthree}
ds^2=(\sigma_1)^2 +  (\sigma_2)^2 + (\sigma_3)^2 
\end{equation}
where $\sigma_i$ are right invariant oneforms \footnote{This means that they are annihilated by the Lie derivatives 
corresponding to $SU(2)_R$ generators. 
This fact will, however, play no essential role in our computation.} defined in \eqref{oneforms1} and \eqref{phipsi}.
It follows that the metric of $AdS_5$, in global coordinates, can be written as 
\begin{equation}\label{metric}
ds_{AdS_5}^2=-(1+r^2)dt^2+\frac{dr^2}{1+r^2}+r^2\left( (\sigma_1)^2 +  (\sigma_2)^2 + (\sigma_3)^2 \right)
\end{equation} 

\subsection{Bulk Stress tensor in terms of right invariant oneforms}

The bulk stress tensor for our bulk gas was presented in \eqref{frst}. 
In the large $\gamma$ limit of interest to this paper, the second term in \eqref{frst} (the term proportional to $g_{\mu\nu})$) is subdominant compared to the first term (the term proportional to $u_\mu u_\nu$. 
Retaining only the leading term, we find the stress tensor \eqref{bst}. 
It is not difficult to check that this stress tensor can be succinctly rewritten in the form

\begin{equation}\label{tform2}
T= 4{\mathcal F}(x) \gamma^4(\theta)  \left( dt - \sigma_3
\right) \left( dt - \sigma_3
\right)
\end{equation} 

where $x$ (see \eqref{reo1o2}) is defined as 
\begin{equation}\label{xdef}
 x= \frac{r}{\sqrt{2}\gamma(\theta)}
 \end{equation}

\subsection{Computation of the backreaction}

As we have explained in subsection \ref{absagg}, the computation of the back reaction to the bulk stress tensor is a simple exercise because it can be performed in a manner that is `point by point' from a boundary viewpoint. In this subsection, we perform this computation. In comparing the discussion below with that of subsection \ref{absagg}, it is useful to keep in mind that we are working at values of $\omega_1$ and $\omega_2$ such that $1-\omega_1$ and $1-\omega_2$ are both $\sim \frac{1}{N^\frac{2}{3}}$, so that 
$\gamma \sim N^\frac{1}{3}$.

As we see from \eqref{tform}, the stress the bulk gas stress tensor is a nontrivial function of the coordinate $x= \frac{r}{\sqrt{2}\gamma(\theta)}$ 
\footnote{The coordinate $x$ is proportional (with an order one proportionality constant) to $r'$ in \eqref{sctcc}.}Recall that, in the limit of interest, $\gamma(\theta)$ is parametrically large. 
For this reason (and as in 
\cite{Kim:2023sig}) it is convenient to 
rewrite the metric of $AdS_5$ \eqref{metric} in this coordinate. In order to remove all explicit factors of 
$\gamma(\theta)$ from the resultant metric, we also focus on the neighborhood of some time $t_0$, and some angular coordinates $\phi_0, \psi_0$ and $\theta_0$, work with the scaled coordinates (for deviations away from our central coordinate)
\begin{equation}\label{scaledcoord}
\begin{split} 
&t= t_0 + \frac{\tau}{ \sqrt{2}\gamma(\theta_0) }\\
& \phi= \phi_0 + \frac{\phi'}{ \sqrt{2}\gamma(\theta_0) }\\
&\psi= \psi_0+ \frac{\psi'}{ \sqrt{2}\gamma(\theta_0) }\\
&\theta =  \theta_0+ \frac{\theta'}{ \sqrt{2}\gamma(\theta_0) }\\
\end{split}
\end{equation} 
Working to leading order in $\frac{1}{ \sqrt{2}\gamma(\theta_0) }$, the metric of $AdS_5$ becomes 
\begin{equation}
    \begin{split}
        ds^2 =&-x^2d\tau^2+\frac{dx^2}{x^2}+x^2\left( (\sigma'_1)^2 +  (\sigma'_2)^2 + (\sigma'_3)^2 \right) +d\Omega_5^2\\
    \end{split}
\end{equation}
where $\sigma'_i= \sqrt{2}\gamma(\theta_0) \sigma_i $. The explicit factor of $ \sqrt{2}\gamma(\theta_0)$ disappears when these forms are written in terms of differentials of primed coordinates: explicitly 
\begin{equation}\label{oneforms}
    \begin{split}
    \sigma'_1 &= \frac{1}{2}\left(\sin(\phi_0 + \frac{\phi'}{ \sqrt{2}\gamma(\theta_0) }) d (2\theta') - \cos(\phi_0 + \frac{\phi'}{ \sqrt{2}\gamma(\theta_0) })\sin(2\theta_0 + \frac{2\theta'}{ \sqrt{2}\gamma(\theta_0) })d\psi'\right)\\ \sigma'_2 &=\frac{1}{2}\left(\cos(\phi_0 + \frac{\phi'}{\sqrt{2}\gamma(\theta_0)}) d(2\theta') + \sin(\phi_0 + \frac{\phi'}{ \sqrt{2}\gamma(\theta_0) })\sin(2\theta_0 + \frac{2\theta'}{ \sqrt{2}\gamma(\theta_0) })d\psi'\right)\\\sigma'_3 &=\frac{1}{2}(d\phi' - \cos(2\theta_0 + \frac{2\theta'}{ \sqrt{2}\gamma(\theta_0) })d\psi')\\
    \end{split}
\end{equation}

We would now like to define the coordinates 
$w, y$ and $z$ s.t. 
\begin{equation}\label{coorddef}
dw= \sigma_1', ~ dy=\sigma_2', ~dz=\sigma_3'
\end{equation}
Locally, this is possible if and only if 
$d \sigma_i'$ vanish. In the current context (always working in rescaled coordinates), $d\sigma_i'$ are of order $\frac{1}{ 2\gamma(\theta_0) }$, and so vanish at leading order. It follows that we can (in leading order) find coordinates 
that satisfy \eqref{coorddef}. In these coordinates
the metric of $AdS_5$ can be rewritten 
in the `Poincare Patch form'
\begin{equation}\label{ads5rescale}
ds^2= \frac{dx^2}{x^2} + x^2  \left( -d\tau^2
+ dw^2+dy^2 + dz^2 \right) 
\end{equation} 

The bulk stress tensor can be rewritten as
\begin{equation}\label{bstn}
T=\mathcal{F}(x)2\gamma^2(\theta_0) \left(d \tau-d z \right)\left(d \tau-d z \right) 
\end{equation} 

We would now like to compute the backreaction on $AdS$ space by solving the Einstein equation 
\begin{equation}\label{eieq0}
 R_{\mu\nu} -\frac{1}{2}R g_{\mu\nu} - 6 g_{\mu\nu} = 8 \pi G^{(10)}_N  T_{\mu \nu}
\end{equation} 

In order to make the RHS of \eqref{eieq0}
concrete, we recall that the 10-dimensional Newton constant equals $G^{(10)}= 2^3 \pi^6 g_s^2 \alpha^{'4}$, \footnote{So that the coefficient of 
$\int \sqrt{g} R$ equals $\frac{1}{(2 \pi)^7 g_s (\alpha')^4}= \frac{1}{16 \pi G_N}$}
while 
$R_{AdS}$ (in the case of ${\cal N}=4$ Yang Mills theory) is $R_{AdS}^4=4 \pi g_s N \alpha^{'2}$. It follows that 
\begin{equation}\label{Gnew}
G^{(10)}= \frac{\pi^4 R_{AdS}^8}{2 N^2}
\end{equation} 
so that 
\begin{equation}\label{gfive}
G^{(5)}= \frac{G^{(10)}}{\Omega_5 R_{AdS}^5} = \frac{\pi^4 R_{AdS}^3}{2 N^2 \pi^3} = \frac{\pi R_{AdS}^3}{ 2N^2} 
\end{equation}
\footnote{We have used the fact that the volume of a unit $S^5$ is $\pi^3$.}
We are working in units in which $R_{AdS}=1$. Using \eqref{bst} and \eqref{Gnew} we see that \eqref{eieq}
can be rewritten as 
\begin{equation}\label{eieq}
\begin{split}
 \left( R_{\mu\nu} -\frac{1}{2}R g_{\mu\nu} - 6 g_{\mu\nu} \right) d x^\mu dx^\nu&  =  8 \pi  \
 \left( \frac{\pi^4}{2N^2} \right) 
 \mathcal{F}(x)(2\gamma^2(\theta_0)) \left(d \tau-d z \right)\left(d \tau-d z \right)\\
 &=\left( \frac{8\pi^5\gamma^2(\theta_0)}{N^2} \right) 
  \mathcal{F}(x) \left(d \tau-d z \right)\left(d \tau-d z \right)\\
\end{split}
\end{equation}

We see that the RHS of \eqref{eieq} is of 
order $\frac{\gamma^2(\theta_0)}{N^2 }$ . As we 
have explained in earlier sections, we will be interested in values of $\omega$ 
so that $\gamma^2(\theta_0) \sim N^\frac{2}{3}$. Consequently, the RHS of \eqref{eieq}
is of order $\frac{1}{N^\frac{4}{3}}$ and so is parametrically small in the large $N$
limit. Consequently, we can solve this backreaction in a linearized order. In this order,  the correction to bulk $AdS_5$ metric takes the form 
\begin{equation}\label{corradsmet}
\delta ds^2= \frac{8\pi^5\gamma^2(\theta_0)}{N^2}x^2  f(x) 
\left(d \tau-d z \right)\left(d \tau-d z \right)
\end{equation}
where the function $f(x)$ obeys the sourced minimally coupled scalar equation 
\begin{equation}\label{smfe}
\frac{1}{x^3} \partial_x \left( x^5 \partial_x f\right)  
= -2\frac{x\mathcal{F}(x)} {x^3} 
\end{equation}
This equation can be solved by integration. We find 
\begin{equation} \label{solne}
f(x)=    2\int_x^\infty  dy \left( \frac{1}{y^5} \int_0^y dz~ z {\mathcal F}(z)\right)   
\end{equation}
where, we have fixed the integration constants by demanding normalizability of our solution (at $x=\infty$) and regularity of the solution at $x=0$. 

We conclude that our final bulk metric is given by 
\begin{equation}\label{ptmetric}
ds_{AdS_5}^2=-(1+r^2)dt^2+\frac{dr^2}{1+r^2}+r^2\left( (\sigma_1)^2 +  (\sigma_2)^2 + (\sigma_3)^2 \right)
+ \frac{8\pi^2\gamma^2(\theta)}{N^2}r^2  f\left(\frac{r}{\gamma(\theta)}\right) 
\left(d t-\sigma_3 \right)\left(d t-\sigma_3 \right)
\end{equation} 
where the function $f(x)$ is given by \eqref{solne} (and ${\mathcal F}(x)$ was listed in \eqref{fxdef}).

We will find it useful, below, to know the behaviour of $f(x)$ at large and small values of $x$. When $x$ is large, $y$ in 
\eqref{solne} is also everywhere large. 
Consequently the quantity 
\begin{equation}\label{largexint}
\begin{split} 
\int_0^y dz~z {\mathcal F}(z)
&= \int_0^\infty dz ~z {\mathcal F}(z)
- \int_y^\infty dz~z {\mathcal F}(z)\\
&= C - \frac{B}{6y^6}  + {\cal O}(1/y^8)
\end{split}
\end{equation}
where $B$ was defined in \eqref{cflargex} and 
\begin{equation}\label{cexp} 
\begin{split}
C &= \int_0^\infty dz~z{\mathcal F}(z)\\
&=\sum_{q=1}^{\infty}\frac{\text{csch}^6\left(\frac{\beta  q}{2}\right) \text{csch}(\beta  q)}{128 \pi^5\beta ^3 q^3}
\end{split} 
\end{equation}
(We obtain the second line of  \eqref{cexp} by substituting \eqref{fxdef} into the first line.)
\footnote{The fact that integral that evaluates $C$ (first line of \eqref{cexp}) converges follows from \eqref{cfsmallx}
 and \eqref{cflargex}. }  
Comparing with \eqref{hphi} we see that 
\begin{equation}\label{chphi}
C= \frac{h_\phi(\beta)}{\beta \pi^5} 
\end{equation} 
 Inserting \eqref{cflargex} into \eqref{solne} yields the large $x$ expansion 
\begin{equation}\label{fxlarge}
f(x)=\frac{h_\phi(\beta)}{2 \pi^5 x^4 \beta} - \frac{B}{60 x^{10}}  + {\cal O}\left(\frac{1}{x^{12}} \right) 
\end{equation} 
Note that $C$ and $B$ are both constants independent of all coordinates (they are also independent of the chemical potentials $\omega_i$, but are functions of the temperature).

In order to obtain an expansion of $f(x)$ at small $x$ it is useful to rewrite \eqref{solne} (after an integration by parts) as 
\begin{equation}\label{solneinp}
\begin{split}
f(x)=& \frac{1}{2} \int_x^\infty dy \frac{{\mathcal F}(y)}{ y^3} + \frac{1}{2 x^4} \int_0^x dy~y {\mathcal F}(y) \\
=& \frac{1}{2} \int_0^\infty dy \frac{{\mathcal F}(y)}{ y^3} -
\frac{1}{2} \int_0^x dy \frac{{\mathcal F}(y)}{ y^3} 
+ \frac{1}{2 x^4} \int_0^x dy~y {\mathcal F}(y) \\
=&\frac{1}{2} \int_0^\infty dy \frac{{\mathcal F}(y)}{ y^3} -\frac{D}{6} x^2 + {\cal O}(x^4) \\
\end{split}
\end{equation}
where $D$ was defined in \eqref{cfsmallx}.
\footnote{ The integral in the first term in the last line of \eqref{solneinp} is 
convergent (this follows because ${\cal F}(y) \sim y^4$ at small $y$ (see \eqref{cfsmallx}) and $\sim \frac{1}{y^8}$
at large $y$ (see \eqref{cflargex}).}

\subsection{Boundary Stress Tensor}

We can now use the expansion of $f(x)$ at large $x$ to evaluate the boundary stress tensor of our solution. 
Inserting \eqref{fxlarge} into \eqref{corradsmet}, we see that for 
$ r \gg \gamma(\theta)$, our bulk metric simplifies to 
\begin{equation}\label{largeasmet}
    ds_{AdS_5}^2=-(1+r^2)dt^2+\frac{dr^2}{1+r^2}+r^2\left( (\sigma_1)^2 +  (\sigma_2)^2 + (\sigma_3)^2 \right)
+r^2   
\left(d t-\sigma_3 \right)\left(d t-\sigma_3 \right)\frac{16 h_{\phi}(\beta)\gamma^6(\theta)}{N^2 r^4 \beta}
\end{equation}
The usual formulae \cite{deHaro:2000vlm} allow us to read off the boundary stress tensor from 
\eqref{largeasmet}; we find 
\begin{equation}\label{bdryt}
\begin{split}
T_{\rm bdry}&= \left(\frac{4}{16 \pi G_{5}}\right) \frac{16 h_{\phi}(\beta) \gamma^6(\theta)}{\beta N^2}
\left(d t-\sigma_3 \right)\left(d t-\sigma_3 \right)\\
&= \frac{8 h_{\phi}(\beta) \gamma^6(\theta)}{\pi^2\beta}  \left(d t-\sigma_3 \right)\left(d t-\sigma_3 \right)
\end{split}
\end{equation} 
The boundary stress tensor can be rewritten as follows. Let us define the four-velocity vector 
\begin{equation}\label{veldef}
u^\mu \partial_\mu = \gamma(\theta) \left( 
\partial_t - \omega_1 \partial_{\phi_1} - \omega_2 \partial_{\phi_2} \right) 
\end{equation}
Note that $u^\mu$ has been defined in a way that ensures that $u^\mu u_\mu=-1$. 
To leading order in the large $\gamma(\theta)$ expansion, 
\begin{equation}\label{veldef}
u^\mu \partial_\mu = \gamma(\theta) \left( 
\partial_t - \partial_{\phi_1} -  \partial_{\phi_2}  +{\cal O} \left( \frac{1}{\gamma(\theta)} \right)  \right) 
\end{equation}
Ignoring the subleading correction, it follows that velocity one one-form (obtained by lowering indices on the velocity vector field)  takes the form 
\begin{equation}\label{veloneform}
u_udx^\mu= \gamma(\theta) \left( -dt + \sin^2 \theta d\phi_1 +  \cos^2 \theta 
d\phi_2\right)  = \gamma(\theta) \left(dt -\sigma_3 \right)
\end{equation}
(where $\sigma_3$ was defined in \eqref{oneforms}). 
Consequently, the stress tensor \eqref{bdryt} can be rewritten as
\footnote{The term in \eqref{bdrytn} proportional to $g_{\mu\nu}$ has been added by hand to make the stress tensor traceless: this can be done because this term is proportional to $\gamma^4(\theta)$, and so is subleading compared to the term proportional to $u_\mu u_\nu$ (which is proportional to $\gamma^6(\theta)$). }
\begin{equation}\label{bdrytn}
(T_{\rm bdry})_{\mu\nu}= \frac{2h_\phi(\beta)\gamma^4(\theta)}{\beta \pi^2}  \left(4  u_\mu u_\nu + g_{\mu\nu} \right) 
\end{equation}

\eqref{bdrytn} has precisely the spatial dependence of the equilibrium configuration of a four-dimensional conformal fluid rotating with the velocity \eqref{veldef}  (see equations 28 and 14 of \cite{Bhattacharyya:2007vs}). This is a satisfying result. 

We can re-evaluate the energy of the gas by integrating the $(00)$ component of \eqref{bdrytn} over the boundary sphere. Using the fact that $u_0^2= \gamma^2$ we find  
\begin{equation}\label{efromst}
\begin{split}
        E&= \frac{h_\phi(\beta)}{\beta \pi^5} \pi^3 \int_{S^3} 
       8 \gamma^6(\theta)\\&=\frac{8h_{\phi}(\beta)}{\beta}\left(\frac{1}{(1-\omega^2_1)(1-\omega^2_2)^2}+\frac{1}{(1-\omega^2_1)(1-\omega^2_2)^2}\right)
\end{split}
\end{equation}
The final expression in \eqref{efromst} is in perfect agreement with \eqref{efin}. 

\section{Conjectured Boundary Stress Tensor for general bulk matter} \label{cfbtst}

In sections \ref{section5} and \ref{section6} we have computed the detailed bulk solution corresponding to a rank 4 Grey Galaxy supported by the bulk gas from all bulk fields emerging from the Kaluza Klein decomposition of the 10d dilaton. 
It is natural to ask how the computations of the previous two sections would be modified if the bulk matter that supported the Grey Galaxy was different (for instance if the bulk matter consisted of all the fields of IIB supergravity in 10 dimensions). 

We expect that any reasonable choice for bulk matter would give rise to a bulk stress tensor of the form 
\eqref{bst}, but with the function ${\mathcal F}(x)$ replaced by some new function 
${\mathcal F}'(x)$, whose detailed form depends on the details of the bulk matter content. We have explicitly verified that this is the case, both for 5d scalar field of any mass (see \eqref{Ka} in Appendix \ref{summing}) 
as well as for a massive 5d vector field or arbitary mass 
(see \eqref{bstn} in Appendix \ref{spin1}), and see no obstruction to this result holding for matter of arbitrary spin. While the precise form of ${\mathcal F}'(x)$ depends on the details of the bulk matter content and requires a computation, one particular moment of this function can be determined on general grounds. We now illustrate this point (which applies -with minor modifications-  for any bulk matter content) in the context of an important example, namely the case in which the bulk is IIB Supergravity, the bulk dual to 
${\cal N}=4$ Yang-Mills theory on $AdS_5 \times S^5.$

In the particular example of IIB supergravity on $AdS_5\times S^5$, we have already computed the thermodynamics of the bulk gas in \eqref{fflnz}. The energy of this gas is given by the first of \eqref{thermo}, with $h_\Delta(\beta)$ replaced by $h_{YM}(\beta)$. But this total energy may also be obtained by integrating over the bulk stress tensor, which is given in terms of ${\mathcal F}_{YM}$ (the name we give to ${\mathcal F}'$ in this particular case). This constraint tells us (see around \eqref{xint}) that 
\begin{equation}\label{xintn}
\begin{split}
 \int_0^\infty  dx x {\mathcal F_{YM}}(x)&=\frac{h_{YM}(\beta)}{\pi^5 \beta}\\
\end{split}
\end{equation}

The analysis of section \ref{section6} now continues to apply with minor modifications. The back reaction of the metric to the full gas continues to be given by \eqref{ptmetric}, but with 
$f(r/\gamma) \rightarrow f_{YM} (r/\gamma)$, with 
\begin{equation} \label{solneym}
f_{YM}(x)=    2\int_x^\infty  dy \left( \frac{1}{y^5} \int_0^y dz~ z {\mathcal F}_{YM}(z)\right)   
\end{equation}
Repeating the analysis around \eqref{largexint} and \eqref{cexp} we find that $f_{YM}(x)$ admits the large $x$ expansion

\begin{equation}\label{fxlargeYM}
f_{YM}(x)=\frac{h_{YM}(\beta)}{2 \pi^5 x^4 \beta} - \frac{B_{YM}}{60 x^{10}}  + {\cal O}\left(\frac{1}{x^{12}} \right) 
\end{equation} 
Here $B_{YM}$ is a constant whose precise value we do not know but will not need. \footnote{Determination of this quantity - the analogue of \eqref{cflargex} - which requires more knowledge of the distribution of the bulk gas stress tensor than we have without performing a computation.} It follows that the leading deviation of the metric from the metric of pure $AdS_5$ is 
given by \eqref{largeasmet} with $h_\phi(\beta) \rightarrow h_{YM}(\beta)$. Consequently, the gas component of the boundary stress tensor dual to the full Grey Galaxy (sourced by all gas fields in IIB supergravity on $AdS_5 \times S^5$) is given by \eqref{bdrytn} with $h_\phi(\beta) \rightarrow h_{YM}(\beta)$, i.e. by
\begin{equation}\label{bdrytnym}
(T_{\rm bdry}^{YM})_{\mu\nu}= \frac{2h_{YM}(\beta)\gamma^4(\theta)}{\beta \pi^2}  \left(4  u_\mu u_\nu + g_{\mu\nu} \right) 
\end{equation}

In summary, our final prediction is that the boundary stress tensor of the Grey Galaxy in IIB theory on $AdS_5\times S^5$  is the sum of the black hole boundary stress tensor (described in detail in subsection \ref{bhstr}) and the gas boundary stress tensor \eqref{bdrytnym}. A similar result applies to any bulk theory on $AdS_5 \times M$ where, the internal manifold $M$ could, for instance, be a $Y^{p,q}$.

\section{Conclusion and Discussion}\label{disc}

Grey Galaxies in $AdS_D$ were studied exhaustively in \cite{Kim:2023sig} for the special case $D=4$. In this paper, 
we have investigated Grey Galaxies in $AdS_D$ for $D \geq 5$, 
with a special focus on the case $D=5$. In this dimension, rotating black holes are parameterized by two angular velocities, 
$\omega_1$ and $\omega_2$. We find Grey Galaxy phases when either 
$\omega_1$ or $\omega_2$ or both are taken parametrically near unity. When only one of $\omega_1$ or $\omega_2$ is near unity, 
we obtain a Grey Galaxy of rank 2. The gas contribution to the boundary stress tensor of this solution turns out to be sharply localized around a boundary equator (a similar behavior was noticed in $AdS_4$). 

When both $\omega_i$ are parametrically near unity, the Grey Galaxy is qualitatively different from its four-dimensional cousin. In this case, the gas contribution to the boundary stress tensor is smooth. A key result of this paper is that this stress tensor takes the fluid dynamical form \eqref{bcssn}, where the `fluid velocity' $u_\mu$ is given in \eqref{umusp}, and the function $h_{YM}(\beta)$ can be read off from the partition function of the bulk gas by comparison with \eqref{fflnz}. In the special case of $AdS_5 \times S^5$ (the bulk dual to ${\cal N}=4$ Yang-Mills at strong coupling) we have explicitly computed $h_{YM}(\beta)$ (see \eqref{hym}). We emphasize that the gas component of the boundary stress tensor takes a fluid dynamical form even at values of the temperature of order unity (in units of the radius of the boundary $S^3$). 
\footnote{As we take, say, $\omega_2$ away from unity, the fluid dynamical form of the stress tensor becomes increasingly localized around an equator (over angular scale $1-\omega_1 =\frac{\delta^2}{4}$) in qualitative agreement with the boundary stress tensor of a rank 2 Grey Galaxy. This follows because, in this limit, 
$$ \gamma = \frac{1}{\sqrt{ 1-\omega^2 \sin^2 \theta}} 
\approx \frac{\sqrt{2}}{\sqrt{\delta^2 + \delta \theta^2}}, \quad\quad 1-\omega^2\approx\frac{\delta^2}{2}$$
While the fluid model gets the localization scale right, it should not be expected to (and does not) get the details of the localization function correct (see equations 5.33-5.39 of \cite{Kim:2023sig}) as dynamics at these length scales are not properly captured by hydrodynamics (from a boundary point of view) or by the `point by point' method of obtaining the bulk solution of this paper (contrast with the more complicated bulk 
evaluation in \cite{Kim:2023sig})}

Although we have mainly focused on the study of rank 4 Grey Galaxies in $AdS_5$, the structure of these solutions strongly suggests that the (obvious generalization of) form \eqref{bcssi} always accurately captures the form of the boundary stress tensor for an odd $D$ Grey Galaxy of maximum rank (of course with  $h_{\rm YM}(\beta)$ replaced by the function read off from the partition function of the relevant theory).  In the case of Grey Galaxies of non-maximal rank (or of maximal rank but in even $D$) the boundary fluid is sharply localized on some part of the boundary. Even in these cases, we continue to expect the variation of the boundary stress tensor {\it along} the localizing manifold to be captured by the fluid model. It would be interesting to perform computations to verify these guesses. 

From the boundary point of view, the black hole represents the bulk dual to a `gluon plasma'. The gas of gravitons in the Grey Galaxy is a gas of `glueballs' (or, more accurately, single trace fields). The Grey Galaxy can be thought of as a phase in which 
the fast-rotating gluon plasma dynamically spits out an order one fraction of its energy into glueballs; it does so because the entropy it loses (due to loss of energy) is more than compensated for by the entropy it gains (due to the angular momentum loss). The final boundary stress tensor is effectively a sum of the boundary stress tensor of gluon plasma and the glueball gas. We suspect that the glueball gas takes a precisely fluid form (even at temperatures of order unity) due to relativistic boost effects, which make the effective rest volume of the sphere much larger than the lab frame rest volume, and so 
make the temperature effectively large in units of the effective rest frame size of the sphere. It would be interesting to investigate this point further. 

Along similar lines, in  \S \ref{impbulk} we have also noted that - like its boundary counterpart - the ${\it bulk}$ stress tensor of the bulk gas also takes a fluid dynamical form, despite the fact that the `local temperature' (proper length of the thermal $S^1$), in the regime of interest, is of the same order as the bulk spacetime curvature. 
It would be interesting to understand why this is the case, and whether we should expect it to continue to hold in a regime in which interactions of the bulk gas can no longer be ignored. 

All Grey Galaxies in $AdS_D$ for even $D$ - and all Grey Galaxies of non maximal rank in $AdS_D$ at odd values of $D$- have a striking feature. Their boundary stress tensor is sharply localized on a submanifold of the boundary $S^{D-2}$. This feature can be used as an order parameter that sharply distinguishes these solutions from other phases like vacuum black holes. At odd values of $D$ and maximal rank, however, the gas contribution to the boundary stress tensor is as smooth as the black hole contribution. It would be interesting to find a sharp order parameter that distinguishes such Grey Galaxy phases from vacuum black holes. The key physical difference between the two phases lies in the fact that Grey Galaxies host a high energy gas at parametrically large values of the bulk radial coordinate (such a gas is absent in black hole phases). It should be possible to detect this gas using the boundary two point functions evaluated at parametrically small distances. It would be interesting to make this expectation precise. 

It is, of course, a well-known result of the fluid gravity correspondence that the gluon plasma is a rather perfect fluid,
that obeys the equations of relativistic hydrodynamics with 
a relatively small viscosity \cite{Bhattacharyya:2008ji}. We 
emphasize that the glueball fluid studied in this paper is
entirely different. It is composed of glueballs not gluons, is weakly interacting, and is characterized by parametrically large 
thermalization time scales (in comparison, the thermalization time scale of the gluon fluid is of the order of the inverse temperature).
Moreover, a gas of gluons (on the boundary sphere) is accurately 
described by hydrodynamics only at large temperatures. This point can be seen even in equilibrium in our Grey Galaxy solutions. 
In any of our solutions, the boundary stress tensor of the rotating black hole is accurately captured by the perfect fluid hydrodynamical form only at small $\beta$. On the other hand 
the stress tensor of the glueball gas takes the fluid form 
at every value of $\beta$. 

In the bulk, the graviton gas and the black hole are 
parametrically well separated in the radial coordinate. For this reason the black hole and the gravitons (i.e. the gluon `fluid' 
and the glueball fluid) are essentially non-interacting. This fact would lead to rather striking phenomena from the boundary point of view. For instance, we could excite or force a Grey Galaxy by turning on a time dependent boundary source for any single trace operator - say the stress tensor -  over distance and time scales of order unity. Such a forcing would 
excite the black hole in the usual manner: at late times all excitations would die down in a manner governed by the usual black hole quasi-normal modes. Through all of this, the bulk graviton gas (gas of glueballs) would carry on its perfect hydrodynamical motion, serenely unaffected, even though it lives in the same field theory location as the excited plasma.

It would, of course, be fascinating to understand the existence of Grey Galaxies - and the details of all these fascinating phenomena - from a direct analysis of the field theory. While Grey Galaxies are only sharply defined at large $N$, we see no reason for them not to exist at all values of $\lambda$, even down to $\lambda=0$, and so may be accessible in perturbation theory, suggesting an exciting research program for the future. 

In section \ref{pd} of this paper, we have used the Grey Galaxies constructed in this paper to present a fully quantitative conjecture for one part of the phase diagram of ${\cal N}=4$ Yang-Mills theory as a function of energy and its two angular momenta. As we have explained in the introduction, the low energy part of the relevant phase diagram is dominated by small 10d black holes. It would be interesting to put 10d black holes (e.g. along the lines of \cite{Dias:2015pda}) and Grey Galaxies together, to construct a complete quantitative phase diagram for 
${\cal N}=4$ Yang Mills theory as a function of energy and angular momenta. It would also be interesting to generalize these results (and their counterparts for a charge without angular momentum, \cite{Choi:2024xnv}) to a conjectured phase diagram for ${\cal N}=4$ Yang Mils theory as a function of all 6 Cartan charges (energy, two angular momenta, and three $SO(6)$ charges). Finally, it would be interesting to search for checks of this phase diagram, perhaps from exact computations performed in the BPS limit. 

Our construction of the phase diagram in section \ref{pd} proceeded under the assumption that all relevant phases in the phase diagram are usual black holes and Grey Galaxies. We have seen above that this assumption fails when our background has internal manifolds (like $S^5$), in addition to the $AdS_D$ solutions. In subsection \ref{glins} (and Appendix \ref{appins}) we have briefly investigated the question of whether other purely AdS higher topology black solutions - and associated phenomena - might be relevant for our phase diagram even in the absence of internal manifolds. Our tentative conclusion was that these solutions and phenomena appear irrelevant from the viewpoint of leading order thermodynamics, but this important question deserves a more careful investigation. 

In this paper we have presented a complete bulk solution for rank 4 $AdS_5$ Grey Galaxies for only one choice of bulk matter, namely a single massless scalar field in $AdS_5 \times S^5$. 
Our physical interpretation of the results of this computation allowed us to come up with an educated guess for the boundary stress tensor for a Grey Galaxy in a theory with arbitrary bulk matter, for instance in full IIB supergravity in $AdS_5\times S^5$. It would nonetheless be of interest to perform an honest computation 
of the full bulk solution with this matter content. Such a computation would allow us to verify 
our guesses, and would also give us the full bulk solution
rather than just its boundary stress tensor.
This exercise may be less daunting than it first sounds, as (in the limit of interest) the bulk gas stress tensor is completely determined from a (sum over) two-point functions between points that are almost lightlike separated. It seems likely that the relevant propagators simplify at these short proper distances, 
simplifying the computation.  We leave further investigation of this point to future work.
\acknowledgments
We would like to thank S. Choi, A. Gadde, D. Jain, S. Kim, V. Krishna, S. Kundu, G. Mandal, O. Parrikar, and S. Trivedi and especially E. Lee and C. Patel for the very useful discussions. The work of K.B., V.K., S.M., J.M. and A.R. was supported by the J C Bose Fellowship JCB/2019/000052
and the Infosys Endowment for the study of the Quantum Structure of Spacetime. We would also like to
acknowledge the debt to the people of India for their steady support to the study of the basic science.

\appendix 
\section{Black Rings and Gregory Laflamme Instabilities}\label{appins}

In this Appendix, we investigate the role of Gregory Laflamme-type instabilities and black rings (and other similar solutions) in the phase diagram of the relevant gravitational theory. We argue that Gregory-Laflamme type instabilities only occur for black holes that are already super radiant unstable: the black hole components of Grey Galaxies are never super radiant unstable. As a consequence, Gregory-Laflamme's instabilities play no role in the phase diagram of the theory. Our analysis for black rings is less conclusive because these solutions are not analytically known except in the case that they are small (in units of the $AdS$
radius). In this case (i.e. in the case of small black rings) we demonstrate that these solutions always have $\omega>1$ and
so are super radiant unstable. While we suspect that this conclusion also holds for large black rings, the lack of solutions for these objects stops us from reaching a definite conclusion here. 

\subsection{Gregory Laflamme Type instabilities}

\subsubsection{GL instabilities in flat space}

Spinning black holes in bulk dimensions $D\geq 6$ are qualitatively different from their lower dimensional cousins in one important respect. This difference can already be seen in flat space as we now explain. 

Let us consider a spinning flat space black hole in $D=4$ or  $D=5$. Let us imagine that we hold the angular momentum (or angular momenta, in $D=5$) of this black hole fixed and lower its energy. Both in  $D=4$ and $D=5$ we find that this process cannot be continued indefinitely: there is a minimum energy (determined by the black hole angular momenta: roughly $G M \sim (G J)^\frac{D-3}{D-2}$) at which the black hole becomes extremal. Black holes 
with energies lower than this extremal black hole do not exist \footnote{The corresponding solutions have a naked singularity, and so are unphysical}. It is also interesting to track the temperature of these black holes as their energy is decreased at fixed angular momentum. At large values of the energy the black holes in question are similar to Schwarzschild black holes and have negative specific heat. Starting at large energies, it follows that the black hole temperature increases (from zero) as we lower their energy. At a critical value of the energy (once again roughly of order $ G M \sim (G J)^\frac{D-3}{D-2}$), this temperature reaches a maximum. Upon further lowering the energy, the temperature of the black hole decreases, until it approaches zero at extremality. 

In the case $D \geq 6$, on the other hand, Kerr black holes at fixed angular momentum exist at all energies $\geq 0$. Let us suppose we have turned on only one of the angular momenta (let's say $J_1$) and that all other angular momenta are zero. As we lower the energy down to zero (at fixed $J$), the horizon of the black hole is a thin but large disk (at the center, the horizon is approximately $R^2 \times S^{D-4}$). As we approach zero energy (more precisely at energies $G M \ll (G J)^\frac{D-3}{D-2}$) the radius of the disk increases, and its angular velocity decreases (in such a manner so that the product of the radius and angular velocity is roughly constant). The thickness of the disk - i.e. the radius of the $S^{D-4}$ also decreases
\footnote{Quantitatively, the ratio of the thickness 
to the radius of the disk scales like 
$\frac{(GM)^{\frac{D-2}{D-3}}}{(GJ)}$ when $M$ is small. The thickness scales like $r^{D-3}= G M$, and therefore the size of the disk scales like $\frac{J}{M}$.}- reaching zero when $M=0$. 
\footnote{Upon lowering the mass, the temperature of the black increases monotonically, from zero at very large mass, to infinity at very small mass. This statement may be verified as follows. Recall that the combined transformation $g_{\mu \nu} \rightarrow \lambda^2 g_{\mu \nu}$, $x^\mu \rightarrow \frac{x^\mu}{\lambda} $ is an invariance of the classical Einstein action. Physically, this transformation represents a stretching (by scale factor $\lambda$) of all spacetime features of the solution: for instance, it takes a Schwarzschild black hole of radius $r_0$ to a black hole or radius $\lambda r_0$. It follows from the usual ADM formulae (and the Beckenstein formula) that, under this transformation the mass, angular momenta, temperature and entropy transform like $M \rightarrow \lambda^{D-3} M$, $J \rightarrow \lambda^{D-2} J$, $T \rightarrow \frac{T}{\lambda}$ and $S \rightarrow \lambda^{D-2}S$. Consequently, for flat space black holes with a single rotational angular momentum $J$, 
the temperature as a function of mass, at fixed $J$
can be read off from the single-scale invariant 
relationship
$$T J^{\frac{1}{D-2}}= f\left(\frac{M}{J^{\frac{D-3}{D-2}}}\right)$$
We have computed the function $f$ numerically in $D=6$ and verified it decreases monotonically. We believe that this monotonic decrease is true (of black holes with a single rotation) for all $D \geq 6$.} 

The important point here is the following. As we have mentioned above, the black hole solution at low mass begins, locally,  to resemble the product of 
a Schwarzschild black hole in $D-2$ dimensions times 
a disk. It is, however, well known (\cite{Emparan_2003},\cite{Dias:2010gk}) that configurations this sort are Gregory Laflamme unstable whenever the radius of the disk is much larger than the radius of the $S^{D-4}$. In other words, the black holes described above become unstable before they reach zero mass. 

\subsubsection{Gregory Laflamme instabilities in $AdS$ space}

The flat space discussion above has an analouge in 
$AdS$ space. Once again the moduli space of spinning black holes in $AdS_D$ is qualitatively different in $D=4,5$ and $D\geq 6$. In $D=4, 5$, black holes at every fixed $J$ become extremal\footnote{i.e. have zero temperature} at an energy $E_{\rm ext}$, whose value is such that $R_{AdS} E_{\rm ext}>J$. Black holes do not exist at sub extremal energies, and so do not exist all the way down to the unitarity bound. In $D\geq 6$, however, black holes with a single rotational angular momentum exist all the way down to the unitarity bound $R_{AdS} E= J$(\cite{Dias:2010gk}). At small values of $GJ$ ($GJ \ll R_{AdS}^{D-2}$), the approach to this limiting black hole may be understood as follows. As we have seen above, for $GM \ll (GJ)^\frac{D-3}{D-2}$, the black hole horizon behaves like the thin flat space pancake of radius $ \frac{J}{M}$. As we further decrease the mass of the black hole, its radius increases (and angular velocity decreases, proportional to the radius) until the radius of the pancake becomes $\sim R_{AdS}$ at $
R_{AdS} E \sim J$. At this energy $AdS$ space significantly modifies the flat space-like behaviour. The energy of the black hole can be further lowered all the way down to the unitary bound $R_{AdS}E= J$, at which point the thickness of the pancake shrinks to zero, while, somewhat surprisingly, its radius continues to grow without bound (notwithstanding the AdS potential barrier). 

As in flat space, pancake like black holes in $AdS_D$ $(D\geq 6)$ suffer from a Gregory-Laflamme like instability whenever the thickness of the pancake is much smaller than the minimum radius of the pancake and the AdS radius. When $GJ \ll R_{AdS}^{D-2}$, this happens for $GM \lessapprox (GJ)^\frac{D-3}{D-2}$. 
Note that the ratio of the critical energy for the Gregory Laflamme instability, $(GJ)^\frac{D-3}{D-2}$ to the unitarity bound $\frac{J}{R_{AdS}}$ equals 
$\frac{R_{AdS}}{\left(GJ\right)^{\frac{1}{D-2}}}$. As we have assumed $GJ \ll R_{AdS}^{D-2}$, this ratio is very large, indicating that the Gregory Laflamme instability happens over a large range of energies. 

We will now argue that the Gregory Laflamme instability never shows up in the phase diagram 
of the relevant gravitational theories. This happens because black holes go super radiant unstable much before they go Gregory Lafllamme unstable - and the endpoint of the superradiant instability - namely a black hole with $\omega=\frac{1}{R_{AdS}}$ in equilibrium with a rotating gas - is never itself 
Gregory-Laflamme unstable.

When $J$ is small, scale invariance tells us that the angular velocity of spinning black holes must be given by a scale-invariant relation very similar to the temperature 
\begin{equation}\label{tempsma}
\omega J^{\frac{1}{D-2}} =g\left(\frac{M}{J^{\frac{D-3}{D-2}}}\right)
\end{equation} 
It turns out that $g$ is a linear function, so that 
\begin{equation}\label{linom}
\omega \sim \frac{M}{J} 
\end{equation} 
In the regime that the approximations above are valid, however, we have already argued that $M$ is parametrically larger than the unitarity bound 
$\frac{1}{R_{AdS}}$. At least when $GJ \ll R_{AdS}^{D-2}$, consequently, it follows that black holes only suffer from the Gregory Laflamme instability when 
$\omega$ is parametrically larger than the superradiant bound. 

From a more general analysis as depicted in Figure 2 of\cite{Dias:2010gk}, one can see that for singly spinning black holes in AdS, the Gregory Laflamme instability sets in after the superradiant instability, and the surface is ultraspinning regime is well separated from the superradiant curve, even for spinning black holes whose size is comparable to the $AdS$ radius or 
larger. 
\subsection{Black Rings}

Another new feature in $D \geq 5$, with no known analogue in $D=4$, is the existence of black solutions with (spatial) horizon topologies more complicated than $S^{D-2}$. The simplest of these solutions is a black ring(\cite{Emparan_2006}), with horizon topology 
$S^{D-3} \times S^1$. It is natural to wonder
whether the large `centrifugal forces' associated with highly spinning black holes tend to thermodynamically favour black rings over black holes at large angular momenta. 

At large values of the energy, a detailed investigation of whether black rings dominate over black holes is complicated by the fact that the relevant black ring solutions are not (yet) analytically known. At small values of the energy, however, the situation is better. At these values of the energy, all relevant solutions are characterized by a length scale $l$ with $l \ll R_{AdS}$. At these length scales, $AdS$ is well approximated by flat space and the length scale $R$ is irrelevant for local analysis. Notice, however, that black holes 
suffer from the superradiant instability at 
$\omega = \frac{1}{R}$, i.e. when $l \omega \ll 1$. 
In other words, the superradiant solution already occurs at values of $\omega$ that are extremely small in units of the proper size of the black holes. On the other hand, (effectively) flat space black rings only exist for $\omega \gtrapprox \frac{1}{l}$ (as might have been anticipated on either physical or dimensional grounds). It follows, 
in other words, all black rings (at small energies) are themselves super radiant unstable. 
Since black rings with $\omega= \frac{1}{R}$ do not exist (at small energies) they never play a dominant role in the thermal ensemble. 

As we have mentioned above, the situation is less clear at energies of order (or greater than) $1/R$. 
It is possible  that black ring solutions play a key role in the phase diagram at these energies, and even that `grey black ring' solutions exist at these energies. We leave clarification of this point to future work. 

\section{Mass and horizon radius of extremal Kerr-$AdS_5$ black holes}\label{horapp}

If $r_1, r_2, r_3$ are the three roots (not necessarily distinct nor real) of the cubic $ax^3 + b x^2 + c x + d =0$
 then the discriminant of this equation is
defined as
\begin{equation}\label{deltadef}
\Delta = a^{4}(r_{1}-r_{2})^{2}(r_{1}-r_{3})^{2}(r_{2}-r_{3})^{2}.
\end{equation}

This definition immediately tells us that, $\Delta >0 $ if and only if the cubic has three distinct real roots. 
\footnote{The `only if' part of this statement may be verified as follows. Let the three roots be $\alpha+i \beta$, $\alpha -i \beta$ and $\gamma$ where $\alpha$, $\beta$ and $\gamma$ are all real. In this case $\Delta= a^4 (i \beta)^2 |\alpha -\gamma + i \beta|^4$ and so is negative. } Similarly, $ \Delta <0$ if and only if cubic has one real root and two complex conjugate roots, and $\Delta =0$ if and only if it has degenerate roots. 

Using standard manipulations it is easy to verify (and well known) that the discriminant of the cubic equation   
$ ax^{3}+bx^{2}+cx+d$ is given in terms of coefficients $a,b,c,d$ by  
\begin{equation}\label{deltanew}
\Delta = 18\,abcd-4\,b^{3}d+b^{2}c^{2}-4\,ac^{3}-27\,a^{2} d^{2}
\end{equation}

Recall, now, that a black hole is extremal when its inner and outer horizon coincide, i.e. when the equation that determines the location of its horizon has a double root. 
The relevant equation is \eqref{horizoneq}
which we reproduce here for convenience: 
\begin{equation}\label{horizoneqagain}
    (r^2 + a^2)(r^2 + b^2)(r^2+ 1) - 2m r^2= 0
\end{equation}
Note, of course, that \eqref{horizoneqagain} is a cubic in $r^2$. Using \eqref{deltanew}, we find that the discriminant of \eqref{horizoneqagain}
 is given by,
\begin{equation}\label{discspc}
    \begin{split}
       \Delta &=-27 a^4 b^4-18 a^2 b^2 \left(a^2+b^2+1\right) \left(a^2 \left(-b^2\right)-a^2-b^2+2 m\right)+4 \left(a^2 \left(-b^2\right)-a^2-b^2+2 m\right)^3\\&+\left(a^2+b^2+1\right)^2 \left(a^2 \left(-b^2\right)-a^2-b^2+2 m\right)^2-4 a^2 b^2 \left(a^2+b^2+1\right)^3 
    \end{split}
\end{equation}
When $\Delta =0$ our black hole is extremal. From \eqref{discspc}, we see that the equation $\Delta=0$ is itself a cubic equation in $m$ (one of the roots of this equation equals $m_{ext}$, the mass of the extremal black hole). Once again by studying the discriminant of this new cubic equation, we can find the nature of its roots. 

Once again using \eqref{deltanew}, we find that the discriminant of the equation $\Delta =0$ (viewed as a cubic equation in $m$) is given by 
\begin{equation}\label{disctwo}
    \frac{1}{16} a^2 b^2 \left(a^6+3 a^4 \left(b^2+1\right)+3 a^2 \left(b^4-7 b^2+1\right)+\left(b^2+1\right)^3\right)^3\geq 0
\end{equation}
(we have checked that the inequality in \eqref{disctwo} holds for all $|a|,|b|\leq 1$ by 3d plotting the expression on the LHS on Mathematica). 

It thus follows that \eqref{disctwo} has three real roots for all physically allowed values of $a$ and $b$, i.e. $|a|,|b|\leq 1$. We have evaluated and plotted the three real roots of this equation for all values of $|a|,|b|\leq 1$, and verified that exactly one of these 
is positive (while the other two are negative or zero) for all values of $a$ and $b$
in the allowed range. The formula for the positive real root is 
\begin{equation}\label{formmext}
    \begin{split}
        m_{ \rm ext} &=\frac{1}{2} \left(\frac{1}{6} \sqrt{\left(a^2+b^2+1\right)^4+216 a^2 b^2 \left(a^2+b^2+1\right)} \cos \left(\frac{\theta }{3}\right)+a^2 \left(b^2+1\right)-\frac{1}{12} \left(a^2+b^2+1\right)^2+b^2\right)\\
    \end{split}
\end{equation}
where $\theta$ is given by
\begin{equation}\label{formtheta}
    \begin{split}
        \cos(\theta) &=\left(\frac{1}{\left(a^2+b^2+1\right)^4+216 a^2 b^2 \left(a^2+b^2+1\right)}\right)^{3/2}\\&\times \left(5832 a^4 b^4-\left(a^2+b^2+1\right)^6+540 a^2 b^2 \left(a^2+b^2+1\right)^3\right)\\
    \end{split}
\end{equation}
as reported in \eqref{mextremal}.

While the formula for $m_{ext}$ is complicated in general, it simplifies when $a=b$. In this case \eqref{discspc} simplifies to 
\begin{equation}
   \Delta_{a=b}= 4 m \left(2 a^2 (-1 + a^2)^3 + m - 4 a^2 (5 + 2 a^2) m + 8 m^2\right)
\end{equation}
In this case the equation $\Delta_{a=b} =0$ is effectively quadratic, and its unique 
the positive root is given by 
\begin{equation}\label{m_ext_a=b}
    \begin{split}
        m_{\rm ext}&= \frac{1}{128} \left(\sqrt{8 a^2+1}-1\right) \left(\sqrt{8 a^2+1}+3\right)^3\\
    \end{split}
\end{equation}
in agreement with Eq 5.28 of \cite{Hawking:1998kw}. 
It is possible, after some work, to check that the complicated expressions \eqref{formmext} and \eqref{formtheta} 
reduce to \eqref{m_ext_a=b} when $a=b$. 

In this simple special case, we can evaluate the value of the radius of the extremal black hole as follows. We differentiate \eqref{horizoneqagain} w.r.t. $r^2$, substitute the value of 
$m=m_{\rm ext}$ from \eqref{m_ext_a=b}, and equate the result to zero (the derivative must vanish because the LHS of 
\eqref{horizoneqagain} has a double root at extremality). This procedure gives us the following expression for the horizon radius of the extremal black hole,
\begin{equation}
\begin{split}
    {(r_+)}_{\rm ext}^2& = \frac{1}{4}\left( \sqrt{1 +8a^2}-1\right)
\end{split}
\end{equation}
again in agreement with Eq. 5.28  of \cite{Hawking:1998kw}.

\section{Partition function of a single scalar in $AdS_5$}\label{partapp}
In this brief appendix, we supply a detailed derivation of \eqref{zexact}.
We see from \eqref{gaspf}
that 
\begin{align}\label{so4qnum}
\ln Z&=
-\sum_{n=0}^{\infty}\sum_{J=0}^{\infty}\sum_{m_L,m_R=-\frac{J}{2}}^{\frac{J}{2}}\ln (1-e^{-\beta(\Delta + 2n + J) + \beta\left(\omega_1 +\omega_2\right)m_L + \beta \left(\omega_1-\omega_2\right) m_R})\nonumber\\
&=-\sum_{n=0}^{\infty}\sum_{J=0}^{\infty}\sum_{a,b=0}^{J}\ln (1-e^{-\beta(\Delta + 2n +a \left(\omega_1 +\omega_2\right)+b \left(\omega_1 -\omega_2\right)) + \beta\left(1-\omega_1 \right)J})\nonumber\\
&=\sum_{n=0}^{\infty}\sum_{J=0}^{\infty}\sum_{a,b=0}^{J}\sum_{q=1}^{\infty}\frac{e^{-q\beta(\Delta + 2n + a \left(\omega_1 +\omega_2\right)+b \left(\omega_1 -\omega_2\right)) + q\beta\left(1-\omega_1 \right)J}}{q}\nonumber\\
\end{align}
In the final line, we use the Taylor expansion of logarithm.
We can now perform sum over $a$ and $b$ which simplifies to
\begin{align}\label{C2}
\ln Z&=\sum_{n,J=0}^{\infty}\sum_{q=1}^{\infty}\frac{1}{q}\Big[-\frac{\exp \left(\beta  (J+1) q \omega _1-\beta  q \left(\Delta +J \omega _1+J \omega _2+J+2 n\right)\right)}{\left(e^{\beta  q \omega _1}-e^{\beta  q \omega _2}\right) \left(e^{\beta  q \left(\omega _1+\omega _2\right)}-1\right)}\nonumber\\
&+\frac{\exp \left(\beta  (J+1) q \omega _2-\beta  q \left(\Delta +J \omega _1+J \omega _2+J+2 n\right)\right)}{\left(e^{\beta  q \omega _1}-e^{\beta  q \omega _2}\right) \left(e^{\beta  q \left(\omega _1+\omega _2\right)}-1\right)}\nonumber\\
&+\frac{\exp \left(-\beta  q \left(\Delta +J \omega _1+J \omega _2+J+2 n\right)+\beta  (J+1) q \omega _1+\beta  (J+1) q \left(\omega _1+\omega _2\right)\right)}{\left(e^{\beta  q \omega _1}-e^{\beta  q \omega _2}\right) \left(e^{\beta  q \left(\omega _1+\omega _2\right)}-1\right)}\nonumber\\
&-\frac{\exp \left(-\beta  q \left(\Delta +J \omega _1+J \omega _2+J+2 n\right)+\beta  (J+1) q \omega _2+\beta  (J+1) q \left(\omega _1+\omega _2\right)\right)}{\left(e^{\beta  q \omega _1}-e^{\beta  q \omega _2}\right) \left(e^{\beta  q \left(\omega _1+\omega _2\right)}-1\right)}\Big]
\end{align}
We can now perform sum over $n$ and $J$ easily and obtain
 \begin{align}\label{C3}
     \ln Z&=\sum_{q=1}^{\infty}\frac{1}{q}\Big[-\frac{(\coth (\beta  q)-1) e^{\beta  q \left(-\Delta +\omega _1+\omega _2+4\right)}}{2 \left(e^{\beta  q \left(\omega _1+1\right)}-1\right) \left(e^{\beta  q \left(\omega _2+1\right)}-1\right) \left(e^{\beta  q \left(\omega _1+\omega _2\right)}-1\right)}\nonumber\\
     &+\frac{(\coth (\beta  q)-1) e^{\beta  q \left(-\Delta +\omega _1+\omega _2+4\right)}}{2 \left(e^{\beta  q}-e^{\beta  q \omega _1}\right) \left(e^{\beta  q}-e^{\beta  q \omega _2}\right) \left(e^{\beta  q \left(\omega _1+\omega _2\right)}-1\right)}\Big]
 \end{align}
 The first term of \eqref{C3} is obtained after summing over $n$ and $J$ of the first two terms of \eqref{C2} which is clearly finite when both $\omega_1$ and $\omega_2$ approach to unity. However, the last term of \eqref{C3} obtained from the last two terms of \eqref{C2} exhibits divergences when both $\omega_i$ approaches to unity.

 To extract the divergences, we explicitly expand \eqref{C2} and \eqref{C3} in the limit $\omega_1\rightarrow 1$ and $\omega_2\rightarrow 1$.   
 
 \begin{align}
     \ln \mathcal{Z}&=\sum_{q=1}^{\infty}\frac{1}{q}\Big[-\frac{e^{\beta  (\Delta -6) (-q)}}{\left(e^{2 \beta  q}-1\right)^4}+\frac{e^{\beta  (\Delta -4) (-q)}}{\beta ^2 q^2 \left(\omega _1-1\right) \left(\omega _2-1\right) \left(e^{2 \beta  q}-1\right)^2}+O\left(\frac{1}{1-\omega_1}\right)+O\left(\frac{1}{1-\omega_2}\right)]\nonumber\\
      &\approx\frac{e^{-q\beta  (\Delta -2) }}{4 \beta ^2 q^3 \left(1-\omega _1\right) \left(1-\omega _2\right) \sinh ^2(\beta  q)}
 \end{align}
 
\section{The Gas partition function in $AdS_5 \times S^5$}\label{fullgas}

In this Appendix, we compute the partition function (at the free 
level) of a gas of gravitons on $AdS_5 \times S^5$. As a consistency check on our computation, we use our answer to compute the superconformal index of the gas and verify that it reproduces the answer previously computed in \cite{Kinney:2005ej}. As the last step, we shut off all $SO(6)$
chemical potentials, and set $\omega_1 \approx 1$, $\omega_2 \approx 1$, to obtain the partition function in the fluid form of interest to this paper. 

Our computation of the gas partition function proceeds as follows. In subsection \ref{conf} we recall the character formulae for long and short representations of the conformal group $SO(4,2)$. In subsection 
\ref{su4} we recall the Schur formula for $SU(N)$ characters, specialized to $SU(4)$. In subsection \ref{results} we combine these results with the listing of particles in $AdS_5 \times S^5$ of \cite{Gunaydin:1984fk} to obtain the full Bosonic single particle partition function, as well as the full Fermionic 
single particle partition function in $AdS_5 \times S^5$. 
We present our answer as a function of all 6 chemical potentials. In subsection \ref{check} (as a check on our answer) we restrict our single particle partition function to the values of chemical potentials that compute the superconformal index and find the result previously reported in \cite{Kinney:2005ej}. 
Finally, in subsection \ref{so6off} we shut off all $SO(6)$ chemical potentials, set $\omega_1 \approx 1$, $\omega_2 \approx 1$, and multiparticle the answer, obtaining a formula for the partition function that takes the fluid form.

\subsection{Review of conformal characters} \label{conf}

 Recall that the isometry group of the $AdS_5\times S^5$ is $SO(2,4)\times SO(6)$.  So states in this background space-time are labeled by the eigenvalues of the Cartans of the compact subgroup $SO(2)\times SO(4)\times SO(6) $, which can also be written as, $SO(2)\times SU_L(2)\times SU_R(2)\times SO(6) =SO(2)\times SU_L(2)\times SU_R(2)\times SU(4)$. The labels for the quantum states are then conformal dimension $\Delta$, corresponding to $SO(2)$ Cartan, $j_1/j_2$ angular momentum corresponding to the Cartans of $SU_L(2)/SU_R(2)$, and the $SU(4)$ Cartan charges, $[R_1, R_2,R_3]$ (as in \cite{Kinney:2005ej}, $R_i$ are eigenvalues under the diagonal $SU(4)$ matrices with $1$ in the $i^{th}$ 
 diagonal entry, and $-1$ in the $(i+1)^{th}$ diagonal).

 In this subsection, we focus on the $SO(4,2)$ part of this algebra. Single particle states in the bulk dual transform in given irreducible representations of $SO(4,2)$. These representations are conveniently labeled by the scaling dimension $\Delta$ and angular momenta $j_1 /j_2$ (see above).
 Let us recall that generic (or long)  unitary representations representations of $SO(4,2)$ occur for scaling dimensions that obey \cite{Mack:1975je}
 \begin{equation}\label{longcond}
\Delta > j_1 + j_2 + 2-\delta_{j_1,0} -\delta_{j_2,0} +\delta_{j_1,0}\delta_{j_2,0}
\end{equation}
Modules of this sort have no null states. Their character
\begin{equation}\label{trch}
{\rm Tr } \left( s^H x^{j^z_1} y^{j^z_2} \right) 
\end{equation}
\footnote{Note the $s$, $x$ and $y$ are related to the chemical temperature and angular velocities in other parts of this paper by $s=e^{-\beta},~ x= e^{\beta(\omega_1 +\omega_2)},~y= e^{\beta(\omega_1 -\omega_2)}$.}
is obtained by the free action of derivatives on the primary, and is given by 
\begin{equation} \label{longrep}
    z^{\Delta j_1 j_2}_{B/F}(s,x,y) = \frac{s^{\Delta}\chi^{L}_{j_1}(x)\chi^{R}_{j_2}(y)}{\left(1-sx^{\frac{1}{2}} y^{-\frac{1}{2}}\right) \left(1-s x^{-\frac{1}{2}} y^{\frac{1}{2}}\right) \left(1-s x^{-\frac{1}{2}} y^{-\frac{1}{2}}\right) \left(1-s x^{\frac{1}{2}}y^{\frac{1}{2}}\right)}
\end{equation}
where, $\chi_j(x)$ is $SU(2)$ characters in the spin $j$ representation.

In addition to long representations, $SO(4,2)$ also admits 
three families of nontrivial short representations. The most special of these is the single representation at $\Delta=1, j_1=j_2=0$. 
This `free scalar' representation has character 
\begin{equation} \label{fsrep}
    z^{100}_{B/F}(s,x,y) = \frac{s(1-s^2) }{\left(1-sx^{\frac{1}{2}} y^{-\frac{1}{2}}\right) \left(1-s x^{-\frac{1}{2}} y^{\frac{1}{2}}\right) \left(1-s x^{-\frac{1}{2}} y^{-\frac{1}{2}}\right) \left(1-s x^{\frac{1}{2}}y^{\frac{1}{2}}\right)}
\end{equation}
(the subtraction in the numerator is a consequence of the free equation of motion). 
A less special family short representation has $j_2=0$, $j_1 \neq 0$  $\Delta=j_1+1$. The character of this family of representations is given by \cite{Dolan:2005wy}
\begin{equation}\label{ssem}
    z^{j_1 + 1 ,j_1 0}_{B/F}(s,x,y) = \frac{s^{j_1 +1}\left(\chi^{L}_{j_1}(x) -s\chi^{L}_{j_1 -1/2}(x) \chi^{R}_{1/2}(y) + s^2\chi^{L}_{j_1-1}(x)\right)}{\left(1-sx^{\frac{1}{2}} y^{-\frac{1}{2}}\right) \left(1-s x^{-\frac{1}{2}} y^{\frac{1}{2}}\right) \left(1-s x^{-\frac{1}{2}} y^{-\frac{1}{2}}\right) \left(1-s x^{\frac{1}{2}}y^{\frac{1}{2}}\right)}
\end{equation}
(the subtraction in the numerator is a consequence of a `Bianchi Identity' type null state). 
Of course the analogeous formula for short representations with $j_1=0$, 
$j_2 \neq 0$ are obtained by acting with $1 \leftrightarrow 2$ on 
\eqref{ssem}.

Finally, the character of short representations with $j_1 \neq 0$, $j_2 \neq 0, \Delta = j_1 + j_2 + 2$ is given by 

\begin{equation}
    z^{j_1 + j_2 + 2, j_1j_2}_{B/F}(s,x,y)= \frac{s^{j_1+j_2 +2}\left(\chi^{L}_{j_1}(x)\chi^{R}_{j_2}(y) -s\chi^{L}_{j_1 -1/2}(x)\chi^{R}_{j_2 -1/2}(y)\right)}{\left(1-sx^{\frac{1}{2}} y^{-\frac{1}{2}}\right) \left(1-s x^{-\frac{1}{2}} y^{\frac{1}{2}}\right) \left(1-s x^{-\frac{1}{2}} y^{-\frac{1}{2}}\right) \left(1-s x^{\frac{1}{2}}y^{\frac{1}{2}}\right)}
\end{equation}

\subsection{$SU(4)$ character formula} \label{su4}

In this subsection, we briefly review the Schur formula for $SU(4)$ 
characters in terms of determinants. Consider the $SU(4)$ character
\begin{equation}\label{formalcharacter}
 \rm{Tr} e^{i(\theta_1 R_1+ \theta_2 R_2+ \theta_3R_3)} =\rm{Tr} (p^{R_1}q^{R_2}r^{R_3}) 
\end{equation}
where the Cartan charges $R_i$ were defined above. \footnote{When evaluated on highest weight states, $R_i$ equals the number of columns 
of length $i$ in the Young Tableaux corresponding to the given 
representation.} Let us note, in particular, that the weights $(R_1, R_2, R_3)$ of the four vectors in the fundamental representation are, respectively
$$[1,0,0], [-1,1,0],[0,-1,1],~\rm{and}~[0,0,-1].$$

The contribution of these states  to \eqref{formalcharacter}, respectively, are 
\begin{equation}\label{replacement}
v_1 = e^{i \theta_1},~~~v_2 = e^{i (\theta_2-\theta_1)},~~~ v_3 =e^{i (\theta_3-\theta_2)},~~~ v_4= e^{-i\theta_3}.
\end{equation} 
\footnote{Note the product $v_1v_2v_3v_4=1$ carries zero charge under all three Cartans; this is 
a consequence of the fact that the contraction of four vectors with the 
$\epsilon$ tensor is a singlet.} 

Let us now recall that the Schur formula that asserts that 
\begin{equation}
    \chi  = \frac{{\rm det}[v_i^{l_j + n -j}]}{{\rm det}[v_i^{n -j}]}
\end{equation}
where $v_i$ are the `weight vectors' of the fundamental representation 
listed in \eqref{replacement}, and $l_i$ are the number of boxes in the 
$i^{th}$ row of the Young Tableaux (with the convention $l_4$=0).

While representations of $SU(4)$ are conveniently specified by the row lengths $l_i$, a second specification (the Dynkin specification) is \
commonly used in the literature. In this convention, representations 
are specified by the values of $R_1$, $R_2$ and $R_3$ acting on the highest weight state of the representation. Let these values be denoted by $r_i$.  A little thought will convince the reader that $r_i$ denotes the number of columns of length 
$i$ in the Young Tableaux \footnote{The argument goes as follows. Recall that the Young Tableaux can be thought of as a tensor product of fundamentals, in a particular symmetry channel. The highest weight state associated with a Young is obtained by occupying the first row by the highest weight state in the fundamental (namely the column with $1$ in the first row and zero every where else), the second highest weight 
state in the second row, and the third highest wt state in the third row. By the definition of this Cartan element, the action of $R_1$ on this state equals the number of boxes in the first row minus the number of boxes in the second row, i.e. the number of columns of length $1$. The action of $R_2$ equals the number of boxes in the second row minus the number of boxes in the third row, i.e. the number of columns of length $2$, and so on.}. The relationship between $l_i$ (the length of the $i^{th}$ row) and $r_i$ (the number of columns of length $i$) is clearly given by 
\begin{equation}
    \begin{split}
        l_1&=r_1+r_2+r_3
        \\l_2&=r_2+r_3 
        \\l_3&=r_3\\
        l_4&=0\\
    \end{split}
\end{equation}
Rewritten in terms of the column lengths or Dynkin labels $r_i$, it 
follows that 
\begin{equation}\label{dynkinchar}
    \chi_{(r_1, r_2,r_3)}(p,q,r)=\frac{{\rm det}\left(
\begin{array}{cccc}
 v_1^{r_1+r_2+r_3+3} & v_2^{r_1+r_2+r_3+3} & v_3^{r_1+r_2+r_3+3} & v_4^{r_1+r_2+r_3+3} \\
 v_1^{r_2+r_3+2} & v_2^{r_2+r_3+2} & v_3^{r_2+r_3+2} & v_4^{r_2+r_3+2} \\
 v_1^{r_3+1} & v_2^{r_3+1} & v_3^{r_3+1} & v_4^{r_3+1} \\
 1 & 1 & 1 & 1 \\
\end{array}
\right)}{{\rm det}\left(
\begin{array}{cccc}
 v_1^3 & v_2^3 & v_3^3 & v_4^3 \\
 v_1^2 & v_2^2 & v_3^2 & v_4^2 \\
 v_1 & v_2 & v_3 & v_4 \\
 1 & 1 & 1 & 1 \\
\end{array}
\right)}
\end{equation}
As a check, the fundamental representation is labeled by $(r_1, r_2, r_3)=(1,0,0)$. It is easy to check that plugging these values into  \eqref{dynkinchar} yields $v_1 + v_2 + v_3 + v_4$ as expected.

\subsection{Full Single Particle Partition Function}\label{results}

The spectrum of supergravitons in $AdS_5 \times S^5$ is the spectrum of 
chiral primaries ${\rm Tr} Z^p$ $p=1 \ldots \infty$ and their descendents. Each of the chiral primaries described above can be decomposed into a finite number of primaries of the conformal algebra. 
A full listing of this decomposition is provided in table 1 of 
\cite{Gunaydin:1984fk}. All conformal multiplets that appear in this listing for $p\geq 3$ are long representations. When $p=1$ and 
$p=2$, on the other hand, some of the conformal representations that appear in this listing are short. We list all these short representations in Table \ref{p1table} and Table \ref{p2}. 

Taking the products of $SO(4,2)$ and $SU(4)$ characters (described in detail in the previous two subsections) we obtain a definite character formula for each field listed in Table 1 of \cite{Gunaydin:1984fk}. 
It is then a simple matter to sum over the full spectrum of \cite{Gunaydin:1984fk} and find the full single-letter partition function. We now present our final (unfortunately rather unwieldy) answer
for this partition function, separately for bosonic and fermionic letters. 

\begin{table}[h!]

\centering
\begin{tabular}{|c|c|c|c|}
\hline
$(j_1,j_2)$ &  $2\Delta$ & Fields & $SU(4)$ Dynkin labels \\ 
\hline
$(0, 0)$ & $2$     & $\varphi^{(1)}$      & $(0, 1, 0)$ \\
$(\frac{1}{2}, 0)$ & $3$ & $\lambda^{(1)}$ & $(0, 0, 1)$ \\
$(0, \frac{1}{2})$ & $3$ & $\lambda^{(1)}$ & $(1, 0, 0)$ \\
$(1, 0)$ & $4$ & $A_{\mu \nu}^{(1)}$ & $(0, 0, 0)$ \\
$(0, 1)$ & $4$ & $\tilde{A}_{\mu \nu}^{(1)}$ & $(0, 0, 0)$ \\
\hline
\end{tabular}
\caption{$p=1$ multiplet}
\label{p1table}
\end{table}

\begin{table}[h!]
\centering

\begin{tabular}{|c|c|c|c|}
\hline
$SO(4)$ Labels & $2\Delta$ & Fields & $SU(4)$ Dynkin Labels \\ 
\hline
$\left( \frac{1}{2}, \frac{1}{2} \right)$ & $2p+2 = 6$ & $A^{(1)}_{\mu}$ & $(1, p-2, 1) = (1, 0, 1)$ \\ 
\hline
$\left( 1, \frac{1}{2} \right)$ & $2p+3 = 7$ & $\psi^{(1)}_{+\mu}$ & $(1, p-2, 0) = (1, 0, 0)$ \\ 
\hline
$\left( \frac{1}{2}, 1 \right)$ & $2p+3 = 7$ & $\psi^{(1)}_{-\mu}$ & $(0, p-2, 1) = (0, 0, 1)$ \\ 
\hline
$\left( 1, 1 \right)$ & $2p+4 = 8$ & $h_{\mu\nu}$ & $(0, p-2, 0) = (0, 0, 0)$ \\ 
\hline
\end{tabular}
\caption{$ p=2$ multiplet}
\label{p2}
\end{table}

\begin{equation}\label{zb}
    \begin{split}
        z_B&=\mathcal{D}s (s (-s^2 (s^2-1) x y r^4+(s (s-q) (q s-1) (x+1) \sqrt{x} y^{3/2}+s^3 (1-q s) x y^2-(q (s-q) x^2 s^3\\&+q (s-q) s^3+(s^2+1) ((q^2+1) s^3+q s^2-2 (q^2+1) s+q) x) y+s^3 (1-q s) x\\&+s (s-q) (q s-1) (x+1) \sqrt{x} \sqrt{y}) r^2-q^2 s^2 (s^2-1) x y) p^4+(s^2 (s-q) (q s-1) (r^4+q (s^2+1)^2 r^2\\&+q^2) (x+1) \sqrt{x} y^{3/2}+q s (q^3 x s^3+r^2 (-s r^2+s^2+1) x s^3+q^2 (r^2 (s^2+1)-s) x s^3\\&+q r^2 ((-2 x s^4-(x (x+3)+1) s^2+r^2 x s+x^2+1) s^2+x)) y^2+(s^2 (2-s (r^2+s) (s^2-1)) x q^4\\&+(r^2 (s^2+1)-s) ((x^2+x+1) s^4+2 x s^2+x) q^3+s (s (-s^4+s^2+2) x r^4-(x s^8+(x (2 x+3)+2) s^6\\&+(x (3 x+7)+3) s^4+3 x s^2-(x-1)^2) r^2+s^3+s x (-s^4+(x+1) s^2+2)) q^2\\&+r^2 (-s r^2+s^2+1) ((x^2+x+1) s^4+2 x s^2+x) q+r^4 s^4+r^4 s^4 x^2+r^2 s^2 (-s^3+s+\\&r^2 (-s^4+s^2+2)) x) y+q s (q^3 x s^3+r^2 (-s r^2+s^2+1) x s^3+q^2 (r^2 (s^2+1)-s) x s^3\\&+q r^2 ((-2 x s^4-(x (x+3)+1) s^2+r^2 x s+x^2+1) s^2+x))+s^2 (s-q) (q s-1) (r^4+q (s^2+1)^2 r^2\\&+q^2) (x+1) \sqrt{x} \sqrt{y}) p^2-q r (r^2+q) (s^2 (s^2+1) (q (s^3+s-q)-1) (x+1) \sqrt{x} y^{3/2}\\&-s^4 (s-q) (q s-1) x y^2+(q-s) (q s-1) ((x^2+x+1) s^4+2 x s^2+x) y-s^4 (s-q) (q s-1) x\\&+s^2 (s^2+1) (q (s^3+s-q)-1) (x+1) \sqrt{x} \sqrt{y}) p-p^3 r (r^2+q) (s^2 (s^2+1) (q (s^3+s-q)\\&-1) (x+1) \sqrt{x} y^{3/2}-s^4 (s-q) (q s-1) x y^2+(q-s) (q s-1) ((x^2+x+1) s^4+2 x s^2+x) y\\&-s^4 (s-q) (q s-1) x+s^2 (s^2+1) (q (s^3+s-q)-1) (x+1) \sqrt{x} \sqrt{y})+q^2 s (-s^2 (s^2-1) x y r^4\\&+(s (s-q) (q s-1) (x+1) \sqrt{x} y^{3/2}+s^3 (1-q s) x y^2-(q (s-q) x^2 s^3+q (s-q) s^3+(s^2+1) ((q^2+1) s^3\\&+q s^2-2 (q^2+1) s+q) x) y+s^3 (1-q s) x+s (s-q) (q s-1) (x+1) \sqrt{x} \sqrt{y}) r^2-q^2 s^2 (s^2-1) x y))\\
        \end{split}
\end{equation}

\begin{equation}\label{zf}
    \begin{split}
        z_F&=-\mathcal{D}(s^{3/2} (s^2+1) (p^4 r s^2 \sqrt{x} \sqrt{y} (q+r^2) (\sqrt{x} (y+1) (q s-1)+x \sqrt{y} (s-q)+\sqrt{y} (s-q))\\&+p^3 (y^{3/2} (q-s) (q^2 s^2 x (r^2 s-1)-q r^2 (s^2 (x (s^2+x+4)+1)+x)+r^2 s x (1-r^2 s))\\&+\sqrt{y} (q-s) (q^2 s^2 x (r^2 s-1)-q r^2 (s^2 (x (s^2+x+4)+1)+x)+r^2 s x (1-r^2 s))\\&+(x+1) \sqrt{x} y (q s-1) (q^2 s (s-r^2)+q r^2 (s^4+4 s^2+1)+r^2 s^2 (r^2-s)) +q r^2 s^2 (x+1) \sqrt{x} y^2 (q s-1)\\&+q r^2 s^2 (x+1) \sqrt{x} (q s-1))+p^2 r (q+r^2) (-(y^{3/2} (q s-1) (q^2 s x-q (s^2 (x (s^2+x+4)+1)+x)\\&+s^3 x))-\sqrt{y} (q s-1) (q^2 s x-q (s^2 (x (s^2+x+4)+1)+x)+s^3 x)+(x+1) \sqrt{x} y (q-s) (q^2 s^3\\&-q (s^4+4 s^2+1)+s)-q s^2 \sqrt{x} (x+1) y^2 (q-s)-q s^2 (x+1) \sqrt{x} (q-s))+p q (y^{3/2} (q-s) (q^2 s^2 x (r^2 s-1)\\&-q r^2 (s^2 (x (s^2+x+4)+1)+x)+r^2 s x (1-r^2 s))+\sqrt{y} (q-s) (q^2 s^2 x (r^2 s-1)\\&-q r^2 (s^2 (x (s^2+x+4)+1)+x)+r^2 s x (1-r^2 s))+(x+1) \sqrt{x} y (q s-1) (q^2 s (s-r^2)+q r^2 (s^4+4 s^2\\&+1)+r^2 s^2 (r^2-s))+q r^2 s^2 (x+1) \sqrt{x} y^2 (q s-1)+q r^2 s^2 (x+1) \sqrt{x} (q s-1))\\&+q^2 r s^2 \sqrt{x} \sqrt{y} (q+r^2) (\sqrt{x} (y+1) (q s-1)+x \sqrt{y} (s-q)+\sqrt{y} (s-q))).\\
    \end{split}
\end{equation}
where $\mathcal{D}$ equals 
\begin{equation}
    \frac{1}{\mathcal{D}} = (q-s) (q s-1) (p s-r) (r s-p) \left(s \sqrt{x}-\sqrt{y}\right) \left(s \sqrt{y}-\sqrt{x}\right) \left(s-\sqrt{x} \sqrt{y}\right) \left(s \sqrt{x} \sqrt{y}-1\right) (q s-p r) (q-p r s)
\end{equation}

\subsection{Matching with the superconformal index}\label{check}

The single particle superconformal index is defined in \cite{Kinney:2005ej} as,
\begin{equation}
    i_{s.p} = {\rm Tr}_{s.p}[(-1)^F u^{\delta} t^{2(E+j_1)}\bar{y}^{2j_2} v^{R_2}w^{R_3}]
\end{equation}

where, $\delta = E - 2 j_1 -\frac{1}{2}(3R_1 + 2R_2+ R_3).$

We can obtain the single letter superconformal index from \eqref{zb} 
\eqref{zf} by computing $z_B-z_F$ with the replacements
 \begin{equation} \label{repla}
    s \rightarrow u t^2, x \rightarrow  u^{-2} t^2, y  \rightarrow  \bar{y}^2, p  \rightarrow u^{-3/2},  q\rightarrow  u^{-1}v, r\rightarrow u^{-1/2} w
\end{equation}
Magically, we find that the ugly expressions \eqref{zb} and \eqref{zf} 
collapse into the extremely simple
\begin{equation}
    i_{s.p} = -\frac{t^3}{\bar{y} \left(1-\frac{t^3}{\bar{y}}\right)}-\frac{t^3\bar{y}}{1-t^3 \bar{y}}+\frac{t^2 w}{v \left(1-\frac{t^2 w}{v}\right)}+\frac{t^2 v}{1-t^2 v}+\frac{t^2}{w \left(1-\frac{t^2}{w}\right)}, 
\end{equation}
an expression that agrees perfectly with the single letter superconformal index computed in  \cite{Kinney:2005ej}(eqn. 4.14). We view this agreement as a highly nontrivial consistency check of \eqref{zb}
and \eqref{zf}.

\subsection{Turning $SU(4)$ chemical potentials off}\label{so6off}

Our rather complicated single-letter partition functions, 
\eqref{zb} and \eqref{zf} simplify somewhat when we set all $SU(4)$
fugacities to unity. Upon setting $p,q,r=1$ we find 
\begin{align}\label{zboff}
    \begin{split}
       z_{B}(s,x,y,1,1,1) &=\mathcal{N} s \left(-s^3 (s+1) x^2-s^3 (s+1)+s^2 \left((s ((s-5) s-2)-14) s^2+s-13\right) (x+1) \sqrt{x} y^{3/2}\right. \\&\left.+s^2 \left((s ((s-5) s-2)-14) s^2+s-13\right) (x+1) \sqrt{x} \sqrt{y}-y \left(s \left(s \left(s \left(s \left(s \left(s \left(s^2+s+4\right)+6\right)\right.\right.\right.\right.\right.\right.\\&\left.\left.\left.\left.\left.\left.+31\right)+5\right)+26\right) x+s \left((2 (s-2) s+11) s^2+s+1\right)+\left((2 (s-2) s+11) s^3+s^2+s+1\right) x^2\right.\right.\right.\\&\left.\left.\left.-8 x+1\right)+6 x\right)-s \left((2 (s-2) s+11) s^3+s^2+s+1\right) x\right. \\&\left.-s y^2 \left(s \left(s \left((s+1) x^2+s (2 (s-2) s+11) x+s+x+1\right)+x\right)+x\right)\right)\\z_{F}(s,x,y,1,1,1) &=-4\mathcal{N} s^{3/2} \left(s^2+1\right) \left(\sqrt{x} (y+1)+x \sqrt{y}+\sqrt{y}\right) \left(s^2 (x+1) y+s^2 (x+1)\right. \\&\left.+(s (s ((s-1) s+4)-1)+1) \sqrt{x} \sqrt{y}\right)\\ 
    \end{split}
\end{align}
where $\mathcal{N}$ equals,
$$\mathcal{N}=\frac{1}{(s-1)^5 \left(s \sqrt{x}-\sqrt{y}\right) \left(s \sqrt{y}-\sqrt{x}\right) \left(s-\sqrt{x} \sqrt{y}\right) \left(s \sqrt{x} \sqrt{y}-1\right)}$$
\subsubsection{Partition function at $\omega_1\rightarrow 1,~\omega_2 \rightarrow 1$ }\label{fluidym}

We can multi-particle the single-particle partition functions 
\eqref{zboff} (with Bose and Fermi statistics respectively) 
using the formulae
\begin{equation}
\begin{split}
    \ln Z_{B} &= \sum_{n=1}^{\infty} \frac{1}{n} z_{B}(s^n,x^n,y^n,p^n,q^n,r^n)\\\ln Z_{F} &= \sum_{n=1}^{\infty} \frac{(-1)^{n+1}}{n} z_{F}(s^n,x^n,y^n,p^n,q^n,r^n)
\end{split}
\end{equation}
In the limit $\omega_1\rightarrow 1,\omega_2\rightarrow 1$ and putting $p,q,r=1$ we get,
\begin{equation}\label{fluidpf}
    \begin{split}
        \ln Z_{B} &=\sum_{n=1}^{\infty}\frac{(22 \cosh (\beta  n)+17 \cosh (2 \beta  n)+6 \cosh (3 \beta  n)+\cosh (4 \beta  n)+18) \text{csch}^7\left(\frac{\beta  n}{2}\right) \text{sech}\left(\frac{\beta  n}{2}\right)}{32 \beta ^2 n^3 \left(1-\omega _1^2\right) \left(1-\omega _2^2\right)}\\\ln Z_{F}&=\sum_{n=1}^{\infty} \frac{(-1)^{n+1}\cosh (\beta  n) (\cosh (\beta  n)+\cosh (2 \beta  n)+2) \text{csch}^7\left(\frac{\beta  n}{2}\right)}{2 \beta ^2 n^3 \left(1 -\omega _1^2\right) \left(1 -\omega _2^2\right)}\\
    \end{split}
\end{equation}
\eqref{fluidpf} is the final result of this Appendix. 
 
\section{The Dilaton Propagator in $AdS_5 \times S^5$} \label{propagator}

In this Appendix we review the elegant derivation of \cite{Dorn:2003au, Dai:2009zg} of the dilaton propagator in $AdS_5 \times S^5$.

To start with let us work in coordinates in which the metric of $AdS_5 \times S^5$ takes the form 
\begin{equation}\label{metric1}
\begin{split}
ds^2&= R_{AdS_5}^2 \left( 
\frac{dz^2}{z^2}+ \frac{dx_i^2}{z^2} 
+ d \Omega_5^2 \right) \\
&=  \frac{R_{AdS_5}^2}{z^2} \left( dx_i^2 + dz^2 + z^2 d\Omega_5^2\right) \\
&=  \frac{R_{AdS_5}^2}{z^2} \left( dx_i^2 + dy_j^2 \right) \\
\end{split}
\end{equation} 
where $i=1 \ldots 4$, $j=1 \ldots 6$, 
and $y_j$ are 6 Cartesian coordinates in the space whose polar coordinate line element is given by $dz^2+z^2 d \Omega_5^2$. We see that the metric $AdS_5 \times S^5$ is Weyl equivalent to flat 10-dimensional spacetime. 

Now recall that, in $D$ spacetime dimensions, the conformally coupled 
Laplacian 
\begin{equation}\label{}
\nabla^2_{\rm conf, g_{\mu \nu} } \ \equiv \left( g^{\mu \nu} \nabla_\mu \nabla_\mu  
-  \frac{D-2}{4(D-1)}R \right)
\end{equation} 
obeys the property
\begin{equation}\label{confme}
\nabla^2_{x,~\rm conf,~e^{2\chi} g_{\mu \nu}}
\left( e^{- \chi \frac{D-2}{2}} \phi \right) 
= e^{-\frac{(D +2)\chi }{2}} \nabla^2_{\rm conf, ~g_{\mu \nu} } \left( \phi \right) 
 \end{equation} 

Now let us suppose we have found a Green's function $G$ for the operator $\nabla^2_{\rm conf, ~g_{\mu \nu} }$. By definition, $G$ obeys the equation 
\begin{equation}\label{gfG1}
 \nabla^2_{\rm conf, ~g_{\mu \nu} }G(x, y)=
\frac{1}{\sqrt{-g}} \delta^D(x -y)
\end{equation} 
Using \eqref{confme}, it follows that 
\begin{equation}\label{gfG}
\begin{split}
     \nabla^2_{\rm conf, ~e^{2 \chi} g_{\mu \nu} } \left(  e^{- \chi(x) \frac{D-2}{2}}  G(x, y) \right) &= \frac{ e^{-\frac{(D +2)\chi(y) }{2}}  \times e^{ \chi(y) D}} {\sqrt{-ge^{2 \chi(x)}}} \delta^D(x-y)\\&=\frac{ e^{\frac{(D -2)\chi(y) }{2}}  } {\sqrt{-ge^{2 \chi(x)}}} \delta^D(x-y)\\
\end{split}
\end{equation} 
The key point here is that while the 
derivatives on the LHS are derivatives w.r.t. $x_\mu$, the $\delta$ function allows us to view every $\chi$ on the RHS as a function of $y^\mu$. As functions of 
$y^\mu$ are just constants as far as the differential operator on the RHS is concerned. As a consequence, it follows that 
\begin{equation}\label{gfGn}
\begin{split}
     \nabla^2_{\rm conf, ~e^{2 \chi} g_{\mu \nu} } \left(  e^{- \chi(x) \frac{D-2}{2}}  G(x, y) e^{- \chi(y) \frac{D-2}{2}} \right) &= \frac{1} {\sqrt{-ge^{2 \chi(x)}}} \delta^D(x-y)
\end{split}
\end{equation} 
and so we conclude that \cite{Dorn:2003au,Dai:2009zg}
\begin{equation}
    G_{g e^{2\chi}}(x,y) = e^{- \chi(x) \frac{D-2}{2}}  G_{g_{\mu\nu}}(x, y)e^{- \chi(y) \frac{D-2}{2}}
\end{equation}
is the Greens function of the operator 
$\nabla^2_{\rm conf}$ in the metric 
$e^{2 \chi} g_{\mu\nu}$. 

In summary
\begin{equation}\label{onh}
g_{\mu \nu} \rightarrow e^{2 \chi} g_{\mu\nu} \implies G(x, y) \rightarrow
e^{- \chi(x) \frac{D-2}{2}}  G(x, y)e^{- \chi(y) \frac{D-2}{2}}
\end{equation}
Viewing the propagator as a two-point function $\langle \phi(x) \phi(y) \rangle$, we see that this two-point function transforms exactly as we would expect of a field of Weyl weight $\frac{D-2}{2}$, which, of course, is also the scaling dimension of $\phi$. 

Now recall that the scalar curvature 
of $AdS_5 \times S^5$ vanishes (because the positive scalar curvature of $S^5$ exactly cancels the negative scalar curvature of $AdS_5$). The scalar curvature of 10-dimensional flat space also, of course, vanishes. So in the particular case of flat space and $AdS_5 \times S^5$, it follows that 
$\nabla^2_{\rm conf}$ is simply the (minimally coupled) Laplacian operator, the kinetic term of the dilaton. It follows, therefore, that in the coordinate 
\eqref{metric1}, the propagator dilaton propagator is given by 
\begin{equation}\label{dilmet}
\langle \phi(x) \phi (y) \rangle= z_x^4 G(x, y) z_y^4
\end{equation}
where $G(x, y)$ is the well known propagator in flat space given by 
\begin{align}\label{fsprop}
    G(x,x')&=\frac{\Gamma[4]}{4\pi^5} \left(\frac{1}{(x_i-x'_j)^2}\right)^{4}\nonumber\\
\end{align}
Given \eqref{fsprop}, we evaluate
\begin{align}\label{fuv}
    \langle \phi(\tilde{x}_1) \phi (\tilde{x}_2) \rangle&=z_1^4 G(\tilde{x}_1,\tilde{x}_2) z_2^4\\
    &=\frac{\Gamma[4]}{4\pi^5}z_1^4 \left(\frac{1}{(x_1-x_2)^2+(y_1-y_2)^2}\right)^{4}z_{2}^4\nonumber\\
    &=\frac{\Gamma[4]}{2^6\pi^5}\left( \frac{1}{\frac{(x_1-x_2)^2}{2z_1z_2}+\frac{(y_1-y_2)^2}{2z_1z_2}}\right)^{4}\nonumber\\&=\frac{\Gamma[4]}{2^6\pi^5}\left( \frac{1}{\frac{(x_1-x_2)^2}{2z_1z_2}+\frac{(z_1y'_1-z_2y'_2)^2}{2z_1z_2}}\right)^{4}\nonumber\\&=\frac{\Gamma[4]}{2^6\pi^5}\left( \frac{1}{\frac{(x_1-x_2)^2}{2z_1z_2}+\frac{(z_1 -z_2)^2 +2z_1z_2 - 2z_1z_2y'_1 \cdot y'_2}{2z_1z_2}}\right)^{4}\nonumber\\&=\frac{\Gamma[4]}{2^6\pi^5}\left( \frac{1}{\frac{(x_1-x_2)^2}{2z_1z_2}+\frac{(z_1 -z_2)^2}{2z_1z_2} +1  - y_1^{'}\cdot {y_2^{'}}}\right)^{4}\nonumber \\&=\frac{\Gamma[4]}{4\pi^5}\left( \frac{1}{u + v}\right)^{4}\nonumber 
\end{align}

where $y'_i=\frac{y_i}{z}$
\footnote{Recall $y_i$'s are embedding space Cartesian coordinates (in $\mathbb{R}^6$) of a sphere of radius $z$ so that 
$$\sum y_i^2 = z^2$$
It follows that $ y'_i$ lie on a unit sphere.} and 
\begin{equation}\label{uvdef}
v=2\left(1-y_1^{'}. y_2^{'}\right), ~~~~~u = 2\left( \frac{(x_1-x_2)^2 +(z_1 -z_2)^2}{2z_1z_2}\right)
\end{equation}

We will now explain the geometric meaning of the quantities $v$ and $u$. We will verify that $v$ is the chordal
distance between two points on the unit sphere, while $u$ is the chordal distance between two points on the unit $AdS_5$
\footnote{The chordal distance is defined to be the distance between the points as measured (by the usual Pythagorean formula) 
in the relevant $\mathbb{R}^{4,2}$ or $\mathbb{R}^{6,0}$ embedding space. It gets its name because it is the distance of the chord 
- the shortest straight line between the two points - in embedding space. } 

That $v$ is the chordal distance on the unit sphere follows immediately from 
\begin{equation}
(y'_1 -y'_2)^2=2\left(1-y_1^{'}. y_2^{'}\right) = v
\end{equation}
As we have parameterized $AdS_5$ in coordinates adapted to the Poincare Patch, it takes a little more effort to verify that $u$ is the $AdS_5$ chordal distance.  Recall that the Poincare patch coordinates $x_i$ and $u$ are related to embedding space coordinates
\begin{equation}
    -X_{-1}^2 - X_{0}^2  + \sum_{1}^{d-1}X_{i}^2 = -1
\end{equation}
via 
\begin{equation}
    \begin{split}
        X_i &= \frac{x_i}{z},~~~i= 0,... d-2\\X_{-1} &=\frac{1}{2}\left( \frac{(z^2 +x^2)}{z} + \frac{1}{z}\right)\\ X_{d-1} &=\frac{1}{2}\left( \frac{(z^2 +x^2)}{z} - \frac{1}{z}\right)
    \end{split}
\end{equation}
It follows that
\begin{equation}\label{udef}
    \begin{split}
       (X_1 - X_2)^2=-2-2X_1\cdot X_2=-2 +2\left(1+ \frac{(x_1-x_2)^2}{2z_1z_2}+\frac{(z_1 -z_2)^2}{2z_1z_2}\right)=u\\
    \end{split}
\end{equation}

\section{Sum over KK modes}\label{summing}

In this Appendix, we study the relationship between the 10-dimensional dilaton propagator and the propagators of its five-dimensional descendants. We also demonstrate that the bulk stress tensor, computed using the 10-dimensional dilaton propagator, agrees with the sum of the bulk stress tensors (computed from the various 5-dimensional propagators). 

\subsection{Decomposition of the 10d propagator}

Let us consider the metric of $AdS_5\times S^5$ with equal radius (unity)
\begin{align}\label{space}
    ds^2=-(1+r^2)dt^2+\frac{dr^2}{1+r^2}+r^2(d\theta^2+\sin^2\theta d\phi_1^2+\cos^2\theta d\phi_2^2)+d\Omega_5^2.
\end{align}
Consider a massless minimally coupled scalar field propagating on this space. We can KK decompose this field on $S^5$. The KK masses so obtained are the eigenvalues of $-\nabla^2$ (acting on $S^5$). Now the eigenstates of $-\nabla^2$ on $S^5$ are 
the spherical harmonics $Y_l^\alpha$, with eigenvalue $l(l+4)$ and degeneracy labels running over $d_l=\frac{(2l+4)(l+1)(l+2)(l+3)}{24} $ values. Consequently, the 
10 dimensional Greens function $G^{(10)}$ on the space \eqref{space} can be 
\begin{equation}\label{gteno}
G^{(10)}\left( (x, \theta) , (x', \theta')  \right) 
= \sum _{l=0}^\infty  \sum_{\alpha=1}^{d_l} G_{l}^{(5)}(x, x') Y_l^\alpha(\theta) Y_l^{*\alpha}(\theta') 
\end{equation} 
where 
\begin{equation} 
-\nabla_{10}^2 G^{(10)}\left( (x, \theta) , (x', \theta')  \right) = \delta^{(10)}\left( x-x', \theta-\theta'\right)
\end{equation} 
\begin{equation}\label{norm5}
-\nabla_{5}^2 G_{l}^{(5)}\left( x , x' \right) = \delta^{(5)}\left( x-x'\right)
\end{equation} 
and our spherical harmonics are normalized so that 
\begin{equation}\label{sphnorm}
\int_{S^5} Y_l^\alpha (\theta) Y_l^{*\alpha}(\theta)=1
\end{equation}
\footnote{\eqref{gteno} can be established by acting on both sides of this equation 
with $-\nabla^2_{10}$, and using the completeness relation 
$$ \sum_{l, \alpha} Y_l^\alpha (\theta) Y_l^{*\alpha}(\theta') = \delta^{(5)}(\theta-\theta')$$}
As in the main text, we can obtain the bulk stress tensor by acting on \eqref{gteno}
with the appropriate derivatives. In the limit of interest to this paper, the variation of $g$ in factors of $\sqrt{g}$
don't contribute to the stress tensor and the 5 and 10-dimensional stress tensors
are computed via the same derivative operations (all derivatives are in the $AdS_5$ direction). It thus follows from 
\eqref{gteno} that 
\begin{equation}\label{gtstress}
T_{\mu\nu}^{(10)}
= \sum _{l=0}^\infty  \sum_{\alpha=1}^{d_l} (T_l)_{\mu\nu}^{(5)} Y_l^\alpha(\theta) Y_l^{*\alpha}(\theta) 
\end{equation}
(In this section we interchangeably use angles on the sphere, and unit vectors in 
embedding space, as coordinates on $S^5$). 

In order to further process this relation, we use the completeness formula 
\begin{equation}\label{complete}
\sum_\alpha Y_l^\alpha({\hat n}) Y_l^{*\alpha}({\hat n}')= \frac{d_l} {\Omega_5 P^l({\hat n}.{\hat n}'=1)}  P^l({\hat n}.{\hat n}')
\end{equation}
\footnote{We obtain the proportionality constant on the RHS by setting ${\hat n}= {\hat n}'$.}where $P^l({\hat n}.{\hat n}')$ is the unique $l^{th}$ spherical harmonic that preserves $SO(5)$ (and is normalized to obey \eqref{sphnorm}). Setting ${\hat n}={\hat n}'$ in \eqref{complete}, and substituting in \eqref{gtstress} gives 
\begin{equation}\label{gtstressn}
T_{\mu\nu}^{(10)}
= \sum _{l=0}^\infty d_l ~ \frac{ (T_l)_{\mu\nu}^{(5)} }{\Omega_5} 
\end{equation}
Integrating the LHS over $S^5$ (recall the stress tensor is independent of the coordinates on the $S^5$) gives 
\begin{equation}\label{gtstressnn}
\int_{S^5} T_{\mu\nu}^{(10)}
= \sum _{l=0}^\infty d_l ~ (T_l)_{\mu\nu}^{(5)} 
\end{equation}
\eqref{gtstressnn} tells us that the 
energy computed using the full 10 dimensional propagator the sum over energies of each of the 5 dimensional particles. 

\subsection{Explicit check of equality of 10d and 5d stress tensors}

In this subsection, we explicitly perform the summation on the RHS of \eqref{gtstressn}, and demonstrate that the result equals the LHS (already computed in  \eqref{stress10}). 

The mass of a minimally coupled scalar in $AdS_5\times S^5$ is related to its conformal dimension can be obtained 
\begin{align}
   M^2 =\Delta(\Delta-4)=m^2_{KK}=l(l+4), 
\end{align}
The particles of interest to this paper have $\Delta=l+4$ and $M^2= l(l+4)$. 

The bulk to bulk propagator (defined by  \eqref{norm5}) ( for a scalar in $AdS_5$ with any given value of $\Delta$) is given by  well known \cite{Alday:2020eua}
\begin{align}\label{b2b}
  G^{(5)}_{l}(y,y')&=  \frac{\Gamma(\Delta)}{2\pi^{2}\Gamma(\Delta-1)}\frac{1}{u^\Delta}{}\,_2F_1 (\Delta,\Delta-\frac{3}{2},2\Delta-3,-\frac{4}{u})\nonumber\\
   &=\frac{\Gamma(l+4)}{2\pi^{2}\Gamma(l+3)}\frac{1}{u^{l+4}}{}\,_2F_1 (l+4,l+5/2,2l+5,-\frac{4}{u})
\end{align}

Consequently \eqref{gteno} takes the explicit form 
\begin{align}\label{expsplit}
G^{(10)}(x,x')&=\sum_{l=0}^{\infty}\sum_{\alpha=1}^{d_l}\frac{\Gamma(l+4)}{2\pi^{2}\Gamma(l+3)}\frac{1}{u^{l+4}}{}\,_2F_1 (l+4,l+5/2,2l+5,-\frac{4}{u}) Y_{l}^{\alpha} (\Omega)Y_{l}^{*\alpha}(\Omega')
\end{align}
As explained in the main text, the stress tensor is obtained by taking appropriate derivatives on ${\tilde G}(x, x')$ 
as in \eqref{olmju1}. As in the main text (see around \eqref{deru}), when working at leading order, we need only retain those terms $(\partial_\mu u)^2$ (and can ignore terms proportional to $\partial_\mu \partial_\nu u)$. Acting on \eqref{expsplit} we find an answer of the form 
\begin{align}
    \langle T_{\mu\nu}^{(10)}\rangle=-2\sum_{q=1}^{\infty}\sum_{l=0}^{\infty}\sum_{\alpha=1}^{d_l}K_{l,\mu\nu}^q  Y_{l}^{\alpha} (\Omega)Y_{l}^{*\alpha}(\Omega)
\end{align}
Let us first explicitly compute $\langle T^{(10)}_{00}\rangle$. 
    \begin{align}\label{Ka}
    -K_{l,\tau\tau}^q&=-\frac{4x^4\,\gamma^4(\theta)\sinh^2(q\beta)}{2\pi^2}\frac{\Gamma[l+4](l+4)}{\Gamma[l+3](4+u_q)u_q^{l+6}}\Big[5 (u_q+2) \, _2F_1\left(l+\frac{5}{2},l+5;2 l +5;-\frac{4}{u_q}\right)\nonumber\\
    &+(l u_q) \, _2F_1\left(l+\frac{5}{2},l+4 ;2 l+5;-\frac{4}{u_q}\right)\Big]\nonumber\\
    &=-\frac{4x^4\,\gamma^4(\theta)\sinh^2(q\beta)}{2\pi^2}\frac{\Gamma[l+5]}{\Gamma[l+3]u_q^{l+6}}\Big[(l+5)  \, _2F_1\left(l+\frac{5}{2},l+6;2l+5;-\frac{4}{u_q}\right)\Big]
\end{align}
In the last line, we use (15.2.11) identity of \cite{Abramowitz} that relates derivatives of the relevant hypergeometric functions to linear combinations of other such functions. 
Since \eqref{Ka} only depends on the quantum number $l$, we can sum over other quantum numbers on $S^5$ using the `addition theorem of spherical harmonics'
\begin{align}\label{add2}
    \sum_{l=0}^{\infty} \sum_{\alpha=1}^{d_l} Y_{l}^{\alpha}(\Omega)Y_{l}^{*\alpha}(\Omega')& = \sum_{l=0}^{\infty}\frac{(2l + 4) \Gamma\left(2\right)}{4\Omega_5} C_l^{(2)}(\cos \Theta), \quad \cos\Theta=1-\frac{v}{2}.
\end{align}
where $C_l^{(2)}(\cos \Theta)$ are the Gegenbauer
polynomials and $v$ is the chordal distance on $S^5$.
Note that, in the limit $\Theta \rightarrow 0$, the Gegenbaur polynomial $C_l^{(2)}(1)$ becomes $\frac{(l+3)(l+2)(l+1)}{6}$. Therefore the addition theorem in \eqref{add2} precisely matches with \eqref{complete}.
Substituting the addition formula in \eqref{Ka}, we obtain
\begin{align}
      \langle T_{00}^{(10)}\rangle&=\lim_{\Theta \rightarrow 0}\sum_{q=1}^{\infty}\sum_{l=0}^{\infty}\frac{4x^4\gamma^4(\theta)\sinh^2(q\beta)(2l+4)}{8\pi^5}\frac{\Gamma[(l+6)]}{\Gamma[l+3]u_q^{l+6}}\nonumber\\
      &~~~~~~~~~~~~~~~\times\Big[\, _2F_1\left(l+\frac{5}{2},l+6;2l+5;-\frac{4}{u_q}\right)C_l^{(2)}(\cos \Theta)\Big]\nonumber\\
      &=\lim_{\Theta \rightarrow 0}\sum_{q=1}^{\infty}\sum_{l=0}^{\infty}\frac{x^4\,\gamma^4(\theta)\sinh^2(q\beta)}{\pi^5}\frac{\Gamma[(l+6)]}{\Gamma[l+2](u_q+2)^{l+6}}\nonumber\\
      &~~~~~~~~~~~~~~~\times\Big[ \, _2F_1\left(\frac{l}{2}+3,\frac{l}{2}+\frac{7}{2};l+3;\frac{4}{(u_q+2)^2}\right)C_l^{(2)}(\cos \Theta)\Big]
\end{align}
To obtain the last line, we use  the identity given in equation (15.3.16) of \cite{Abramowitz}
\footnote{$_2F_1\left(a,b,2b,z\right)=(1-\frac{z}{2})^{-a}\,_2F_1\left(a/2,a/2+1/2,b+1/2,\frac{z^2}{(2-z)^2}\right)$}

We now perform sum over $l$ using the following identity given in (B.9) of \cite{Dorn:2003au} 
\begin{equation}
\sum_{l=0}^{\infty} \frac{\Gamma(l + \alpha)}{\Gamma(l + \beta)} \left( \frac{z}{2} \right)^l
F\left( \frac{l}{2} + \frac{\alpha}{2}, \frac{l}{2} + \frac{\alpha}{2} + \frac{1}{2}; l + \beta + 1; z^2 \right)
C_l^{(\beta)}(\cos\Theta) = \frac{\Gamma(\alpha)}{\Gamma(\beta)} \frac{1}{(1 - z\cos\Theta)^{\alpha}}, \end{equation}
In our case, $\alpha=6$ , $\beta=2$ and $z=\frac{2}{u_q+2}$.
\begin{align}
      \langle T_{00}^{(10)}\rangle&=\lim_{\Theta \rightarrow 0}2\sum_{q=1}^{\infty}\frac{x^4\,\gamma^4(\theta)\sinh^2(q\beta)\Gamma[6] }{\Gamma[2]\pi^5(u_q+2)^6\left(1-\frac{2}{u_q+2}\cos\Theta\right)^6}
\end{align}
Since, all the angles on $S^5$ have to identified with each other, $\Theta\rightarrow 0$ and finally we obtain
\begin{align}\label{sts}
     \langle T_{00}^{(10)}\rangle&=\sum_{q=1}^{\infty}\frac{2x^4\,\gamma^4(\theta)\sinh^2(q\beta)\Gamma[6] }{\Gamma[2]\pi^5 \,u_q^6}
\end{align}

Similarly, we compute all other components of the stress tensor and they agree with \eqref{stress10}.

\section{Thermodynamics of the bulk gas from equilibrium partition functions}\label{eqflu}

In \S \ref{bulkstrten} we pointed out that fluid dynamics was more than an inspiration for guessing the correct ansatz; in fact the improved  bulk stress tensor \eqref{frst} takes precisely the ideal fluid (bulk) fluid dynamical form described in \cite{Banerjee:2012iz}. In this Appendix we elaborate on this point. In \S \ref{reviewmat} below we review the formalism of \cite{Banerjee:2012iz}. In \ref{tech} then demonstrate that the bulk stress tensor of the gas - as presented in \eqref{frst}  - fits perfectly into the framework of  \cite{Banerjee:2012iz}, provided we make the identifications \eqref{txpt}. In \S \ref{totalnfd} and \ref{toten} we report a couple of consistency checks of this formalism (by checking that it correctly reproduces the correct total - hence boundary - partition function and energy of the gas).

\subsection{Review of `Equilibrium Partition Functions'}\label{reviewmat}

In this brief subsection we provide a brief review of the leading order results of \cite{Banerjee:2012iz}. 

Consider a (Lorentzian) spacetime with an everywhere timelike killing vector $k= k^\mu \partial_\mu$. Let the (space dependent) norm of this vector be given by 
\begin{equation}\label{normvect}
g_{\mu\nu}k^\mu k^\nu \equiv - e^{2 \sigma} 
\end{equation}
(\eqref{normvect} defines the variable $\sigma$), and define the `velocity vector' field $u^\mu$ by 
\begin{equation}\label{vvf}
k= e^{\sigma} u, 
\end{equation}
It follows, of course, from \eqref{normvect} and \eqref{vvf} that $u^2=-1$. In the situation described above, it is always possible - and often useful - to choose our coordinates, $(\tau, x_i)$ so that curves of constant $x_i$ are tangent to the killing vector $k$, and the normalizations are chosen so that $k=\partial_\tau$.

The question posed in \cite{Banerjee:2012iz}
was the following. Consider quantum field theory living on a spacetime of the form described above. Let $H$ be the conserved charge that generates the translation $\delta x^\mu = k^\mu$ in this QFT. Question: what can we say on general grounds about the thermal system - defined by $e^{-\beta H}$ - in this system (here the inverse temperature $\beta=\frac{1}{T}$ is an arbitrary constant).

In order to address this question, \cite{Banerjee:2012iz} performed the analytic continuation to Euclidean space
$\tau=- it$, and compactified the resultant Euclidean time coordinate $t$ 
s.t. $t =t +\beta$. The resultant spacetime has a metric of the form 
\begin{equation}
    ds^2 = -e^{2\sigma(x)} \left( dt + a_i(x)\, dx^i \right)^2 + g_{ij}(x)\, dx^i dx^j
    \label{eq:stationary_metric}
\end{equation}
In these coordinates, the killing vector, velocity vector field and  local temperature (i.e. inverse of the proper size of the thermal circle) are given by
\begin{equation}
k^\mu=(1, {\vec 0})~~~~    u^\mu = e^{-\sigma}(1, {\vec 0}), \qquad T(x) = T_0 e^{-\sigma(x)}
    \label{eq:local_temp_velocity}
\end{equation}
With all these conventions in place, the authors of \cite{Banerjee:2012iz} then proceeded to present a general local expression for the partition function 
${\rm Tr} e^{-\beta H}$ in an expansion in derivatives. At leading (or perfect fluid) order in this expansion they found 
\begin{equation}\label{eq:partition_function}
    W = \ln Z = \int d^p x \sqrt{g_p}\, \frac{P\big(T(x)\big)}{T(x)}\, 
\end{equation}
where $P(T)$ is an arbitrary function
which we will now interpret. A standard thermodynamical relation asserts that the free energy density of a system is the negative of its pressure, i.e. that 
$F= -V P$. It follows, therefore, that $\ln Z= \frac{P V}{T}$. Applying this relationship patch by patch (as is appropriate in a slowly varying situation where the derivative expansion should apply) and integrating gives \eqref{eq:partition_function}, provided we interpret the function $P(T)$ as the 
thermodynamical `pressure' as a function of temperature.

The stress tensor that follows from varying 
\eqref{eq:partition_function} w.r.t. the metric turns out to be given by 
\begin{equation}\label{eq:perfect_fluid_tensor}
 \begin{split}   
 &T_{\mu\nu} = \left(\epsilon \left(T(x) \right)+ {P}\left(T(x) \right) \right)  u_\mu u_\nu + {P}\left(T(x) \right) g_{\mu\nu} \\
  &\epsilon(T) =  -P + T 
  \frac{\partial P}{\partial T}
\end{split}
\end{equation}
Note that the first of \eqref{eq:perfect_fluid_tensor} is precisely the form of the stress tensor of a fluid with pressure $P$ and energy density $\epsilon$, while the second the same equation (\eqref{eq:perfect_fluid_tensor}) is 
precisely the thermodynamical relation that relates the pressure and the energy 
density. 

\subsection{The improved bulk stress tensor from perfect fluid dynamics}\label{tech}

We now apply this formalism reviewed earlier in this subsection to the situation studied in this paper. The metric of interest to us is simply global $AdS_5$ space. The killing vector of interest is $k=\partial_t+\omega_1 \partial_{\phi_1} + \omega_2 \partial_{\phi_2}$. The norm of this vector is easily computed: we find 
\begin{equation}\label{normvec}
-k^2= e^{2 \sigma}= r^2+1 - r^2 \left(\omega_1^2 \cos^2 \theta+ \omega_2^2 \sin^2 \theta \right) = \frac{r^2}{\gamma^2} +1=
1+ 2x^2
\end{equation}
(recall $x^2 \equiv \frac{r^2}{2\gamma^2(\theta)}$), and so 
\begin{equation}
    u^{\mu} = \frac{k^\mu}{\sqrt{1 + 2x^2}} 
    \label{eq:fluid_velocity}
\end{equation}
\footnote{We can move from the usual global $AdS_5 $ coordinates to coordinates of the form 
\eqref{eq:stationary_metric} via the coordinate change
\begin{equation}
    \phi'_1 = \phi_1 - \omega_1 t, \qquad \phi'_2 = \phi_2 - \omega_2 t
    \label{eq:coord_transform}
\end{equation}
This is the coordinate system in which the fluid is at rest. The metric takes the form \eqref{eq:stationary_metric} with 
\begin{equation}
\begin{split}
    e^{2\sigma(x)} &= 1 + \frac{r^2}{\gamma^2(\theta)} = 1 + 2x^2 \\
    e^{2\sigma(x)} a_1 &= r^2 \omega_1 \sin^2\theta \\
    e^{2\sigma(x)} a_2 &= r^2 \omega_2 \cos^2\theta \\
    g_{ij} dx^i dx^j &= \frac{1}{1 + r^2} dr^2 + r^2 d\theta^2 
    + \left( r^2 \sin^2\theta + e^{2\sigma} a_1^2 \right) d{\phi'_1}^2 \\
    &\quad + \left( r^2 \cos^2\theta + e^{2\sigma} a_2^2 \right) d{\phi'_2}^2 + 2e^{2\sigma(x)}a_1 a_2 d\phi'_1 d\phi'_2
\end{split}
\label{eq:metric_components}
\end{equation}}
It is easily verified that \eqref{frst}
takes the form \eqref{eq:perfect_fluid_tensor}, with the fluid velocity given by \eqref{eq:fluid_velocity}, 
and the fluid pressure function  
and the temperature given by 
\begin{equation}
\begin{split}
    P(x) &= \int_x^\infty dz\, \frac{2\mathcal{F}(z)}{z^3} \\
    T(x)&=\frac{T_0}{\sqrt{1+2x^2}}
\end{split}
\label{eq:bulk_pressure_energy}
\end{equation}
\footnote{Using \eqref{eq:bulk_pressure_energy}, it follows from the second of \eqref{eq:perfect_fluid_tensor} that 
\begin{equation}\label{enden}
\begin{split}
\epsilon(x) &= \frac{\mathcal{F}(x)(1 + 2x^2)}{x^4} - \int_x^\infty dz\, \frac{2\mathcal{F}(z)}{z^3}\\
(\epsilon+P)(x)&=\frac{\mathcal{F}(x)(1 + 2x^2)}{x^4}
\end{split}
\end{equation}
so that the stress tensor in the first of \eqref{eq:bulk_pressure_energy} matches \eqref{frst}.}

As a check on our algebra, we verify in 
Appendix \ref{totalnfd} that the partition function obtained by integrating \eqref{eq:partition_function}
over a spatial slice of $AdS_5$ - and the energy obtained by integrating the  appropriate components of stress tensor \eqref{eq:perfect_fluid_tensor} - in both cases with $P$ given by \eqref{eq:bulk_pressure_energy} -   reproduce, respectively, the previously computed system partition function (see \eqref{hphi}) as well as the previously computed energy of the ensemble (first of \eqref{thermo}, but with $h_\Delta$ replace by $h_\phi$).  

\subsection{Reproducing boundary thermodynamics from bulk fluid dynamics}
\label{totalnfd}

The total partition function is given by \eqref{eq:partition_function}, where we substitute the metric given in \eqref{eq:metric_components} and the pressure and temperature given in \eqref{eq:bulk_pressure_energy}. The volume element $\sqrt{g_p}$ compute from \eqref{eq:metric_components}
is \begin{equation}
    \sqrt{g_p}=\frac{r^3\sin\theta\cos\theta}{\sqrt{(1+r^2)}}\frac{\sqrt{(1+r^2)}}{\sqrt{(1+2x^2)}}
\end{equation}

We observe that, in the fluid rest frame, the spatial volume element of $AdS_5$ at a constant time slice acquires an additional factor of $\frac{\sqrt{(1+r^2)}}{\sqrt{(1+2x^2)}}$. This factor at large $x\gg 1$, asymptotically approaches to $\gamma(\theta)$. We can interpret this factor as a relativistic Lorentz boost factor.
Now, substituting this measure factor into the integral \eqref{eq:partition_function}, we have 
\begin{equation}
    W =\ln Z=\int d\Omega_3\gamma(\theta)^4\int \frac{x^3 P(x)}{T_0}dx
\end{equation}
Using the expression for $\mathcal{F}(x)$ given in \eqref{fxdef} we can compute the pressure which is given by
\begin{equation}\label{pressure}
    P(x)=\sum_{q=1}^{\infty}\frac{3  \text{csch}^4(\beta  q)}{4 \pi ^5 \beta  q \left(\tanh \left(\frac{\beta  q}{2}\right)+\beta  q x^2\right)^5}
\end{equation}
We perform the $x$ integral and obtain the following,
\begin{equation}
    \ln Z= \sum_{q=1}^{\infty}\frac{4\text{csch}^6\left(\frac{\beta  q}{2}\right) \text{csch}(\beta  q)}{256 \pi ^5 \beta ^3 q^3 T_0}\int ~d\Omega_3\gamma(\theta)^4
\end{equation}
Finally, the integral on $S_3$ gives, 
\begin{equation}
    \ln Z=\frac{1}{(1-\omega_1^2)(1-\omega_2^2)}\sum_{q=1}^{\infty}\frac{4\text{csch}^6\left(\frac{\beta  q}{2}\right) \text{csch}(\beta  q)}{128 \pi ^3 \beta ^3 q^3 T_0}
\end{equation}
This is exactly the expression for the partition function we derived from the thermodynamic consideration in \eqref{hphi}, provided we identify $T_0$ with $\frac{1}{\beta}$.



The pressure given in \eqref{pressure}
in the large temperature limit becomes,
\begin{equation}
    \begin{split}
    P(x)&=\sum_{q=1}^{\infty}\frac{24}{\pi ^5 \beta ^{10} q^{10} \left(2 x^2+1\right)^5}+\mathcal{O}\left(\beta^{-9}\right)\\ & \approx \frac{8 \pi ^5 T(x)^{10}}{31185 \beta ^{10} {T_0}^{10}}
    \end{split}
\end{equation}

In the above expression, by substituting $T_0 = \frac{1}{\beta}$, we get the following high temperature limit of the equation of state of the fluid,
\begin{equation}
    P(x) = \frac{8 \pi ^5 }{31185}T(x)^{10}
\end{equation}




\subsection{Reproducing the total energy from bulk fluid dynamics}\label{toten}

 The total energy $E$ contained a bulk stress tensor is simply the charge associated with the conserved current obtained by contracting the stress tensor with the killing vector $\zeta^\alpha$ associated with time translations, i.e. with  the conserved current $J_\mu= T_{\alpha \mu} \zeta^\alpha$.  In global $AdS_5$ space 
$\zeta^t=1$, with all other $\zeta^\mu=0$. Consequently, $J_\mu=T_{\mu 0}$ and so 
$J^\mu = g^{\mu \alpha} T_{\alpha 0}$. The charge associated with this current equals the integral of $\sqrt{-g} g^{0 \alpha} T_{\alpha 0}$ over a (spatial) slice of constant time. As the metric of global $AdS_5$ is diagonal, it follows that 
\begin{equation}
\begin{split}
      E &= \int d^{10}x \sqrt{-g} g^{00}T_{00}\nonumber\\
      &= \Omega_5   \int d\Omega_3 dr r \times4\gamma^4(\theta) 
       {\mathcal F}(x)\\
      &= \Omega_5   \int_{S_3} \gamma^6(\theta) 
      \int_0^\infty  dx x \times 8{\mathcal F}(x)\\
      \end{split}
      \end{equation} 
\footnote{In going from the first to the second line we have used that $\int r dr = 2\gamma^2(\theta) \int x dx$ .}
The integral over $S^3$ is easily evaluated 
\begin{equation}\label{intovers}
\begin{split}
& \int_{S_3} \gamma^6(\theta) = 
 4 \pi^2 \int_{0}^{\frac{\pi}{2}
}\frac{ d\theta\,\sin{\theta}\cos{\theta}}{\left(1-\omega_1^2\sin^2\theta-\omega_2^2\cos^2\theta\right)^3}\\
 &= \pi^2 \left(\frac{1}{(1-\omega_1^2)(1-\omega_2^2)^2}+\frac{1}{(1-\omega_1^2)(1-\omega_2^2)^2}\right)
 \end{split}
 \end{equation} 
The integral over $x$ can also be computed
\begin{equation}\label{xint}
\begin{split}
 &\int_0^\infty  dx x {\mathcal F}(x)\\
&=\sum_{q=1}^{\infty}\int_0^{\infty}\frac{2 \Gamma(6)\sinh^2{\beta q}dx\,x^5}{\pi^5\left(2 \left(\cosh (\beta  q)+\beta  q x^2 \sinh (\beta  q)-1\right)\right)^6}\\
&=\sum_{q=1}^{\infty}\frac{\text{csch}^6\left(\frac{\beta  q}{2}\right) \text{csch}(\beta  q)}{128 \pi^5\beta ^3 q^3}\nonumber\\&=\frac{h_{\phi}(\beta)}{\pi^5 \beta}\\
\end{split}
\end{equation}

Putting it all together and using 
$\Omega_5= \pi^3$ we find 
\begin{equation}\label{efin}
\begin{split}
E&=\left(\frac{1}{(1-\omega_1^2)(1-\omega_2^2)^2}+\frac{1}{(1-\omega_1^2)(1-\omega_2^2)^2}\right)\sum_{q=1}^{\infty}\frac{\text{csch}^6\left(\frac{\beta  q}{2}\right) \text{csch}(\beta  q)}{16 \beta ^3 q^3}\\& = \frac{8 h_{\phi}(\beta)}{\beta}\left(\frac{1}{(1-\omega_1^2)(1-\omega_2^2)^2}+\frac{1}{(1-\omega_1^2)(1-\omega_2^2)^2}\right)\\
\end{split}
\end{equation} 
in perfect agreement with \eqref{thermo} with $h_\Delta(\beta) \rightarrow h_\phi(\beta)$.

\section{Stress tensor of conformally coupled free scalar on $S^1\times S^3$ }\label{s1s3}
In this section, we evaluate the one-point function of the stress tensor of a free conformally coupled scalar field on $S^3$. After presenting a formal result for this object, we specialize to two limits 
\begin{itemize}
\item The limit `$\omega_1 \to 1$, $\omega_2 \to 1$'.
\item The high temperature limit $(\beta \to 0)$.
\end{itemize}
We demonstrate that the one-point function of the stress tensor takes the fluid dynamical form in both these limits, and also 
verify that these two limits commute with each other.

\subsection{Two Point function on $S^3 \times$ time}

Consider a conformally coupled scalar on $S^3$ times time in Euclidean space. The metric $S^3 \times $ time can be rewritten as
\begin{align}\label{sthrewe}
    ds^2&=dt^2+d\Omega_3^2\nonumber\\
    &=\frac{dr^2}{r^2}+d\Omega_3^2\nonumber\\
    &=\frac{1}{r^2}\left(dr^2+r^2d\Omega_3^2\right)
\end{align}
(where we have defined $r=e^{t}$) demonstrating the (well known) fact that $S^3 \times$ time is Weyl equivalent to $R^4$. 

By construction, a conformally coupled scalar is Weyl invariant, 
where the scalar has Weyl weight one. This means that we get the same correlators from the `couples' 
$(\phi, g_{\mu\nu})$ and $(e^\sigma \phi, e^{2 \sigma }g_{\mu\nu})
$ \footnote{The first term in the couple is the field, while the second term is base space on which it propagates.}. 

We can apply this rule to the current situation. Let us parameterize points on $S^3$ by the unit vector ${\hat n}$ (which, itself, can be written as a function of any convenient angular coordinates on 
$S^3$). Consequently, our coordinates on $S^3 \times$ time are 
${\hat n}$ and $t$. Let us parameterize points on $R^4$ by the four-dimensional position vector ${\vec r}$ from the origin. We see that from \eqref{sthrewe} the map from $S^3 \times $ time to 
$R^4$ is given by 
\begin{equation}
{\vec r}= e^{t} {\hat n}
\end{equation}
Let us now apply the rule of the previous paragraph. Let $g_{\mu\nu}$ be the flat metric. Let $e^{\sigma} g_{\mu\nu}$ be 
the metric on $S^3 \times$ time. it follows from \eqref{sthrewe}
that $e^{\sigma}=\frac{1}{r} = e^{-t}$. We conclude that 
\begin{equation}\label{nds}
\begin{split}
&e^{-t} G_{S^3 \times R}\left(t, {\hat n}; t^{\prime}, {\hat n^{\prime}}\right) e^{-t^{\prime}}= \frac{{\mathcal N}}{|e^{t} {\hat n_1} -e^{t^{\prime} } {\hat n^{\prime}}|^2} \\
& \implies G_{S^3 \times R}\left(t, {\hat n}; t^{\prime}, {\hat n^{\prime}} \right)
= \frac{{\mathcal N} e^{t} e^{t^{\prime}} }{|e^{t} {\hat n} -e^{t^{\prime} } {\hat n^{\prime}}|^2} = \frac{{\mathcal N}  }{\left|e^{\frac{t-t^{\prime}}{2}} {\hat n} -e^{\frac{t^{\prime}-t}{2}  } {\hat n^{\prime}}\right|^2}
\end{split}
\end{equation} 
where 
\begin{equation}\label{nval}
{\mathcal N}=\frac{1}{4\pi^2}.
\end{equation}
\footnote{We have used the fact that the Greens for a massless scalar on $R^4$ = $\frac{{\mathcal N}}{r^2}$.}
In what follows we will use coordinate on $S^3$ s.t. 
\begin{equation}
{\hat n}= \left( \sin \theta \cos \phi_1, \sin\theta \sin \phi_1, \cos\theta \cos \phi_2,\cos\theta \sin\phi_2 \right) 
\end{equation}
In these coordinates the Greens function takes the form 
\begin{equation}\label{gfer}
G_{S^3 \times R}\left(t, {\hat n}; t^{\prime}, {\hat n^{\prime}}\right) = \frac{{\mathcal N}}{ e^{t-t^{\prime}}+e^{t^{\prime}-t}-2\left(\cos\theta\cos\theta'\cos (\phi_1 -\phi_1') +\sin\theta\sin\theta'\cos (\phi_2 - \phi_2 ') \right)}
\end{equation}

\subsection{Greens function on (twisted) $S^3 \times S^1$}

It is now a simple matter to evaluate the Greens function of our field on $S^3 \times S^1$, twisted with the angular velocities 
$\omega_1$ and $\omega_2$. As in the main text we use the method of images, 
the $q^{th}$ image is displaced by $\delta t= \beta q$, $\delta \phi_1=  -i q \beta \omega_1$, 
and $\delta \phi_2 = - i q \beta \omega_2$. It follows that Green's function is given by summing over images, and so is given by 
\begin{align}\label{imG}
  &  G_{S^3 \times S^1}\left(t, {\hat n}; t^{\prime}, {\hat n^{\prime}}\right)=
  G_{S^3 \times R}\left(t, {\hat n}; t^{\prime}, {\hat n^{\prime}}\right)\nonumber\\
   & +\sum_{q=-\infty; q\neq 0}^{\infty} \frac{\mathcal{N}}{\left(e^{t-t'+q\beta}+e^{t'-t-q\beta}-2\left(\cos\theta\cos\theta^{\prime}\cos(\phi_1-\phi_1'-iq\beta\omega_1)+\sin\theta\sin\theta^{\prime}\cos(\phi_2-\phi_2'-iq\beta\omega_2)\right)\right)}
\end{align}
where the first line of \eqref{imG} is the contribution of the 
$0^{th}$ image, and the second line corresponds to contributions from all the non-zero images.

We are interested in this Green's function in the limit that 
$\theta'=\theta$, $\phi_1'=\phi_1$, and $\phi_2'=\phi_2$. 
Inserting these coordinate choices, and also taking the `near extremal limit' $\omega_1\rightarrow 1$ and $\omega_2 \rightarrow 1$, we find that at leading order in $(1-\omega_1)$ and $(1-\omega_2)$, the two-point function is given by \footnote{It is easy to verify that the denominator of each of the terms in \eqref{om12} simply vanishes when $\omega_1=\omega_2=1$ and $\theta'=\theta$, $\phi_1'=\phi_1$, and $\phi_2'=\phi_2$. When $\omega_1$ and $\omega_2$ are not quite unity (but differ slightly from unity) the denominators do not quite vanish but are small.}
\begin{align}\label{om12}
     G_{S^3 \times S^1}\left(t, {\hat n}; t^{\prime}, {\hat n^{\prime}}\right) - G_{S^3 \times R}\left(t, {\hat n}; t^{\prime}, {\hat n^{\prime}}\right)&=\gamma^2(\theta)\sum_{q=1}^{\infty}\frac{2\mathcal{N}}{\beta  q \sinh (\beta  q) }
\end{align}
where $\gamma(\theta)$ was defined in \eqref{gammafac}. The factor of $\gamma^2(\theta)$ tells us that the propagator is taking the `fluid form'. It is interesting this happens when $\omega_1 \sim 1$ and $\omega_2 \sim 1$, even though we are at a temperature
of order unity (rather than at high temperatures, at which fluid dynamics traditionally emerges). 

As an aside, we can, once again set $\theta'=\theta$, $\phi_1'=\phi_1$, and $\phi_2'=\phi_2$ and evaluate 
the two-point function in the high-temperature limit $\beta\rightarrow 0$. We find 
\begin{align}\label{beta0last}
    \lim_{\beta \rightarrow 0}G_{S^3 \times S^1}\left(t, {\hat n}; t^{\prime}, {\hat n^{\prime}}\right) - G_{S^3 \times R}\left(t, {\hat n}; t^{\prime}, {\hat n^{\prime}}\right)&=2\sum_{q=1}^{\infty}\frac{\mathcal{N}}{q^2\beta^2\left(1-\omega^2 _2 \sin ^2(\theta )-\omega^2 _1 \cos ^2(\theta )\right)}\nonumber\\
    &=2\sum_{q=1}^{\infty}\frac{\mathcal{N}\gamma^2(\theta)}{q^2\beta^2}\nonumber\\
\end{align}
The factors of $\gamma^2(\theta)$ are expected here as 
$T \to \infty$ takes us to the fluid limit. It is not difficult to check that the $\beta \rightarrow 0$ limit of \eqref{om12} agrees with the $\omega_1 \to 1$ and $\omega_2 \rightarrow 1$ limit of \eqref{beta0last}. It follows, therefore, that \eqref{om12} is genuinely fluid dynamical in nature.  

\subsection{Derivative structures required to compute the stress tensor in the $\omega_i\rightarrow 1$ limit}\label{ne}
To compute the stress tensor, we follow the same strategy discussed in \ref{stressf} which is by taking appropriate derivatives on the two-point function  \eqref{imG}.  We first take the derivatives on the Greens function \eqref{imG} and then take the limit $t=t'$,
$\theta'=\theta$, $\phi_1'=\phi_1$, and $\phi_2'=\phi_2$. As discussed in \ref{chir},  we have renormalized our answers by subtracting the zero temperature answer (i.e. by dropping the 
$q=0$ contribution; recall that this is independent of the temperature and $\omega_i$). On performing the computation for the `time time' derivative we find that (to leading order of the  $\omega_i\rightarrow 1$)
\begin{align}\label{dertt}
   & \partial_{t}\partial_{t'}G_{S^3\times S^1}(t,\hat{n},t',\hat{n'})|_{(t,\hat{n})\rightarrow (t',\hat{n'})}\nonumber\\
   &\approx\sum_{q=1}^{\infty}-\frac{8\gamma^6(\theta)\text{csch}(\beta  q) \left(2-\beta  q \coth (\beta  q) \left(\frac{1}{2\gamma^2(\theta)}\right)\right)}{4 \pi ^2 \beta ^3 q^3 }\nonumber\\
   &\approx \sum_{q=1}^{\infty}-\frac{4\gamma^6(\theta)\text{csch}(\beta  q) }{ \pi ^2 \beta ^3 q^3 }
\end{align}
Note that, the leading order term goes as $\gamma^6(\theta)$. 
In a similar manner all (double) derivatives of the propagator 
can be evaluated; at leading order we find 
\begin{align}\label{derphi11}
    & \partial_{\phi_1}\partial_{\phi_1'}G_{S^3\times S^1}(t,\hat{n},t',\hat{n'})|_{(t', n') \to (t, n)}\approx\gamma^6(\theta) \sum_{q=1}^{\infty}\frac{4\omega^2_1\cos ^4(\theta ) \text{csch}(\beta  q)}{ \pi ^2 \beta ^3 q^3}
\end{align}
\begin{align}\label{derphi22}
&\partial_{\phi_2}\partial_{\phi_2'}G_{S^3\times S^1}(t,\hat{n},t',\hat{n'})|_{(t,\hat{n})\rightarrow (t',\hat{n'})}
    \approx\gamma^6(\theta) \sum_{q=1}^{\infty}\frac{4\omega^2_2\sin^4(\theta ) \text{csch}(\beta  q)}{ \pi ^2 \beta ^3 q^3}
\end{align}

\begin{align}\label{derph1ph2} \partial_{\phi_1}\partial_{\phi_2'}G_{S^3\times S^1}(t,\hat{n},t',\hat{n'})|_{(t,\hat{n})\rightarrow (t',\hat{n'})}&=\partial_{\phi_2}\partial_{\phi_1'}G_{S^3\times S^1}(t,\hat{n},t',\hat{n'})|_{(t,\hat{n})\rightarrow (t',\hat{n'})}\nonumber\\
&\approx \gamma^6(\theta)\sum_{q=1}^{\infty}\frac{4\omega_1\omega_2\sin ^2(2 \theta ) \text{csch}(\beta  q) }{ \pi ^2 (\beta  q)^3 } 
\end{align}
\begin{align}\label{dertph2}
\partial_{t}\partial_{\phi_2'}G_{S^3\times S^1}(t,\hat{n},t',\hat{n'})|_{(t,\hat{n})\rightarrow (t',\hat{n'})}&=\partial_{\phi_2}\partial_{t'}G_{S^3\times S^1}(t,\hat{n},t',\hat{n'})|_{(t,\hat{n})\rightarrow (t',\hat{n'})}\nonumber\\
&\approx\gamma^6(\theta)\sum_{q=1}^{\infty} \frac{i 4\omega_2\sin ^2(\theta ) \text{csch}(\beta  q) }{\pi ^2 \beta ^3 q^3 }
\end{align}

    \begin{align}\label{dertph1}
\partial_{t}\partial_{\phi_1'}G_{S^3\times S^1}(t,\hat{n},t',\hat{n'})|_{(t,\hat{n})\rightarrow (t',\hat{n'})} 
&=\partial_{\phi_1}\partial_{t'}G_{S^3\times S^1}(t,\hat{n},t',\hat{n'})|_{(t,\hat{n})\rightarrow (t',\hat{n'})} \nonumber\\
&\approx\gamma^6(\theta)\sum_{q=1}^{\infty} \frac{i 4\omega_1\cos ^2(\theta ) \text{csch}(\beta  q) }{\pi ^2 \beta ^3 q^3 }
\end{align}

\begin{align}
\partial_{\theta}\partial_{\phi_1'}G_{S^3\times S^1}(t,\hat{n},t',\hat{n'})|_{(t,\hat{n})\rightarrow (t',\hat{n'})}  =\partial_{\theta}\partial_{\phi_2'}G_{S^3\times S^1}(t,\hat{n},t',\hat{n'})|_{(t,\hat{n})\rightarrow (t',\hat{n'})} \approx O(\frac{1}{\gamma^2(\theta)})
\end{align}
\begin{align}\label{derth1th2}
&\partial_{\theta}\partial_{\theta'}G_{S^3\times S^1}(t,\hat{n},t',\hat{n'})|_{(t,\hat{n})\rightarrow (t',\hat{n'})}
    \approx\gamma^4(\theta) \sum_{q=1}^{\infty}\frac{\text{csch}^2(\beta  q)}{ \pi ^2 \beta ^2 q^2 }
\end{align}
\begin{align}\label{dertth} \partial_{t}\partial_{\theta'}G_{S^3\times S^1}(t,\hat{n},t',\hat{n'})|_{(t,\hat{n})\rightarrow (t',\hat{n'})}
&\approx O\left(\frac{1}{\gamma^2(\theta)}\right)
\end{align}
\subsection{The stress tensor}
The action of a conformally coupled free scalar on an arbitrary curved spacetime is given by
\begin{align}
    S&=\frac{1}{2}\int \sqrt{-g}d^4x\, \left(-g^{\mu\nu}\nabla_{\mu}\phi\nabla_{\nu}\phi-\xi R\phi^2\right), \quad \xi=\frac{1}{6}.
\end{align}
By varying the action with respect to metric one can determine the stress tensor for this theory to be (see, e.g. equation (3.190) of \cite{Birrell:1982ix})
\begin{align}\label{stress4}
    T_{\mu\nu}&=\nabla_{\mu}\phi\nabla_{\nu}\phi-\frac{1}{2}g_{\mu\nu}\nabla_{\alpha}\phi\nabla^{\alpha}\phi+\xi G_{\mu\nu}\phi^2+\xi \left(g_{\mu\nu}\nabla^{\alpha}\nabla_{\alpha}-\nabla_{\mu}\nabla_{\nu}\right)\phi^2
    \end{align}
Tracelessness of the stress tensor in \eqref{stress4} can be verified immediately (this verification works in any bulk dimension $D$ once we use the appropriate value of $\xi$, namely, $\xi=\frac{D-2}{4(D-1)})$
\begin{align}
    T^{\mu}_{~\mu}&=\left(1-\frac{D}{2}\right)\nabla_{\alpha}\phi\nabla^{\alpha}\phi+\xi R(1-\frac{D}{2})\phi^2+\xi\left(2(D-1)\nabla_{\alpha}\phi\nabla^{\alpha}\phi+2(D-1)\phi\nabla^2\phi\right)\nonumber\\
    &=\left(1-\frac{D}{2}\right)\phi\left(-\nabla^2+\xi R\right)\phi\nonumber\\
    &=0
\end{align}
In the last line, we have used the equation of motion.

 From the derivative structures presented in the \ref{ne}, it is easy to write the non-zero components of the stress tensor in the $\omega_i \rightarrow 1$ in the leading order.
 \begin{align}
     \langle T_{00}\rangle&=\gamma^6(\theta)\sum_{q=1}^{\infty}\frac{4\text{csch}(\beta  q) }{ \pi ^2 \beta ^3 q^3 }\\
     \langle T_{\phi_1\phi_1}\rangle&=\gamma^6(\theta) \sum_{q=1}^{\infty}\frac{4\omega^2_1\cos ^4(\theta ) \text{csch}(\beta  q)}{ \pi ^2 \beta ^3 q^3}\\
     \langle T_{\phi_2\phi_2}\rangle&=\gamma^6(\theta) \sum_{q=1}^{\infty}\frac{4\omega^2_2\sin^4(\theta ) \text{csch}(\beta  q)}{ \pi ^2 \beta ^3 q^3}\\
     \langle T_{\theta\theta}\rangle&=\gamma^4(\theta) \sum_{q=1}^{\infty}\frac{\text{csch}^2(\beta  q)}{ \pi ^2 \beta ^2 q^2 }\\
     \langle T_{0\phi_1}\rangle&=\gamma^6(\theta)\sum_{q=1}^{\infty} \frac{ 4\omega_1\cos ^2(\theta ) \text{csch}(\beta  q) }{\pi ^2 \beta ^3 q^3 }\\
     \langle T_{0\phi_2}\rangle&=\gamma^6(\theta)\sum_{q=1}^{\infty} \frac{4\omega_2\sin ^2(\theta ) \text{csch}(\beta  q) }{\pi ^2 \beta ^3 q^3 }\\
     \langle T_{\phi_1\phi_2}\rangle&=\gamma^6(\theta)\sum_{q=1}^{\infty}\frac{4\omega_1\omega_2\sin ^2(2 \theta ) \text{csch}(\beta  q) }{ \pi ^2 (\beta  q)^3 } 
 \end{align}
These results agree (at leading order in $\gamma$) with the `fluid form' stress tensor
\begin{equation}\label{fformst}
T_{\mu\nu}= \gamma^4(\theta)\left(4 u_{\mu}u_{\nu}+g_{\mu\nu}\right)\sum_{q=1}^{\infty}\frac{1}{\pi^2\beta^3 \sinh(q\beta)q^3 }
\end{equation}

In a similar manner, we can also compute the one-point function of the stress tensor in the small $\beta$ limit (but at finite values of $\gamma$). This is the limit in which we expect to recover the usual results of fluid dynamics. Performing the computation we find
\begin{align}\label{smallbeta}
     \lim_{\beta \rightarrow 0}   \langle T_{\mu\nu}\rangle&=\frac{\pi^2}{90\beta^4}\gamma^4(\theta)\left(g_{\mu\nu}+4u_{\mu}u_{\nu}\right), \quad u_{\mu}=\gamma(-1,0,\omega_1\cos^2\theta,\omega_2\sin^2\theta).
    \end{align}
Using $\sum_{q=1}^{\infty} \frac{1}{q^4}=\frac{\pi^4}{90}$, it is straightforward to verify that the small $\beta$ limit of \eqref{fformst} matches with \eqref{smallbeta}. 
\section{Spin-$1$ contribution to the bulk stress tensor}\label{spin1}

In this Appendix we compute the expectation value of the thermal stress tensor (at nonzero values of $\omega_i$) 
for a five dimensional massive vector field. We work with a bulk field dual to a boundary operator of dimension $\Delta$ (so that the mass of the bulk field is given by 
$M^2=(\Delta -1) (\Delta -3)$). The main qualitative take away from this Appendix is that this thermal stress tensor takes the form \eqref{bst} (for an appropriate choice of the function ${\mathcal F}(x)$), giving evidence for the conjecture made at the beginning of section \ref{cfbtst}.

We remind the reader that the stress tensor of massive vector field is given by
\begin{equation}\label{stresssp1}
    \begin{split}
        T_{\mu\nu}&= g^{\alpha\beta}F_{\mu\alpha}F_{\nu\beta} + M^2A_{\mu} A_{\nu} -g_{\mu\nu}\left( \frac{1}{4} F_{\alpha \beta}F^{\alpha\beta}+\frac{M^2}{2}A_{\alpha}A^{\alpha}\right)\\
    \end{split}
\end{equation}

\subsection{Stress tensor in thermal $AdS_5$}
The expectation value of \eqref{stresssp1} is given by
\begin{align}
    \langle T_{\mu\nu}
    \rangle&=\lim_{x_1\rightarrow x_2}\Big[\langle F_{\mu\alpha}(x_1)F_{\nu}^{~\alpha}(x_2)+M^2\langle A_{\mu}(x_1)A_{\nu}(x_2)\rangle
    -\frac{g_{\mu\nu}}{4}\left(\langle F_{\alpha\beta}(x_1)F^{\alpha\beta}(x_2)\rangle+\langle A_{\alpha}(x_1)A^{\alpha}(x_2)\right)\Big]\nonumber\\
    &=\lim_{x_1\rightarrow x_2}\Big[\partial^{x_1}_{\mu} \partial^{x_2}_{\nu} {\tilde {\Pi}^{~\alpha}_{\alpha}}(x_1,x_2)-\partial^{x_1}_{\mu} \partial^{x_2,\alpha} {\tilde \Pi}_{\alpha,\nu}(x_1,x_2)-\partial^{x_1}_{\alpha} \partial^{x_2}_{\nu} {\tilde \Pi}_{\mu}^{~\alpha}(x_1,x_2)+\partial^{x_1}_{\alpha} \partial^{x_2,\alpha} {\tilde \Pi}_{\mu,\nu}(x_1,x_2)\nonumber\\
    &-\frac{g_{\mu\nu}}{4}\Big(\partial^{x_1}_{\alpha} \partial^{x_2,\alpha} {\tilde \Pi}_{\beta}^{~\beta}(x_1,x_2)+\partial^{x_1}_{\beta} \partial^{x_2,\beta} {\tilde \Pi}_{\alpha}^{~\alpha}(x_1,x_2)-\partial^{x_1}_{\alpha} \partial^{x_2,\beta} {\tilde \Pi}_{\beta}^{~\alpha}(x_1,x_2)-\partial^{x_1}_{\beta} \partial^{x_2,\alpha} {\tilde \Pi}_{\alpha}^{~\beta}(x_1,x_2)\Big)\nonumber\\
    &+M^2 \tilde{\Pi}_{\mu,\nu}(x_1,x_2)-\frac{g_{\mu\nu}}{4}M^2\tilde{\Pi}_{\alpha}^{~\alpha}(x_1,x_2)\Big]
\end{align}
where ${\tilde \Pi}_{\mu\nu}(x_1,x_2)$ is massive vector propagator on $AdS_5 $ with the identifications mentioned in \eqref{bcs}.
Using the method of images, it follows that  ${\tilde \Pi}(x_1,x_2)$ is given by a sum over propagators on ordinary (non-thermal) Euclidean $AdS_5 $: 
\begin{equation}\label{tg1}
{\tilde \Pi}_{\mu,\nu}(x_1, x_2) = \sum_{q=-\infty}^\infty \Pi_{\mu,\nu}\left(x_1, R^{q}(x_2) \right)
\end{equation} 
where $R^q$ was defined under \eqref{tg}. The  expression of the bulk to bulk propagator of massive vector field in Euclidean (non-thermal) $AdS_5 $ is known \cite{Costa:2014kfa}
\begin{align}\label{2ptads}
  \langle A_{\mu}(x_1) A_{\nu}(x_2)\rangle= \Pi_{\mu,\nu}(x_1,x_2)=-\frac{\partial^2 u}{\partial x^{\mu}_1\partial x^{\nu}_2}g_0(u)+\frac{\partial u}{\partial x^{\mu}_1}\frac{\partial u}{\partial x^{\nu}_2}g_1(u)
\end{align}
In $AdS_5$, the expressions of $g_0(u)$ and $g_1(u)$ are given by \cite{Costa:2014kfa}
\begin{align}
    \begin{split}
        g_0(u)&=\frac{\mathcal{N}}{(u)^{\Delta}}\Big[(4-\Delta)\, _2F_1\left(\Delta ,\frac{1}{2} (-3+2 \Delta );-3+2 \Delta ;-\frac{1}{u}\right)\nonumber\\
        &-\frac{2+u}{u}\, _2F_1\left(\Delta +1,\frac{1}{2} (-3+2 \Delta );-3+2 \Delta ;-\frac{1}{u}\right)\Big]\\
        g_1(u)&=\frac{\mathcal{N}}{(u)^{\Delta}}\Big[\frac{2(2+u)(4-\Delta)}{u(u+4)}\,_2F_1\left(\Delta ,\frac{1}{2} (-3+2 \Delta );-3+2 \Delta ;-\frac{1}{u}\right)\nonumber\\
       & -2\frac{4+(u+2)^2}{u^2(u+4)}\,_2F_1\left(\Delta +1,\frac{1}{2} (-3+2 \Delta );-3+2 \Delta ;-\frac{1}{u}\right)\Big]
    \end{split}
\end{align}
where the normalization constant $\mathcal{N}$ for $AdS_5$ is given by
\begin{align}
  \mathcal{N}&=\frac{\Gamma(\Delta+1)}{2\pi^2(3-\Delta)(\Delta-1)\Gamma(\Delta-1)}.  
\end{align}
  We compute the two-point function of the field strengths by acting with appropriate derivatives first on the correlator \eqref{tg1} and then take the coincident limit. As we already discussed in \ref{stressf}, we renormalize the two-point function of the field strengths by removing the temperature independent ($q=0$) term. We present our results for the thermal two-point function of the field strengths in the leading order of $\omega_i\rightarrow 1$.
\begin{align}
  \lim_{x_1\rightarrow x_2}    \langle F_{0\lambda}(x_1)F_{0}^{~\lambda}(x_2)\rangle
  &=4x^4\gamma^4(\theta)f^{(1)}(x)
\end{align}
\begin{align}
  \lim_{x_1\rightarrow x_2}    \langle F_{0\lambda}(x_1)F_{\phi_1}^{~\lambda}(x_2)\rangle &=4x^4\gamma^4(\theta)\sin^2(\theta)f^{(1)}(x)
\end{align}
\begin{align}
  \lim_{x_1\rightarrow x_2}    \langle F_{0\lambda}(x_1)F_{\phi_2}^{~\lambda}(x_2)\rangle &  =4x^4\gamma^4(\theta)\cos^2(\theta)f^{(1)}(x)
\end{align}
\begin{align}
    \lim_{x_1\rightarrow x_2}    \langle F_{\phi_1\lambda}(x_1)F_{\phi_1}^{~\lambda}(x_2)\rangle &  =4x^4\gamma^4(\theta)\sin^4(\theta)f^{(1)}(x)
\end{align}
\begin{align}
    \lim_{x_1\rightarrow x_2}    \langle F_{\phi_2\lambda}(x_1)F_{\phi_2}^{~\lambda}(x_2)\rangle &  =4x^4\gamma^4(\theta)\cos^4(\theta)f^{(1)}(x)
\end{align}
\begin{align}
  \lim_{x_1\rightarrow x_2}    \langle F_{\phi_1\lambda}(x_1)F_{\phi_2}^{~\lambda}(x_2)\rangle &  =4x^4\gamma^4(\theta)\sin^2(\theta)\cos^2(\theta)f^{(1)}(x)
\end{align}
where $f^{(1)}(x)$ is given in terms of the chordal distance $u_q$ (see equation \eqref{fxdef})
\begin{align}
  f^{(1)}(x)=\sum_{q=1}^{\infty}\sinh ^2(\beta  q)\left(g_0''(u_q)+g_1'(u_q)\right) \left(3 \cosh (\beta  q)+\beta  q x^2 \sinh (\beta  q)\right)+2 g_0'(u_q)+2 g_1(u_q) 
\end{align}
We also compute the term which turns out to be subleading
\begin{align}
    g_{\mu\nu}\lim_{x_1\rightarrow x_2}  \langle F_{\alpha\beta}(x_1)F^{\alpha\beta}(x_2)\rangle&=O\left(\gamma^2(\theta)\right)
\end{align}
From the two-point functions of the field strengths, it is now easy to evaluate the stress tensor
\begin{equation}\label{mats1}
   \langle T_{\mu\nu}\rangle= 4\mathcal{F}_{\Delta}^{(1)}(x)\gamma^4(\theta)\left(
\begin{array}{ccccc}
 1 & 0 & 0 & \sin ^2(\theta ) & \cos ^2(\theta ) \\
 0 & 0 & 0 & 0 & 0 \\
 0 & 0 & 0 & 0 & 0 \\
 \sin ^2(\theta ) & 0 & 0 & \sin ^4(\theta ) & \sin ^2(\theta ) \cos ^2(\theta ) \\
 \cos ^2(\theta ) & 0 & 0 & \sin ^2(\theta ) \cos ^2(\theta ) & \cos ^4(\theta ) \\
\end{array}
\right)
\end{equation}
where $\mathcal{F}_{\Delta}^{(1)}(x)$ is given by
\begin{align}\label{f1d}
    \mathcal{F}_{\Delta}^{(1)}(x)&=\sum_{q=1}^{\infty}\Big[\sinh ^2(\beta  q)\left(g_0''(u_q)+g_1'(u_q)\right) \left(3 \cosh (\beta  q)+\beta  q x^2 \sinh (\beta  q)\right)\nonumber\\
    &+2 g_0'(u_q)+2 g_1(u_q)+M^2g_1(u_q)\Big]\nonumber\\
    &=\sum_{q=1}^{\infty}\Big[\sinh ^2(\beta  q)\left(g_0''(u_q)+g_1'(u_q)\right) \left(3 \cosh (\beta  q)+\beta  q x^2 \sinh (\beta  q)\right)\nonumber\\
    &+2 g_0'(u_q)+2 g_1(u_q)+(\Delta-1)(\Delta-3)g_1(u_q)\Big]
\end{align}
In the last line, we use $M^2=(\Delta-1)(\Delta-3)$.

The final form of the stress tensor is manifestly fluid in nature 
\begin{equation}\label{bstn}
T_{\mu\nu}=\mathcal{F}^{(1)}_{\Delta}(x)\gamma^4(\theta) 4 w_\mu w_\nu
\end{equation}
and so, once again, takes the form \eqref{bst} (with 
$\mathcal{F}^{(1)}_{\Delta}(x)$ replacing ${\mathcal F}(x)$).


\bibliographystyle{JHEP}
\bibliography{biblio.bib}

\providecommand{\href}[2]{#2}\begingroup\raggedright\begin{thebibliography}{10}

\bibitem{Kim:2023sig}
S.~Kim, S.~Kundu, E.~Lee, J.~Lee, S.~Minwalla and C.~Patel, \emph{{Grey Galaxies\textquoteright{} as an endpoint of the Kerr-AdS superradiant instability}}, \href{https://doi.org/10.1007/JHEP11(2023)024}{\emph{JHEP} {\bfseries 11} (2023) 024} [\href{https://arxiv.org/abs/2305.08922}{{\ttfamily 2305.08922}}].

\bibitem{Cardoso:2004hs}
V.~Cardoso and O.J.C.~Dias, \emph{{Small Kerr-anti-de Sitter black holes are unstable}}, \href{https://doi.org/10.1103/PhysRevD.70.084011}{\emph{Phys. Rev. D} {\bfseries 70} (2004) 084011} [\href{https://arxiv.org/abs/hep-th/0405006}{{\ttfamily hep-th/0405006}}].

\bibitem{Dias:2015rxy}
O.J.C.~Dias, J.E.~Santos and B.~Way, \emph{{Black holes with a single Killing vector field: black resonators}}, \href{https://doi.org/10.1007/JHEP12(2015)171}{\emph{JHEP} {\bfseries 12} (2015) 171} [\href{https://arxiv.org/abs/1505.04793}{{\ttfamily 1505.04793}}].

\bibitem{Ishii:2018oms}
T.~Ishii and K.~Murata, \emph{{Black resonators and geons in AdS5}}, \href{https://doi.org/10.1088/1361-6382/ab1d76}{\emph{Class. Quant. Grav.} {\bfseries 36} (2019) 125011} [\href{https://arxiv.org/abs/1810.11089}{{\ttfamily 1810.11089}}].

\bibitem{Chesler:2018txn}
P.M.~Chesler and D.A.~Lowe, \emph{{Nonlinear Evolution of the AdS$_4$ Superradiant Instability}}, \href{https://doi.org/10.1103/PhysRevLett.122.181101}{\emph{Phys. Rev. Lett.} {\bfseries 122} (2019) 181101} [\href{https://arxiv.org/abs/1801.09711}{{\ttfamily 1801.09711}}].

\bibitem{Ishii:2020muv}
T.~Ishii, K.~Murata, J.E.~Santos and B.~Way, \emph{{Superradiant instability of black resonators and geons}}, \href{https://doi.org/10.1007/JHEP07(2020)206}{\emph{JHEP} {\bfseries 07} (2020) 206} [\href{https://arxiv.org/abs/2005.01201}{{\ttfamily 2005.01201}}].

\bibitem{Chesler:2021ehz}
P.M.~Chesler, \emph{{Hairy black resonators and the AdS4 superradiant instability}}, \href{https://doi.org/10.1103/PhysRevD.105.024026}{\emph{Phys. Rev. D} {\bfseries 105} (2022) 024026} [\href{https://arxiv.org/abs/2109.06901}{{\ttfamily 2109.06901}}].

\bibitem{Kunduri:2006qa}
H.K.~Kunduri, J.~Lucietti and H.S.~Reall, \emph{{Gravitational perturbations of higher dimensional rotating black holes: Tensor perturbations}}, \href{https://doi.org/10.1103/PhysRevD.74.084021}{\emph{Phys. Rev. D} {\bfseries 74} (2006) 084021} [\href{https://arxiv.org/abs/hep-th/0606076}{{\ttfamily hep-th/0606076}}].

\bibitem{Murata:2008xr}
K.~Murata, \emph{{Instabilities of Kerr-AdS(5) x S**5 Spacetime}}, \href{https://doi.org/10.1143/PTP.121.1099}{\emph{Prog. Theor. Phys.} {\bfseries 121} 1099} [\href{https://arxiv.org/abs/0812.0718}{{\ttfamily 0812.0718}}].

\bibitem{Cardoso:2013pza}
V.~Cardoso, O.J.C.~Dias, G.S.~Hartnett, L.~Lehner and J.E.~Santos, \emph{{Holographic thermalization, quasinormal modes and superradiance in Kerr-AdS}}, \href{https://doi.org/10.1007/JHEP04(2014)183}{\emph{JHEP} {\bfseries 04} (2014) 183} [\href{https://arxiv.org/abs/1312.5323}{{\ttfamily 1312.5323}}].

\bibitem{Choi:2024xnv}
S.~Choi, D.~Jain, S.~Kim, V.~Krishna, E.~Lee, S.~Minwalla et~al., \emph{{Dual Dressed Black Holes as the end point of the Charged Superradiant instability in ${\cal N} = 4$ Yang Mills}},  \href{https://arxiv.org/abs/2409.18178}{{\ttfamily 2409.18178}}.

\bibitem{Hubeny:2002xn}
V.E.~Hubeny and M.~Rangamani, \emph{{Unstable horizons}}, \href{https://doi.org/10.1088/1126-6708/2002/05/027}{\emph{JHEP} {\bfseries 05} (2002) 027} [\href{https://arxiv.org/abs/hep-th/0202189}{{\ttfamily hep-th/0202189}}].

\bibitem{Dorn:2003au}
H.~Dorn, M.~Salizzoni and C.~Sieg, \emph{{On the propagator of a scalar field on AdS x S and on the BMN plane wave}}, \href{https://doi.org/10.1088/1126-6708/2005/02/047}{\emph{JHEP} {\bfseries 02} (2005) 047} [\href{https://arxiv.org/abs/hep-th/0307229}{{\ttfamily hep-th/0307229}}].

\bibitem{Dai:2009zg}
P.~Dai, R.-N.~Huang and W.~Siegel, \emph{{Covariant propagator in AdS(5) x S**5 superspace}}, \href{https://doi.org/10.1007/JHEP03(2010)001}{\emph{JHEP} {\bfseries 03} (2010) 001} [\href{https://arxiv.org/abs/0911.2211}{{\ttfamily 0911.2211}}].

\bibitem{Bhattacharyya:2007vs}
S.~Bhattacharyya, S.~Lahiri, R.~Loganayagam and S.~Minwalla, \emph{{Large rotating AdS black holes from fluid mechanics}}, \href{https://doi.org/10.1088/1126-6708/2008/09/054}{\emph{JHEP} {\bfseries 09} (2008) 054} [\href{https://arxiv.org/abs/0708.1770}{{\ttfamily 0708.1770}}].

\bibitem{Bhattacharyya:2008ji}
S.~Bhattacharyya, R.~Loganayagam, S.~Minwalla, S.~Nampuri, S.P.~Trivedi and S.R.~Wadia, \emph{{Forced Fluid Dynamics from Gravity}}, \href{https://doi.org/10.1088/1126-6708/2009/02/018}{\emph{JHEP} {\bfseries 02} (2009) 018} [\href{https://arxiv.org/abs/0806.0006}{{\ttfamily 0806.0006}}].

\bibitem{Bhattacharyya:2008mz}
S.~Bhattacharyya, R.~Loganayagam, I.~Mandal, S.~Minwalla and A.~Sharma, \emph{{Conformal Nonlinear Fluid Dynamics from Gravity in Arbitrary Dimensions}}, \href{https://doi.org/10.1088/1126-6708/2008/12/116}{\emph{JHEP} {\bfseries 12} (2008) 116} [\href{https://arxiv.org/abs/0809.4272}{{\ttfamily 0809.4272}}].

\bibitem{Bhattacharyya:2008xc}
S.~Bhattacharyya, V.E.~Hubeny, R.~Loganayagam, G.~Mandal, S.~Minwalla, T.~Morita et~al., \emph{{Local Fluid Dynamical Entropy from Gravity}}, \href{https://doi.org/10.1088/1126-6708/2008/06/055}{\emph{JHEP} {\bfseries 06} (2008) 055} [\href{https://arxiv.org/abs/0803.2526}{{\ttfamily 0803.2526}}].

\bibitem{Bhattacharyya_2008}
S.~Bhattacharyya, S.~Lahiri, R.~Loganayagam and S.~Minwalla, \emph{Large rotating ads black holes from fluid mechanics}, \href{https://doi.org/10.1088/1126-6708/2008/09/054}{\emph{Journal of High Energy Physics} {\bfseries 2008} (2008) 054–054}.

\bibitem{Emparan_2003}
R.~Emparan and R.C.~Myers, \emph{Instability of ultra-spinning black holes}, \href{https://doi.org/10.1088/1126-6708/2003/09/025}{\emph{Journal of High Energy Physics} {\bfseries 2003} (2003) 025–025}.

\bibitem{Dias:2010gk}
O.J.C.~Dias, P.~Figueras, R.~Monteiro and J.E.~Santos, \emph{{Ultraspinning instability of anti-de Sitter black holes}}, \href{https://doi.org/10.1007/JHEP12(2010)067}{\emph{JHEP} {\bfseries 12} (2010) 067} [\href{https://arxiv.org/abs/1011.0996}{{\ttfamily 1011.0996}}].

\bibitem{Caldarelli_2008}
M.M.~Caldarelli, R.~Emparan and M.J.~Rodríguez, \emph{Black rings in (anti)-de sitter space}, \href{https://doi.org/10.1088/1126-6708/2008/11/011}{\emph{Journal of High Energy Physics} {\bfseries 2008} (2008) 011–011}.

\bibitem{Emparan_2008}
R.~Emparan and H.S.~Reall, \emph{Black holes in higher dimensions}, \href{https://doi.org/10.12942/lrr-2008-6}{\emph{Living Reviews in Relativity} {\bfseries 11} (2008) }.

\bibitem{Hawking:1998kw}
S.W.~Hawking, C.J.~Hunter and M.~Taylor, \emph{{Rotation and the AdS / CFT correspondence}}, \href{https://doi.org/10.1103/PhysRevD.59.064005}{\emph{Phys. Rev. D} {\bfseries 59} (1999) 064005} [\href{https://arxiv.org/abs/hep-th/9811056}{{\ttfamily hep-th/9811056}}].

\bibitem{Caldarelli:1999xj}
M.M.~Caldarelli, G.~Cognola and D.~Klemm, \emph{{Thermodynamics of Kerr-Newman-AdS black holes and conformal field theories}}, \href{https://doi.org/10.1088/0264-9381/17/2/310}{\emph{Class. Quant. Grav.} {\bfseries 17} (2000) 399} [\href{https://arxiv.org/abs/hep-th/9908022}{{\ttfamily hep-th/9908022}}].

\bibitem{Gibbons:2004ai}
G.W.~Gibbons, M.J.~Perry and C.N.~Pope, \emph{{The First law of thermodynamics for Kerr-anti-de Sitter black holes}}, \href{https://doi.org/10.1088/0264-9381/22/9/002}{\emph{Class. Quant. Grav.} {\bfseries 22} (2005) 1503} [\href{https://arxiv.org/abs/hep-th/0408217}{{\ttfamily hep-th/0408217}}].

\bibitem{Dandekar:2017aiv}
Y.~Dandekar, S.~Kundu, S.~Mazumdar, S.~Minwalla, A.~Mishra and A.~Saha, \emph{{An Action for and Hydrodynamics from the improved Large D membrane}}, \href{https://doi.org/10.1007/JHEP09(2018)137}{\emph{JHEP} {\bfseries 09} (2018) 137} [\href{https://arxiv.org/abs/1712.09400}{{\ttfamily 1712.09400}}].

\bibitem{Bhattacharyya:2015dva}
S.~Bhattacharyya, A.~De, S.~Minwalla, R.~Mohan and A.~Saha, \emph{{A membrane paradigm at large D}}, \href{https://doi.org/10.1007/JHEP04(2016)076}{\emph{JHEP} {\bfseries 04} (2016) 076} [\href{https://arxiv.org/abs/1504.06613}{{\ttfamily 1504.06613}}].

\bibitem{Bhattacharyya:2015fdk}
S.~Bhattacharyya, M.~Mandlik, S.~Minwalla and S.~Thakur, \emph{{A Charged Membrane Paradigm at Large D}}, \href{https://doi.org/10.1007/JHEP04(2016)128}{\emph{JHEP} {\bfseries 04} (2016) 128} [\href{https://arxiv.org/abs/1511.03432}{{\ttfamily 1511.03432}}].

\bibitem{Dandekar:2016fvw}
Y.~Dandekar, A.~De, S.~Mazumdar, S.~Minwalla and A.~Saha, \emph{{The large D black hole Membrane Paradigm at first subleading order}}, \href{https://doi.org/10.1007/JHEP12(2016)113}{\emph{JHEP} {\bfseries 12} (2016) 113} [\href{https://arxiv.org/abs/1607.06475}{{\ttfamily 1607.06475}}].

\bibitem{Bhattacharyya:2016nhn}
S.~Bhattacharyya, A.K.~Mandal, M.~Mandlik, U.~Mehta, S.~Minwalla, U.~Sharma et~al., \emph{{Currents and Radiation from the large $D$ Black Hole Membrane}}, \href{https://doi.org/10.1007/JHEP05(2017)098}{\emph{JHEP} {\bfseries 05} (2017) 098} [\href{https://arxiv.org/abs/1611.09310}{{\ttfamily 1611.09310}}].

\bibitem{Banerjee:2012iz}
N.~Banerjee, J.~Bhattacharya, S.~Bhattacharyya, S.~Jain, S.~Minwalla and T.~Sharma, \emph{{Constraints on Fluid Dynamics from Equilibrium Partition Functions}}, \href{https://doi.org/10.1007/JHEP09(2012)046}{\emph{JHEP} {\bfseries 09} (2012) 046} [\href{https://arxiv.org/abs/1203.3544}{{\ttfamily 1203.3544}}].

\bibitem{Jensen:2012jh}
K.~Jensen, M.~Kaminski, P.~Kovtun, R.~Meyer, A.~Ritz and A.~Yarom, \emph{{Towards hydrodynamics without an entropy current}}, \href{https://doi.org/10.1103/PhysRevLett.109.101601}{\emph{Phys. Rev. Lett.} {\bfseries 109} (2012) 101601} [\href{https://arxiv.org/abs/1203.3556}{{\ttfamily 1203.3556}}].

\bibitem{deHaro:2000vlm}
S.~de~Haro, S.N.~Solodukhin and K.~Skenderis, \emph{{Holographic reconstruction of space-time and renormalization in the AdS / CFT correspondence}}, \href{https://doi.org/10.1007/s002200100381}{\emph{Commun. Math. Phys.} {\bfseries 217} (2001) 595} [\href{https://arxiv.org/abs/hep-th/0002230}{{\ttfamily hep-th/0002230}}].

\bibitem{Dias:2015pda}
O.J.C.~Dias, J.E.~Santos and B.~Way, \emph{{Lumpy AdS$_{5}$\texttimes{} S$^{5}$ black holes and black belts}}, \href{https://doi.org/10.1007/JHEP04(2015)060}{\emph{JHEP} {\bfseries 04} (2015) 060} [\href{https://arxiv.org/abs/1501.06574}{{\ttfamily 1501.06574}}].

\bibitem{Emparan_2006}
R.~Emparan and H.S.~Reall, \emph{Black rings}, \href{https://doi.org/10.1088/0264-9381/23/20/r01}{\emph{Classical and Quantum Gravity} {\bfseries 23} (2006) R169–R197}.

\bibitem{Kinney:2005ej}
J.~Kinney, J.M.~Maldacena, S.~Minwalla and S.~Raju, \emph{{An Index for 4 dimensional super conformal theories}}, \href{https://doi.org/10.1007/s00220-007-0258-7}{\emph{Commun. Math. Phys.} {\bfseries 275} (2007) 209} [\href{https://arxiv.org/abs/hep-th/0510251}{{\ttfamily hep-th/0510251}}].

\bibitem{Gunaydin:1984fk}
M.~Gunaydin and N.~Marcus, \emph{{The Spectrum of the s**5 Compactification of the Chiral N=2, D=10 Supergravity and the Unitary Supermultiplets of U(2, 2/4)}}, \href{https://doi.org/10.1088/0264-9381/2/2/001}{\emph{Class. Quant. Grav.} {\bfseries 2} (1985) L11}.

\bibitem{Mack:1975je}
G.~Mack, \emph{{All unitary ray representations of the conformal group SU(2,2) with positive energy}}, \href{https://doi.org/10.1007/BF01613145}{\emph{Commun. Math. Phys.} {\bfseries 55} (1977) 1}.

\bibitem{Dolan:2005wy}
F.A.~Dolan, \emph{{Character formulae and partition functions in higher dimensional conformal field theory}}, \href{https://doi.org/10.1063/1.2196241}{\emph{J. Math. Phys.} {\bfseries 47} (2006) 062303} [\href{https://arxiv.org/abs/hep-th/0508031}{{\ttfamily hep-th/0508031}}].

\bibitem{Alday:2020eua}
L.F.~Alday, M.~Kologlu and A.~Zhiboedov, \emph{{Holographic correlators at finite temperature}}, \href{https://doi.org/10.1007/JHEP06(2021)082}{\emph{JHEP} {\bfseries 06} (2021) 082} [\href{https://arxiv.org/abs/2009.10062}{{\ttfamily 2009.10062}}].

\bibitem{Abramowitz}
M.~Abramowitz and I.~Stegun, \emph{{Handbook of Mathematical Functions with Formulas, Graphs, and Mathematical Tables}}, Dover Publications (1964).

\bibitem{Birrell:1982ix}
N.D.~Birrell and P.C.W.~Davies, \emph{{Quantum Fields in Curved Space}}, Cambridge Monographs on Mathematical Physics, Cambridge University Press, Cambridge, UK (1982), \href{https://doi.org/10.1017/CBO9780511622632}{10.1017/CBO9780511622632}.

\bibitem{Costa:2014kfa}
M.S.~Costa, V.~Gon\c{c}alves and J.a.~Penedones, \emph{{Spinning AdS Propagators}}, \href{https://doi.org/10.1007/JHEP09(2014)064}{\emph{JHEP} {\bfseries 09} (2014) 064} [\href{https://arxiv.org/abs/1404.5625}{{\ttfamily 1404.5625}}].

\end{thebibliography}\endgroup

\end{document}